\documentclass[11pt]{article}
\pdfoutput=1 
\usepackage{./jheppub}

\usepackage[T1]{fontenc} 

\usepackage{graphicx}
\usepackage{amsfonts,amsmath,amssymb}
\usepackage{verbatim}
\usepackage{array}
\usepackage{mathrsfs}
\usepackage{mathtools}
\usepackage{yfonts}
\usepackage{dsfont}
\usepackage{bbm}
\usepackage{colonequals}
\usepackage{amscd}
\usepackage{slashed}
\usepackage {extarrows}

\usepackage{float}
\usepackage{afterpage}

\usepackage{subcaption}

\usepackage{wrapfig}

\usepackage{suffix,mathtools,cancel,bbm}

\usepackage{multirow}

\usepackage{enumerate}

	\usepackage{tikz-cd}

\newcommand{\bn}{\begin{enumerate}}
\newcommand{\en}{\end{enumerate}}

\def\bZ{\mathbb{Z}}

\def\d{\delta}

\def\gcd{\mathop{\mathrm{gcd}}}

\def \d{\partial}

\newcommand{\beq}{\begin{equation}}
\newcommand{\eeq}{\end{equation}}

\newcommand\nn{\nonumber}

\newcommand{\cA}{\mathcal{A}}
\newcommand{\cB}{\mathcal{B}}
\newcommand{\cC}{\mathcal{C}}
\newcommand{\cD}{\mathcal{D}}
\newcommand{\cE}{\mathcal{E}}
\newcommand{\cF}{\mathcal{F}}
\newcommand{\cG}{\mathcal{G}}
\newcommand{\cH}{\mathcal{H}}
\newcommand{\cI}{\mathcal{I}}

\newcommand{\cK}{\mathcal{K}}
\newcommand{\cL}{\mathcal{L}}
\newcommand{\cM}{\mathcal{M}}
\newcommand{\cN}{\mathcal{N}}
\newcommand{\cO}{\mathcal{O}}

\newcommand{\cR}{\mathcal{R}}

\newcommand{\cU}{\mathcal{U}}
\newcommand{\cV}{\mathcal{V}}
\newcommand{\cW}{\mathcal{W}}
\newcommand{\cX}{\mathcal{X}}

\newcommand{\cZ}{\mathcal{Z}}

\numberwithin{equation}{section}

\def\bea{\begin{eqnarray}}
\def\eea{\end{eqnarray}}

\DeclarePairedDelimiterX\MeijerM[3]{\lparen}{\rparen}%
{\begin{smallmatrix}#1 \\ #2\end{smallmatrix}\delimsize\vert\,#3}

\newcommand\MeijerG[8][]{%
  G^{\,#2,#3}_{#4,#5}\MeijerM[#1]{#6}{#7}{#8}}

\WithSuffix\newcommand\MeijerG*[7]{%
  G^{\,#1,#2}_{#3,#4}\MeijerM*{#5}{#6}{#7}}

\def\bR{\mathbb{R}}
\def\bZ{\mathbb{Z}}
\def\cH{\mathcal{H}}

\def\cN{\mathcal{N}}

\def \beg#1{\begin{#1}} 

\def \bea{\beg{eqnarray}}
\def \eea{\end{eqnarray}}
\def \ee{\end{equation}}

\def \restr#1#2{{\left.\kern-\nulldelimiterspace#1\vphantom{\big|}\right|_{#2}}}

\def \nn{\nonumber}

\def \d{\partial}

\def \sfA{\mathsf{A}}

\def \sfF{\mathsf{F}}

\def \bR{\mathbb{R}}

\def \bZ{\mathbb{Z}}

\usepackage{colortbl}
\definecolor{mygray}{gray}{0.93}

\usepackage{sectsty}
\allsectionsfont{\boldmath}

\usepackage{comment}

\title{\boldmath Non-Invertible Symmetries, Brane Dynamics, and Tachyon Condensation}

\author[a]{Ibrahima Bah,} 
\author[a]{Enoch Leung,} 
\author[a]{and Thomas Waddleton} 
\affiliation[a]{William H.~Miller III Department of Physics and Astronomy, Johns Hopkins University, 3400 North Charles Street, Baltimore, MD 21218, U.S.A.}

\emailAdd{iboubah@jhu.edu, yleung5@jhu.edu, twaddle1@jhu.edu}

\abstract{We study the Symmetry Topological Field Theory in holography associated with 4d $\mathcal{N}=1$ Super Yang-Mills theory with gauge algebra $\mathfrak{su}(M)$. From this, all the bulk symmetry operators are computed and matched to various D-brane configurations. The fusion algebra of the operators emerges from brane dynamics. In particular, we show that the symmetry operators are purely determined from the center-of-mass modes of the branes. We identify the TQFT fusion coefficients with the relative motion of the branes. We also establish the origin of condensation defects, arising from fusion of non-invertible operators, as the consequence of tachyon condensation in brane-anti-brane pairs.}

\usepackage{etoolbox}
\appto\appendix{\addtocontents{toc}{\protect\setcounter{tocdepth}{1}}}

\appto\listoffigures{\addtocontents{lof}{\protect\setcounter{tocdepth}{1}}}
\appto\listoftables{\addtocontents{lot}{\protect\setcounter{tocdepth}{1}}}

\begin{document} 

\setcounter{tocdepth}{2}

\maketitle
\flushbottom



\section{Introduction} \label{sec:introduction}

In recent times, the study of Quantum Field Theory (QFT) has localized to its {\it surprisingly} rich kinematic sector.  This follows from the novel perspective, due to \cite{Gaiotto:2014kfa}, that such aspects of QFT can be characterized by various topological structures acting on them.  It has led to several generalizations of the notion of symmetry such as higher-form symmetries, higher groups, non-invertible symmetries and inevitably to categorical symmetries.  In all of these cases there continues to be rapid development on the nature of topological symmetries, and the many ways they provide a fundamental scaffolding of QFT.  {\it A priori} these emerging symmetries seem exotic, however in the last two years they have been discovered in many ``bread-and-butter'' QFTs such as Yang-Mills theories in four dimensions, and even for the Standard Model of particle physics.  

Two central themes of research now are to understand the scope of topological symmetries, broadly, and the mathematical framework needed to characterize them.  It has come to focus that a useful way to describe topological symmetries associated to a QFT in a $d$-dimensional space, $M_d$, is to consider a topological field theory on a $(d+1)$-dimensional space $X_{d+1}$ with $\partial X_{d+1} = M_d$ \cite{Gaiotto:2014kfa,Gaiotto:2020iye,Freed:2022qnc}.  Such TQFTs are referred to as Symmetry Topological Field Theories (SymTFTs) \cite{Apruzzi:2021nmk}; they naturally emerge via anomaly inflow methods for QFTs engineered from string theory and holography \cite{Bah:2019rgq,Bah:2020uev,Bah:2020jas,Apruzzi:2021nmk}.  From this point of view, string theory provides an important avenue for studying topological symmetries and their role in the dynamics of QFTs.

An important result appeared recently in the context of non-invertible symmetries \cite{Apruzzi:2022rei,GarciaEtxebarria:2022vzq}.  In these works, it was demonstrated how to realize them in string theory and holography from supergravity and brane dynamics.  Such developments seeded many other interesting results where non-invertible symmetries are constructed and analyzed from various D-brane setups \cite{Apruzzi:2022rei,GarciaEtxebarria:2022vzq,Heckman:2022muc,Heckman:2022xgu,Etheredge:2023ler,Apruzzi:2023uma,Cvetic:2023plv,Dierigl:2023jdp,Bashmakov:2022uek,Bashmakov:2022jtl}.

In this paper, we will explore the SymTFT associated to 4d $\mathcal{N}=1$ Super Yang-Mills (SYM) theory with gauge algebra $\mathfrak{su}(M)$.  This QFT emerges in the IR limit of a stack of $N$ $D3$ branes probing the conifold in IIB string theory, with $M$ units of $G_3$ flux where $M$ is a divisor of $N$.  This is famously described by the Klebanov-Tseytlin \cite{Klebanov:2000nc} and Klebanov-Strassler \cite{Klebanov:2000hb} (KT-KS) backgrounds in IIB supergravity; these are holographically dual to cascading field theories.  A relevant review for these backgrounds is provided in the Appendix \ref{sec:app_SymTFT_dualization}. Our main interest is to obtain the full SymTFT as initiated in \cite{Apruzzi:2022rei}, and then construct all the symmetry operators and their fusion rules. We will identify all the brane configurations that are associated to the operators, and then characterize their fusion rules from brane dynamics.  Another objective will be to sharply understand what data captures the topological operators and their associated physics derived from the SymTFT.



We start with a brief discussion of our main results.   First we complete the derivation of the SymTFT of the KT-KS background discussed in \cite{Apruzzi:2022rei}.  Here we identify all the couplings relevant to the TQFT and its associated discrete gauge symmetries.  In particular we emphasize how continuum fields in supergravity can be used to construct and describe the discrete gauge fields in a TQFT.  By using the (classical) Gauss Law constraints, we derive all the topological operators which generate a bulk gauge symmetry from the SymTFT \cite{Apruzzi:2022rei,Belov:2004ht}.  We then match each symmetry generator with a specific D$p$-brane configuration by considering the reduction of the Wess-Zumino action for the branes. The details are the focus of Sections \ref{sec:symtft}, \ref{sec:N_U_operators}, and \ref{sec:O_V_operators}, but a quick summary can be found in Table \ref{tab:brane_config}. 

\renewcommand{\arraystretch}{1.5}
\begin{table}[t!]
\begin{center}
\begin{tabular}{|| c | c | c ||}
\hline
Symmetry Action & Symmetry Generator & Brane Configuration\\
\hline\hline
$\bZ_{2M}^{(0)}$ & $\cN_1(M^3)$ & D5 wrapping $S^3\subset T^{1,1}$\\
$\widehat{\bZ}_{2M}^{(2)}$ & $\cV_1(\gamma^1)$ & D1 wrapping Hopf fiber $S^1\subset S^3$\\
$\bZ_{M}^{(1)}$ & $\cO_1(W^2)$ & D3-D1 bound state wrapping $S^2\subset T^{1,1}$\\
$\widehat{\bZ}_{M}^{(1)}$ & $\cU_1(\Sigma^2)$ & D3 wrapping Hopf base $\widetilde{S^2}\subset S^3$\\
\hline
\end{tabular}
\caption{\label{tab:brane_config} Symmetry generators/defects and their D-brane realizations.\protect\footnotemark}
\end{center}
\end{table}
\footnotetext{We assume a Euclidean spacetime throughout this paper, and use the terminology ``generator'', ``operator'', and ``defect'' interchangeably.}

Once we have the symmetry generators, the next question is to understand their fusion algebra. When considering the parallel fusion of two symmetry generators, the corresponding D$p$-branes become coincident and are subject to non-Abelian dynamics involving the worldvolume gauge fields. In this work we demonstrate that the non-Abelian effects are excluded from the topological data of the symmetry defects; only the Abelian modes in the D$p$-brane worldvolume theory participate in the fusion. It is kinematical information, rather than dynamical, captured in the fusion of the corresponding operators. Specifically, the center-of-mass mode for a given stack of branes yields the symmetry operator. For two stacks brought together, the resulting operator is associated to the {\it total} center-of-mass mode. The modes describing any {\it relative} motion between the two stacks decouple from the defects but survive as nontrivial fusion coefficients in the form of decoupled TQFTs.

Another interesting feature we point out, elaborated upon in the main text, is the non-invertibility of operators can be attributed to induced brane charges due to bulk-brane dynamics. For example, the operator $\cO_1(W^2)$ presented in Table \ref{tab:brane_config} takes the explicit form,
\beq
\cO_1(W^2) = \int\cD\lambda_1\cD\varphi\exp\left(2\pi i\int_{W^2}\hat{c}_2 + \lambda_1\cA_1 + \varphi\cB_2 - M\varphi d\lambda_1\right) \, ,
\eeq
where $\lambda_1$ and $\varphi$ are modes localized on the D3-D1 bound state. The first term in this action encodes all the information contained in the D3- and D1-branes themselves. The second and third terms describe induced charges present on the D3-D1 worldvolume, while the final term is a standard BF-term describing a $\bZ_M\times\bZ_M$ gauge theory. Figure \ref{fig:induced_charge} provides a picture to keep in mind during this discussion. As will be seen later in the paper, the induced charges are associated with D1-branes and F1-strings in the worldvolume. 

\renewcommand{\arraystretch}{1.5}
\begin{table}[t!]
    \begin{center}
        \begin{tabular}{|| c | c ||}
            \hline
            Brane Fusion & Defect Fusion\\
            \hline\hline
            D5 $\otimes$ D5 & $\cN_p(M^3) \otimes \cN_{p'}(M^3) = \cA^{M,pp'(p+p')}\cN_{p+p'}(M^3)$\\
            D5 $\otimes$ $\overline{{\rm D}5}$ & $\cN_p(M^3) \otimes \cN_{-p}(M^3) = \cC(M^3)$\\
            D3 $\otimes$ D3 & $\cU_q(\Sigma^2) \otimes \cU_{q'}(\Sigma^2) = \cU_{q+q'}(\Sigma^2)$\\
            D3 $\otimes$ D5 & $\cU_q(\Sigma^2) \otimes \cN_p(M^3) = \cN_p(M^3)$\\
            D3 $\otimes$ \big(\,$\sum$\,D3\big) & $\cU_q(\Sigma^2) \otimes \cC(M^3) = \cC(M^3)$\\
            D5 $\otimes$ \big(\,$\sum$\,D3\big) & $\cN_p(M^3) \otimes \cC(M^3) = (\mathcal{Z}_M)_0\, \cN_p(M^3)$\\
            \big(\,$\sum$\,D3\big) $\otimes$ \big(\,$\sum$\,D3\big) & $\cC(M^3) \otimes \cC(M^3) = (\mathcal{Z}_M)_0\, \cC(M^3)$\\
            (D3-D1) $\otimes$ (D3-D1) & $\cO_q(W^2)\otimes\cO_{q'}(W^2) = \cX^{M,qq'(q+q')^{-1}}\cO_{q+q'}(W^2)$\\
            (D3-D1) $\otimes$ $(\overline{{\rm D}3\text{-}{\rm D}1})$ & $\cO_q(W^2)\otimes\cO_{-q}(W^2) = \widetilde{\cC}(W^2)$\\
            D1 $\otimes$ D1 & $\cV_p(\gamma^1)\otimes\cV_{p'}(\gamma^1) = \cV_{p+p'}(\gamma^1)$\\
            D1 $\otimes$ (D3-D1) & $\cV_p(\gamma^1) \otimes \cO_q(W^2) = \cO_q(W^2)$\\
            D3 $\otimes$ (D3-D1) & $\cU_q(\Sigma^2) \otimes \cO_{q'}(W^2) = \cO_{q'}(W^2)$\\
            D1 $\otimes$ \big(\,$\sum$\,(D1-F1)\big) & $\cV_p(\gamma^1) \otimes \widetilde{\cC}(W^2) = \widetilde{\cC}(W^2)$\\
            D3 $\otimes$ \big(\,$\sum$\,(D1-F1)\big) & $\cU_q(\Sigma^2) \otimes \widetilde{\cC}(W^2) = \widetilde{\cC}(W^2)$\\
            (D3-D1) $\otimes$  \big(\,$\sum$\,(D1-F1)\big) & $\cO_q(W^2)\otimes\widetilde{\cC}(W^2) = \cX^{M,-1}\cX^{M,-1}\cO_{q}(W^2)$\\
            \big(\,$\sum$\,(D1-F1)\big) $\otimes$  \big(\,$\sum$\,(D1-F1)\big) & $\widetilde{\cC}(W^2)\otimes\widetilde{\cC}(W^2) = \cX^{M,-1}\cX^{M,-1}\widetilde{\cC}(W^2)$\\
            \hline
        \end{tabular}
        \caption{\label{tab:fusion_rules} Fusion rules for the symmetry defects. The left column indicates the schematic fusion in terms of the associated D$p$-branes, while the right column shows the explicit fusion rules. The coefficients $\cA^{M,p}, \cX^{M,q}$, and $(\cZ_M)_0$ denote decoupled TQFTs that are elaborated upon in the main text.}
    \end{center}
\end{table}

In addition to the above fusion, we study further the condensation defect present in many recent examples such as \cite{GarciaEtxebarria:2022vzq,Apruzzi:2022rei,Choi:2022fgx,Choi:2022jqy,Kaidi:2021xfk,Heckman:2022muc}. We find that these defects are described naturally in string theory via tachyon condensation. When a D$p$- and a $\overline{{\rm D}p}$-brane are brought together, these branes will not completely annihilate but leave behind a solitonic D$(p-2)$-brane in the original worldvolume, determined by a choice of tachyon profile. By summing over all possible solitonic D$(p-2)$-branes on the original worldvolume for a specific tachyon profile, we can reproduce exactly the condensation defect that arise when considering the operator fusion. There is a parallel discussion here as in the other cases of fusion; we can see the condensate being given by the center-of-mass mode for the original brane-anti-brane worldvolume. Including the appropriate condensates, we work out the full fusion algebra for the symmetry defects in the KT-KS background. The full list of the bulk fusion rules can be found in Table \ref{tab:fusion_rules}. We also discuss the realization of these fusion rules in the boundary field theory in terms of allowed boundary conditions on the bulk supergravity fields.

This paper is organized as follows. In Section \ref{sec:symtft}, we construct the SymTFT for the KT-KS background, and use the associated Gauss Law constraints to derive the {\it gauge} symmetry generators in this setup. In Section \ref{sec:brane_inflow}, we switch gears and discuss how symmetry generators in a generic string theory background can also be realized using D-branes. In Sections \ref{sec:N_U_operators} and \ref{sec:O_V_operators}, we study in detail the fusion rules for the $(\cN,\cU)$ and $(\cO,\cV)$ defect pairs respectively. These two pairs of defects share a strikingly similar fusion algebra, e.g.~the former undergoes non-invertible fusion to produce a condensation defect containing the latter. We offer an interpretation of these fusion rules in terms of the underlying brane kinematics and dynamics. In Section \ref{sec:cross_fusions}, we derive the mixed fusion rules between all the symmetry defects, as well as their braiding relations. In Section \ref{sec:boundary_conditions}, we analyze the role of the choice of boundary conditions in realizing the bulk {\it gauge} symmetry generators as {\it global} symmetry generators in the boundary field theory. Last but not least, we follow up by a discussion in Section \ref{sec:discussion}.



\section{Bulk Supergravity Gauge Symmetries}\label{sec:symtft}

\subsection{Symmetry Topological Field Theory}

In holographic constructions, one can use anomaly inflow methods to obtain a Symmetry Topological Field Theory (SymTFT) that encodes the symmetry data of a lower-dimensional QFT (e.g.~\cite{Bah:2018gwc,Bah:2019rgq,Bah:2020jas,Bah:2020uev,Apruzzi:2021nmk}). In the case of supergravity solutions in the low energy limit of Type IIB string theory, the SymTFT can be obtained by starting with the anomaly polynomial of the ten-dimensional supergravity theory and dimensionally reducing over the internal space. This anomaly polynomial was derived in \cite{Bah:2020jas}. The resulting topological theory, once augmented with kinetic terms for the relevant fields, is the SymTFT of the boundary field theory.

In the present work, we expand upon the SymTFT of the KT-KS solution as derived in \cite{Apruzzi:2022rei}, and particularly study the effect of additional terms that were previously neglected. A review of the KT-KS solution can be found in Appendix \ref{sec:app_SymTFT_dualization}. We first start with the 11-form anomaly polynomial in Type IIB string theory,\footnote{Unless otherwise specified, we choose the normalization of all field strengths appearing in this work to be such that they have integral periods, and we also work in dimensionless units, i.e.~$2\pi\alpha'=1$ and $g_\text{s}=1$, to avoid cluttering of notation. For the same reason, we will often suppress the wedge/cup products between differential forms/cochains.}
\beq\label{eq:IIB_anomaly_polynomial}
\cI_{11} = \frac{1}{2} \, \cG_5 \wedge d\cG_5 + \cG_5 \wedge H_3 \wedge G_3 \, ,
\eeq
where $G_3$ and $H_3$ are the field strengths for the RR 2-form and NSNS 2-form potentials respectively, and $\cG_5$ is defined such that $G_5 = (1+\star_{10})\cG_5$ is the self-dual RR 5-form. To dimensionally reduce the above polynomial, we perform an expansion of the supergravity field strengths onto the $U(1)_R$\,-equivariant cohomology classes defined in \eqref{eq:omega23} and \eqref{eq:Omega5}, namely,
\beq\label{eq:RR_field_strength_expansion}
H_3 = h_3,\qquad G_3 = M\omega_3 + g_1\omega_2 + g_3 \,,\qquad  \cG_5 = N\Omega_5 + g_2\omega_3\,.
\eeq
The fluctuating fields $h_3, g_1,g_2,$ and $g_3$ are defined on the external space $W^5$, while the coefficients $M$ and $N$ of $\omega_3$ and $\Omega_5$ reflect the nontrivial fluxes threading $S^3$ and $T^{1,1}$ respectively in the KT-KS setup. 

Consistency of the IIB Bianchi identities for the RR field strengths, which in this background take the form $dH_3 = 0, dG_3 = 0$, and $dG_5 = -H_3\wedge G_3$, impose constraints on the external field strengths written above. These extra constraints reduce to the 5d Bianchi identities
\beq\label{eq:Bianchi_ids}
dh_3 = dg_3 = d\sfF_2 = 0\,,\qquad dg_1 = 2M\sfF_2,\qquad dg_2 = -Mh_3\,.
\eeq
We may solve the leftmost equations via
\beq
h_3 = db_2 - \cH_3\,,\qquad g_3 = dc_2 + \cC_3\,,\qquad \sfF_2 = d\sfA_1 + \cF_2\,,
\eeq
where $\cH_3,\cC_3,\cF_2 \in H^*(\cW^5;\bZ)$ and $b_2, c_2$, and $\sfA_1$ are globally well-defined gauge potentials.\footnote{The fact that these are valued in integer cohomology follows from the flux quantization of the RR field strengths \cite{Freed:2000ta}.} In the later computations $\cC_3$ will play no role, so without any loss of generality we will set it to zero. On the other hand, the nontrivial Bianchi identities for $g_1$ and $g_2$ indicate the presence of discrete gauge fields in the bulk theory. Specifically, these putative $U(1)$ gauge fields are respectively Higgsed down to $\bZ_{2M}$ and $\bZ_M$ gauge fields via the St\"{u}ckelberg mechanism. As a result, we may solve the Bianchi identities as 
\beq\label{eq:Bianchi_sols}
 g_1 = dc_0 + 2M\sfA_1 + \cA_1\,,\qquad g_2 = d\beta_1 - Mb_2 + \cB_2\,,
\eeq
where $\cA_1 \in H^1(\cW^5;\bZ_{2M})$ and $\cB_2 \in H^2(\cW^5;\bZ_M)$, while $c_0$ and $\beta_1$ are globally well-defined gauge potentials.\footnote{More precisely, $\cA_1$ and $\cB_2$ are cohomology class representatives. The difference between two representatives of the same class is exact, i.e.~$\cA_1-\cA_1'=d\lambda_0$ for some globally well-defined 0-cochain $\lambda_0$ on $\mathcal{W}^5$. We see that the choice of representative is immaterial since $g_1$ is invariant under the redefinitions $\cA_1 \to \cA_1+d\lambda_0$ and $c_0 \to c_0 - \lambda_0$. Similarly, $g_2$ is invariant under $\cB_2 \to \cB_2 + d\lambda_1$ and $\beta_1 \to \beta_1 - \lambda_1$. See Appendix \ref{sec:app_Bockstein_homomorphism} for a related discussion.} The quantities $g_1$ and $g_2$, containing both free and torsional components, carry the same information as differential cochains. See \cite{Hopkins:2002rd,Freed:2006yc,Cordova:2019jnf,Hsieh:2020jpj} for a collection of reviews.

The coefficients of the cohomology groups in which $\mathcal{A}_1$ and $\mathcal{B}_2$ live can be understood as follows. The Bianchi identites in \eqref{eq:Bianchi_ids} put constraints on the basepoint fluxes $\cF_2$ and $\cH_3$ that we have written above, i.e.~$2M\cF_2$ and $M\cH_3$ must vanish identically for the right-hand side of the relations to be an exact cocycle.\footnote{In general one might demand $2M\cF_2$ to be simply cohomologically trivial rather than to vanish identically, but such an exact piece can be removed via a field redefinition of $\sfA_1$.} This requirement implies that $\cF_2$ and $\cH_3$ are each contained in the image of a Bockstein homomorphism. Specifically, we may express 
\beq\label{eq:torsion_flux_definitions}
\cF_2 = \widetilde{\beta}(\cA_1)\,,\qquad \cH_3 = \widetilde{\beta}'(\cB_2)\,.
\eeq
In the above $\widetilde{\beta}$ and $\widetilde{\beta}'$ are respectively the Bockstein maps\footnote{A brief review of Bockstein homomorphisms can be found in Appendix \ref{sec:app_Bockstein_homomorphism}.}
\beq\label{eq:Bockstein_homomorphisms}
\widetilde{\beta}: H^*(\cW^5;\bZ_{2M})\to H^{*+1}(\cW^5;\bZ) \, ,\qquad \widetilde{\beta}': H^*(\cW^5;\bZ_{M})\to H^{*+1}(\cW^5;\bZ)\,.
\eeq

With this notation we can now compute the 5d SymTFT by reducing \eqref{eq:IIB_anomaly_polynomial} over the conifold $T^{1,1}$. The topological action $S_{\rm top}$ takes the following form,
\beq\label{eq:SymTFT}
S_{\rm top} = 2\pi\int_{W^5}\bigg(Mb_2d\hat{c}_2 - 2Mc_3 d\sfA_1 + b_2\cA_1\cB_2 - \sfA_1\cB_2^2 - \frac{\mathcal{K}}{M}\,\cA_1\beta(\cA_1)\beta(\cA_1)\bigg)\,,
\eeq
where $\mathcal{K}$ is defined in \eqref{eq:anomaly_coefficient}, and we have defined the field $d\hat{c}_2 = kdc_2 + da_2$ in terms of a field $a_2$ that arises when computing the SymTFT.\footnote{Technically, at the level of cochains, the term $\mathcal{B}_2 \cup \mathcal{B}_2$ is more approrpiately written as $\mathcal{P}(\mathcal{B}_2)$, where $\mathcal{P}$ denotes the Pontryagin square operation defined in, e.g.~\cite{Whitehead:1949,Aharony:2013hda,Benini:2018reh,Cordova:2019uob}.} The details of this computation can be found in Appendix \ref{sec:app_SymTFT_dualization}.

\subsection{5d Bulk Gauge Symmetries}

The topological action $S_{\rm top}$ in \eqref{eq:SymTFT} details the symmetry structure of the 5d bulk gauge theory, and it acts as the SymTFT for the 4d QFT living on the boundary. The first two terms of act as BF terms, so for instance, the fluctuating $U(1)$ gauge fields $b_2$ and $\hat{c}_2$ in the former term are Higgsed down to $\bZ_M$ gauge fields at low energies via the St\"{u}ckelberg mechanism \cite{Banks:2010zn}. As a consequence, the holonomies of $\sfA_1$ and $b_2$ in the IR turn into
\begin{equation}
        \exp\left(2\pi i\int_{\gamma^1}\sfA_1\right) \xrightarrow{\text{IR}} \exp\left(\frac{2\pi i}{2M}\int_{\gamma^1}\cA_1\right)\,, \quad \exp\left(2\pi i\int_{\Sigma^2}b_2\right) \xrightarrow{\text{IR}} \exp\left(-\frac{2\pi i}{M}\int_{\Sigma^2}\cB_2\right)\,.\label{eq:A1_B2_holonomies}
\end{equation}
We shall remark that the $U(1)$ gauge symmetry associated with $\sfA_1$ comes from the $S^1$ isometry circle inside $T^{1,1}$. This isometry, as was discussed in the introduction, corresponds to the $R$-symmetry in the 4d $\mathcal{N}=1$ boundary field theory. The Higgsing of the gauge symmetry down to a $\bZ_{2M}$ subgroup is exactly the same as how the $U(1)$ global $R$-symmetry of the boundary theory is broken down to a $\bZ_{2M}$ subgroup by the ABJ anomaly.
Reading off the form degrees of the gauge fields, the first term in $S_{\rm top}$ corresponds to $\bZ_{M}^{(1)}$ and $\widehat{\bZ}_M^{(1)}$ gauge symmetries in the 5d bulk theory, where the hat denotes the Pontryagin-dual group.\footnote{Here the form degree of a symmetry (as labeled by the superscript) is defined with respect to a 4d ``time'' slice in the bulk resulting from Hamiltonian quantization. In this case, a $p$-form gauge field is associated with a $(3-p)$-form symmetry.} Similarly, the latter term corresponds to $\bZ_{2M}^{(0)}$ and $\widehat{\bZ}_{2M}^{(2)}$ gauge symmetries.

The third and fourth terms in $S_{\rm top}$ are interaction terms describing an interplay between the $\bZ_M^{(1)}$ and $\bZ_{2M}^{(0)}$ gauge symmetries present in the bulk theory.\footnote{The reason for $b_2$ being a gauge field for the $\bZ_M^{(1)}$ symmetry rather than the $\widehat{\bZ}_M^{(1)}$ symmetry will be made clear when we consider the boundary field theory later in Section \ref{sec:boundary_conditions}.} The interplay, when considered from the perspective of the 4d boundary theory, is a mixed 't Hooft anomaly between the two symmetries. Similarly, the final term in $S_{\rm top}$ is a cubic interaction term in the bulk gauge theory, and becomes a 't Hooft anomaly purely for the $\bZ_{2M}^{(0)}$ symmetry on the boundary.

\subsection{Gauss Laws and Symmetry Generators}

For each of the discrete gauge symmetries present in the SymTFT, we may construct a symmetry operator that generates the corresponding symmetry. This can be done by treating the radial direction of the external space $\cW^5$ as a time coordinate and performing a Hamiltonian analysis of the 5d bulk action \cite{Witten:1998wy,Gukov:2004id,Belov:2004ht}. In this formalism, the time components of the bulk fields have vanishing conjugate momentum and thus impose (classical) Gauss Law constraints. Each of the fields $b_2, \hat{c}_2, \sfA_1,$ and $c_3$ will impose a Gauss Law constraint, which were originally computed in \cite{Apruzzi:2022rei} and we restate them here for convenience,
\begin{align}\label{eq:Gauss_Laws}
\cG_{b_2} &= Md\hat{c}_2 + \cA_1\cB_2\,, & \cG_{{\sf A}_1} &= 2Mdc_3 + \cB_2^2\nn\,,\\
\cG_{\hat{c}_2} &= Mdb_2\,, & \cG_{c_3} &= 2Md{\sf A}_1\,.
\end{align}
The quantities $\cG_i$ may be thought of as $dP_i$ where $P_i$ is a ``Page charge'' generating the corresponding gauge transformation. For a {\it continuous} symmetry, e.g.~$U(1)$, the operator $\exp\big(2\pi i\int_{M_i} P_i\big)$ acts as a symmetry generator supported on some submanifold $M_i \subset \mathcal{W}^5$ of the appropriate dimension. In contrast, if the symmetry is broken down to a {\it finite} group, e.g.~$\mathbb{Z}_k$, then the proper generator to consider is the ``$k$-th'' root of the aforementioned one, i.e.~$\mathcal{D} \sim \exp\big((2\pi i/k)\int_{M_i} P_i\big)$, such that $\mathcal{D}^{\otimes k} \cong \mathds{1}$ corresponds to the identity element of the finite group (up to condensation defects in the case of non-invertible symmetries).

To illustrate this, let us look at the Gauss Law constraint $\mathcal{G}_{\sfA_1}$. Using the prescription above, one would na\"{i}vely construct an operator of the form,
\begin{equation}
    \mathcal{N}_1(M^3,M^4) = \exp\bigg(2\pi i \int_{M^3} c_3 + \frac{2\pi i}{2M} \int_{M^4} \mathcal{B}_2^2\bigg) \, .
\end{equation}
This operator is well-defined when placed on a 3-manifold $M^3$ that is the boundary of some 4-manifold $M^4$. To define the operator solely on $M^3$, we may rewrite the second term as a 3d action by coupling $\mathcal{B}_2$ to a dynamical $U(1)$ gauge field $a_1$ localized on $M^3$ \cite{Choi:2022jqy,Apruzzi:2022rei}, i.e.
\beq\label{eq:Z2M_generator}
\cN_1(M^3) = \int\cD a_1\exp\bigg(2\pi i\int_{M^3}c_3 + \frac{M}{2}a_1da_1 + a_1\cB_2\bigg)\,.
\eeq
As an aside, one may recognize the latter two terms as an effective action for a fractional quantum Hall state. We claim that such an operator generates a $\mathbb{Z}_{2M}^{(0)}$ symmetry, and as we will show in Section \ref{sec:N_U_operators}, stacking $2M$ copies of \eqref{eq:Z2M_generator} indeed produces a trivial operator up to a condensation defect.

It is instructive to point out that the defect $\cN_1(M^3)$ can be schematically decomposed into two pieces, i.e.
\begin{equation}\label{eq:N1_expression}
    \mathcal{N}_1(M^3) = \exp\left(2\pi i \int_{M^3} c_3\right)\otimes \mathcal{A}^{M,1}[\mathcal{B}_2] \coloneqq \widetilde{\mathcal{N}}_1(M^3) \otimes \mathcal{A}^{M,1}[\mathcal{B}_2] \, ,
\end{equation}
where $\widetilde{\mathcal{N}}_1(M^3)$ generates a $\bZ_{2M}^{(0)}$ $R$-symmetry rotation, and $\mathcal{A}^{M,1}[\mathcal{B}_2]$ is a 3d minimal TQFT associated with the $\widehat{\bZ}_M^{(1)}$ symmetry \cite{Hsin:2018vcg}. Note that the simultaneous appearance of these two operators in $\cN_1(M^3)$ results from the presence of the anomaly term $\sfA_1 \cB_2^2$ in \eqref{eq:SymTFT}. An analogous decomposition of the chiral symmetry defect can be found in \cite{Choi:2022jqy} in the context of 4d massless QED (see also, e.g.~\cite{Kaidi:2021xfk,Cordova:2022ieu} for similar constructions). There, the defect is similarly composed of a na\"ive chiral rotation operator stacked with a minimal TQFT.

Applying the same philosophy, we can construct symmetry defects generating the remaining symmetries encoded by the SymTFT, namely, $\widehat{\bZ}_{2M}^{(2)}, \bZ_{M}^{(1)},$ and $\widehat{\bZ}_M^{(1)}$, using the Gauss Law constraints $\cG_{c_3},\cG_{b_2},$ and $\cG_{\hat{c}_2}$. Explicitly, these defects take the respective forms,
\begin{align}
\cV_1(\gamma^1) &= \exp\left(\frac{2\pi i}{2M}\int_{\gamma^1}\cA_1\right)\,,\label{eq:V_defect}\\
\cO_1(W^2,W^3) &= \exp\left(2\pi i\int_{W^2}(kc_2+a_2) + \frac{2\pi i}{M}\int_{W^3}\cA_1\cB_2\right)\,,\label{eq:O_defect}\\
 \cU_1(\Sigma^2) &= \exp\left(\frac{2\pi i}{M}\int_{\Sigma^2}\cB_2\right)\,.\label{eq:U_defect}
\end{align}
When constructing $\cV_1$ and $\cU_1$, we have made use of the discussion surrounding \eqref{eq:A1_B2_holonomies}.

Analogously to the previous case of $\mathcal{N}_1$, the operator $\cO_1(W^2,W^3)$ as written in \eqref{eq:O_defect} is well-defined when placed on a 2-manifold $W^2$ that is the boundary of some 3-manifold $W^3$. To define the operator solely on $W^2$, we may rewrite the last term in \eqref{eq:O_defect} as a 2d action by coupling $\mathcal{A}_1$ and $\mathcal{B}_2$ respectively to dynamical $U(1)$ gauge fields $\lambda_1$ and $\varphi$ localized on $W^2$, i.e.
\beq
\cO_1(W^2) = \int\cD\lambda_1\cD\varphi\exp\bigg(2\pi i\int_{W^2}\hat{c}_2 + \lambda_1\cA_1 + \varphi\cB_2 - M\varphi d\lambda_1\bigg) \, .
\eeq
For convenience, let us define the shorthand notation,
\begin{equation}
    \mathcal{X}^{M,q}[\mathcal{A}_1,\mathcal{B}_2] \coloneqq \int\mathcal{D}\lambda_1 \mathcal{D}\varphi \exp\bigg(2\pi i q \int \lambda_1 \mathcal{A}_1 + \varphi \mathcal{B}_2 - M \lambda_1 d\varphi\bigg) \, ,
\end{equation}
assuming $\gcd(M,q)=1$. The anomaly label $q \sim q+M$ specifies the braiding statistics between $\lambda_1$ and $\varphi$. Note that in the absence of the coupling to $\mathcal{A}_1$ and $\mathcal{B}_2$, the TQFT $\mathcal{X}^{M,1}$ is an untwisted 2d $\mathbb{Z}_M \times \mathbb{Z}_M$ BF theory.\footnote{\label{footnote:XMq_field_content}By construction, the Wilson-like operators $e^{2\pi i \oint \lambda_1}$ and $e^{2\pi i \varphi}$ generate the two $\mathbb{Z}_M$ symmetries on $W^2$. The cocycles $\mathcal{A}_1$ and $\mathcal{B}_2$ are respectively the BF-duals of $\lambda_1 \in H^1(W^2;\mathbb{Z}_M)$ and $\varphi \in H^0(W^2;\mathbb{Z}_M)$ \cite{Kapustin:2014gua}. This can be seen by rewriting $\mathcal{X}^{M,1}[\mathcal{A}_1,\mathcal{B}_2]$ in terms of continuum fields $\hat{\mathsf{A}}_1 = \mathcal{A}_1/M$ and $b_2 = \mathcal{B}_2/M$, and path-integrating over them instead of $\lambda_1$ and $\varphi$, which results in the equations of motion, $M d\lambda_1 = M d\varphi = 0$.} Similarly to the defect $\mathcal{N}_1(M^3)$, we observe that $\mathcal{O}_1(W^2)$ decomposes into
\beq\label{eq:O1_TQFT}
\cO_1(W^2) = \widetilde{\cO}_1(W^2)\otimes \cX^{M,1}[\cA_1,\cB_2]\,,
\eeq
where $\widetilde{\cO}_1(W^2) = \exp\big(2\pi i \int_{W^2} \hat{c}_2\big)$ is an invertible operator generating a $\bZ_M^{(1)}$ transformation, and the appearance of the 2d TQFT $\cX^{M,1}[\cA_1,\cB_2]$ is a consequence of the interaction term $b_2 \mathcal{A}_1 \mathcal{B}_2$ in \eqref{eq:SymTFT}.



\section{Symmetries from Branes}\label{sec:brane_inflow}

\subsection{Bulk SymTFT vs.~Brane Worldvolume Action}

We derived earlier the SymTFT for the bulk supergravity theory in the KT-KS background, from which the Gauss Law constraints give us a complete set of symmetry generators in the 5d bulk. In this and the subsequent sections, we show that each of these symmetry defects can be engineered as a certain probe D-brane configuration in the KT-KS geometry. As alluded to in Section \ref{sec:introduction}, recently in the literature there have been field-theoretic constructions of non-invertible symmetry defects with dimension $d \geq 2$. The hallmark of such defects is two-fold:
\begin{enumerate}
    \item $\text{(Defect A)} \otimes \text{(Defect B)} = \text{(Decoupled TQFT Coefficient)} \otimes \text{(Defect C)}$,
    \item $\text{(Defect)} \otimes \text{(Oppositely Oriented Defect)} = \sum \text{(Lower-dimensional Defect)}$.
\end{enumerate}
The significance of the brane-defect correspondence in our work is that it allows us to interpret both of these properties purely in terms of brane kinematics and dynamics in the bulk. In fact, as we will discuss in Section \ref{sec:boundary_conditions}, there exist obstructions to realize certain bulk-symmetry-generating branes (or combinations of them) on the boundary field theory. The obstructions are determined by the anomaly terms in the SymTFT and choices of boundary conditions for the 5d bulk gauge fields. A comprehensive analysis of such branes therefore realizes a much richer structure of the topological data encoded by the SymTFT in the bulk, which would otherwise not be appreciated from the perspective of the boundary field theory.

In Type II string theory, the worldvolume action of a general D$p$-brane consists of two parts: a Dirac-Born-Infeld (DBI) action encoding the dynamics of worldvolume gauge fields, and a Wess-Zumino (WZ) action describing the topological coupling of the D$p$-brane to the bulk RR and NSNS fields. To construct a topological operator out of a probe D$p$-brane, one can push it to the conformal boundary of the external spacetime. The dynamical degrees of freedom are frozen out since the brane tension scales as $T_p \sim r^p$, causing the DBI term to decouple. This leaves us with the WZ action (see, e.g.~\cite{Polchinski:1995mt,Polchinski:1996fm}), given by
\begin{equation}
    S_\text{WZ}^\text{D$p$} = \mu_p \int_{\mathcal{M}^{p+1}} I^\text{D$p$}_{p+1} \coloneqq \mu_p \int_{\mathcal{M}^{p+1}} \sum_{q \leq p} C_{q+1} \wedge \text{ch}_B(\cE) \wedge \sqrt{\frac{\hat{\mathcal{A}}(\mathcal{R}_T)}{\hat{\mathcal{A}}(\mathcal{R}_N)}} \, \Bigg|_\text{$(p+1)$-form} \, ,\label{eq:D-brane_WZ_action}
\end{equation}
where $\mu_p$ is the charge of a single D$p$-brane, and $\cE$ is the Chan-Paton bundle over the worldvolume $\mathcal{M}^{p+1}$. We sum over even $q$ in Type IIA and odd $q$ in Type IIB string theory. The quantities $\mathcal{R}_T = 2\pi R_T$ and $\mathcal{R}_N = 2\pi R_N$ denote respectively the (normalized) curvature 2-forms of the tangent and normal bundles of the worldvolume $\mathcal{M}^{p+1}$ of the D$p$-branes, while $\hat{\mathcal{A}}$ is the $A$-roof genus. In the analysis that follows, we will omit the gravitational couplings in the WZ action for simplicity.

For the purpose of our work, we would like to highlight two salient properties of the WZ action. Firstly, a D$p$-brane is not only coupled to the $(p+1)$-form RR field $C_{p+1}$, but also to the lower RR fields $C_{p-1}, C_{p-3},$ etc.~via the (twisted) Chern character $\text{ch}_B(\cE) = e^{B_2+F_2}$. Such couplings are physically interpreted as D$q$-branes ($q<p$) being dissolved in $\mathcal{M}^{p+1}$ \cite{Douglas:1995bn}. The numbers of these D$q$-branes are determined by the worldvolume Chan-Paton flux on the D$p$-brane. As a simple example, suppose $\mathcal{M}^{p+1} = \widetilde{\mathcal{M}}^2 \times \mathcal{M}^{p-1}$, and the Chan-Paton field strength decomposes as $F_2 = \tilde{f}_2 + f_2$ with $\tilde{f}_2 \in H^2(\widetilde{\mathcal{M}}^2;\mathbb{Z})$ and $f_2 \in H^2(\mathcal{M}^{p-1};\mathbb{Z})$. The Chern character factorizes as
\begin{equation}
    e^{B_2+F_2} = e^{\tilde{f}_2} \wedge e^{B_2+f_2} = 1 + \tilde{f}_2 \wedge e^{B_2+f_2} + \cdots \, ,\label{eq:single_Dp-brane_Chern_character_factorization}
\end{equation}
and so the number, $k$, of D$(p-2)$-branes within the D$p$-brane is given by $k = \int_{\widetilde{\mathcal{M}}^2} \tilde{f}_2$.

Let us now consider a single D$p$-brane with zero net D$p$ charge, i.e.~$\int_{\mathcal{M}^{p+1}} C_{p+1} = 0$, but with $k$ units of worldvolume flux through $\widetilde{\mathcal{M}}^2$. As far as the {\it center-of-mass mode} is concerned, such an object can also be thought of as a stack of $k$ coincident D$(p-2)$-branes. To see this, we use the fact that the Chan-Paton bundle $\mathcal{E}$ on the brane stack has structure group $U(k)$. Suppose we parametrize the $\mathfrak{u}(k)$-valued field strength $F_2 = \mathcal{F}_2 + f_2 \mathds{1}_k$, with $\mathcal{F}_2$ being $\mathfrak{su}(k)$-valued and $f_2 \in H^2(\mathcal{M}^{p+1};\mathbb{Z})$, then the Chern character factorizes as
\begin{equation}
    e^{B_2} \, \text{Tr}(e^{F_2}) = \text{Tr}(e^{\mathcal{F}_2}) \wedge e^{B_2 + f_2} = k e^{B_2 + f_2} + \cdots \, .\label{eq:Dp-brane_stack_Chern_character_factorization}
\end{equation}
We therefore recover the second term in \eqref{eq:single_Dp-brane_Chern_character_factorization} with $k = \int_{\widetilde{\mathcal{M}}^2} \tilde{f}_2$. Note that the gauge field $f_2$ here is associated with the center-of-mass mode of the stack of D$(p-2)$-branes, while $\mathcal{F}_2$ is associated with the remaining relative Chan-Paton modes. The relation between these two perspectives will prove useful later as we study the fusion between D$p$-branes.

\subsection{D-brane Anomaly Polynomial}

As for any Chern-Simons action, we can naturally construct an anomaly polynomial associated with the D$p$-brane via descent. It can be defined {\it locally} as\footnote{Formally, $\mathcal{I}^\text{D$p$}_{p+2}$ is defined on a manifold $\mathcal{M}^{p+2}$ such that $\partial\mathcal{M}^{p+2}=\mathcal{M}^{p+1}$ is the worldvolume of the D$p$-brane. We also assume the gauge fields originally defined on $\mathcal{M}^{p+1}$ can be extended to $\mathcal{M}^{p+2}$.}
\begin{equation}
    \mathcal{I}^\text{D$p$}_{p+2} = dI^\text{D$p$}_{p+1} = \sum_{q \leq p} G_{q+2} \, e^{B_2+F_2} \, .\label{eq:Dp_brane_anomaly_polynomial}
\end{equation}
The RR and NSNS field strengths locally take the forms,
\begin{equation}
    G_{q+2} = dC_{q+1} + H_3 C_{q-1} \, , \qquad H_3 = dB_2 \, .\label{eq:RR_field_local_definition}
\end{equation}
In the presence of backgroud D-branes, or when $[H_3] \in H^3(\mathcal{M}^{p+1};\mathbb{Z})$ is nontrivial \cite{Freed:1999vc,Kapustin:1999di}, the RR field strengths $G_{q+2}$ can be topologically nontrivial. Globally, they should be defined by the Bianchi identities,\footnote{Equivalently, the Bianchi identities for the various RR fields $G_{q+2}$ as differential forms can be collected and rewritten in a more suggestive way, $(d+H) G = 0$, which is one of the supporting arguments for the claim that D-brane charges are formally classified by twisted K-theory \cite{Minasian:1997mm,Witten:1998cd}.}
\begin{equation}
    dG_{q+2} = -H_3 G_q \, ,\label{eq:RR_field_Bianchi_identities}
\end{equation}
in addition to the requirement that they are equivariant under all additional gauge symmetries present in the system (e.g.~in our setup we gauged the $R$-symmetry circle in $T^{1,1}$). Throughout the rest of this work, we take the RHS of \eqref{eq:Dp_brane_anomaly_polynomial} as the definition of the D-brane anomaly polynomial, subject to $G_{q+2}$ satisfying the Bianchi identities \eqref{eq:RR_field_Bianchi_identities}. Specifically, we assume the ansatzes for the RR fields as defined in \eqref{eq:RR_field_strength_expansion} to account for global topological data in the KS setup.

\paragraph{Bulk symmetry generators from wrapped D-branes.}

Recall that the 5d topological action $\eqref{eq:SymTFT}$ of the SymTFT is derived by reducing $\mathcal{I}_{11}$ of Type IIB string theory on the conifold base $T^{1,1}$ of the KT-KS background. Similarly, one can reduce $\mathcal{I}^\text{D$p$}_{p+2}$ of a D$p$-brane on some $n$-manifold $X^n \subset T^{1,1}$, resulting in an effective action for a topological defect defined on a $(p+1-n)$-submanifold $M^{p+1-n}$ of the 5d bulk $\mathcal{W}^5$.

We claim that each of the four symmetry generators, $\mathcal{N}_1(M^3)$, $\mathcal{V}_1(\gamma^1)$, $\mathcal{O}(W^2)$, and $\mathcal{U}_1(\Sigma^2)$, derived in Section \ref{sec:symtft} with a Gauss Law constraint admits a D-brane origin. The first symmetry generator we are interested in is $\cN_1 = \widetilde{\cN}_1 \otimes \mathcal{A}^{M,1}[\mathcal{B}_2]$ which generates the $\bZ_{2M}^{(0)}$ gauge symmetry. From the bulk string theory setup, this defect can be realized by reducing a D5-brane on $S^3 \subset T^{1,1}$ \cite{Apruzzi:2022rei}. With this interpretation, the dynamical gauge field $a_1$ in $\mathcal{A}^{M,1}[\mathcal{B}_2]$ can be understood as the Chan-Paton field living on the D5-brane. These localized degrees of freedom thereby naturally encode the stacking of a minimal TQFT in order to construct the symmetry defect. 



\section{$\mathbb{Z}_{2M}^{(0)}$ and $\widehat{\bZ}_M^{(1)}$ Defect Fusion}\label{sec:N_U_operators}

\subsection{$\cN_p \otimes \cN_{p'}$ Fusion and Dielectric Branes}

\paragraph{Defect fusion.} Let us study the parallel fusion between two copies of $\cN_1$ defined on the same 3-manifold $M^3$. Here we would like to neglect the D5-brane origin of $\cN_1$ for a moment and treat them solely as topological defects acting on the Hilbert space of the theory. It is straightforward to deduce that $\widetilde{\mathcal{N}}_1$ fuse invertibly as $\widetilde{\mathcal{N}}_1 \otimes \widetilde{\mathcal{N}}_1 = \widetilde{\mathcal{N}}_2 = \exp\big(4\pi i \int_{M^3} c_3\big)$.\footnote{To avoid notational clutter, we hereafter often omit the dependency of the operators on manifolds.} On the other hand, the minimal TQFT $\mathcal{A}^{M,1}[\mathcal{B}_2]$ satisfies the fusion rule \cite{Choi:2022jqy},
\begin{equation}
	\mathcal{A}^{M,1}[\mathcal{B}_2] \otimes \mathcal{A}^{M,1}[\mathcal{B}_2] = \mathcal{A}^{M,2} \otimes \mathcal{A}^{M,2}[\mathcal{B}_2] \, ,
\end{equation}
assuming $\gcd(M,2)=1$, i.e.~$M$ is odd. Note that the first copy of $\mathcal{A}^{M,2}$ is a decoupled TQFT \cite{Roumpedakis:2022aik}, while the second one couples to the $\mathbb{Z}_M$ gauge field $\mathcal{B}_2$. Combining the results, we conclude that the fusion of two $R$-symmetry defects is given by
\begin{equation}
	\mathcal{N}_1 \otimes \mathcal{N}_1 = \widetilde{\mathcal{N}}_2 \otimes \mathcal{A}^{M,2} \otimes \mathcal{A}^{M,2}[\mathcal{B}_2] \coloneqq \mathcal{A}^{M,2} \otimes \mathcal{N}_2 \, .\label{eq:N1_N1_fusion}
\end{equation}

In general, one may construct an operator $\cN_p$ analogously to how we defined $\cN_2$ above, i.e.~we take the na\"ive operator $\widetilde{\cN}_p$ and stack it with a minimal TQFT $\cA^{M,p}[\mathcal{B}_2]$. The latter has an anomaly label $p \sim p+M$ dictating the braiding statistics of the anyons living on $M^3$. This implies that $\mathcal{A}^{M,M} \cong \mathcal{A}^{M,0}$ is a trivial theory, so the operator $\cN_M \cong \widetilde{\cN}_M$ reduces to an invertible operator when we take $p=M$. The operator $\cN_M$ therefore generates the anomaly-free $\mathbb{Z}_2^{(0)}$ subgroup of the $\mathbb{Z}_{2M}^{(0)}$ $R$-symmetry that is associated with the $(-1)^F$ operator \cite{Klebanov:2000hb,Klebanov:2002gr}. It is important, however, to note that $\widetilde{\cN}_M$ has a periodicity $p \sim p+2M$, and the same applies to the full operator $\cN_p$.

We show in Appendix \ref{sec:app_complete_fusion_rules} that if $\gcd(M,p)=\gcd(M,p')=\gcd(M,p+p')=1$,\footnote{This condition can be satisfied for all possible values of $p,p'=0,\dots,M-1$ if $M>2$ and is prime.} the fusion between general $R$-symmetry defects follows
\begin{equation}
	\mathcal{N}_p \otimes \mathcal{N}_{p'} = \mathcal{A}^{M,pp'(p+p')} \otimes \mathcal{N}_{p+p'} \, ,\label{eq:Np_Np'_fusion}
\end{equation}
which reduces to \eqref{eq:N1_N1_fusion} when $p=p'=1$. It should be emphasized again that the factor $\mathcal{A}^{M,pp'(p+p')}$ does not couple to $\mathcal{B}_2$. Since $\mathcal{A}^{M,pp'(p+p')}$ is symmetric in the labels $p$ and $p'$, the fusion rule \eqref{eq:Np_Np'_fusion} is manifestly commutative.

\paragraph{Dielectric branes.}

The above field-theoretic fusion for the $\cN_p$ operators can alternatively be understood geometrically through the dynamics of probe D$p$-branes in the dual string theory. The $\cN_1\otimes\cN_1$ fusion is modeled by two coincident D5-branes on the $M^3\times S^3$ needed to generate the $\bZ_{2M}^{(0)}$ $R$-symmetry. As the two branes are brought together, the worldvolume theory will undergo an enhancement of the gauge group from $U(1)$ to $U(2)$ as can be seen from the new stringy modes stretching between the two D5-branes. This enhancement will play a significant role in the behavior of the stack of D5-branes.

It is known in the string theory literature that, in the presence of a nontrivial $G_{p+4}$ or $B_2$ background, $k$ coincident D$p$-branes will polarize and ``puff up'' into an $S^2$ transverse to the original stack \cite{Myers:1999ps,Polchinski:2000uf,Apruzzi:2019ecr,Apruzzi:2022rei}, known as the Myers Effect. The resulting configuration is equivalent to that of a D$(p+2)$-brane with vanishing D$(p+2)$-brane charge and $k$ units of D$p$-brane charge. In the case at hand, when two D5-branes wrapping $M^3\times S^3$ are brought close to one another, the configuration becomes indistinguishable from a single D7-brane wrapping $M^3\times T^{1,1}$ with two units of worldvolume flux through the $S^2\subset T^{1,1}$.

There is a subtlety arising from the description of two D5-branes polarizing into a D7-brane. To illustrate the issue, let us reduce a D7-brane on $T^{1,1}$, with two units of worldvolume flux threading the $S^2$ orthogonal to the $S^3 \subset T^{1,1}$ on which the D5-branes are wrapped. The reduction leads to
\begin{equation}
	\exp\bigg(2\pi i \int_{M^4 \times T^{1,1}} \mathcal{I}^\text{D7}_9\bigg) = \exp\bigg(4\pi i \int_{M^3} c_3 + \frac{M}{2}a_1da_1 + a_1\cB_2\bigg) = \mathcal{N}_2(M^3) \, ,
\end{equation}
where $\partial M^4 = M^3$. Note that here $a_1$ is the $U(1)$ Chan-Paton field living on the D7-brane. A quick comparison with \eqref{eq:N1_N1_fusion} reveals the issue: the decoupled TQFT $\mathcal{A}^{M,2}$ seems to be absent from the na\"{i}ve computation above.

To recover this missing information, we can simply model each D5-brane as a ``neutral'' D7-brane with one unit of worldvolume flux (see the discussion following \eqref{eq:Dp-brane_stack_Chern_character_factorization}). Specifically, the anomaly polynomial of each of these neutral D7-branes can be expanded as
\beq\label{eq:I9_dielectric}
\cI_{9,i}^{\rm D7} = G_7(B_2 + da_1^i + f_2^i) + \frac{1}{2}\,G_5(B_2 + da_1^i + f_2^i)^2 + \frac{1}{6}\,G_3(B_2 + da_1^i + f_2^i)^3\, ,
\eeq
where $a_1^i$ denotes the Chan-Paton gauge field along $S^3$, and $f_2^i$ represents the worldvolume flux such that $\int_{S^2} f_2^i = 1$. We can then perform the reduction over $T^{1,1}$ using \eqref{eq:RR_field_strength_expansion} to find
\begin{align}\label{eq:D7D7_fusion}
\int_{M^4 \times T^{1,1}}\cI_{9,1}^{\rm D7} +\cI_{9,2}^{\rm D7} &= \int_{M^4} 2dc_3 + \frac{M}{2}\left(da_1^1da_1^1 + da_1^2da_1^2\right) + d(a_1^1 + a_1^2)\cB_2 \nn\\
&= \int_{M^3} \frac{2^{-1}M}{2} \, \alpha_1' d\alpha_1' + 2\left(c_3 + \frac{M}{2}\,\alpha_1 d\alpha_1 + \alpha_1\cB_2\right)\, ,
\end{align}
where we have made the field redefinitions $2\alpha_1 = a_1^1 + a_2^1$ and $\alpha_1' = -a_1^1 + a_1^2$. Not only have we recovered the action for $\cN_2(M^3)$, but we have also gained an additional term. This term encodes the information of the minimal TQFT $\cA^{M,2^{-1}}$.\footnote{We should be careful as there is an ambiguity in this description. As was discussed in \cite{Hsin:2018vcg}, $U(1)_{mn}\cong \cA^{mn,1}\cong \cA^{m,n}\otimes\cA^{n,m}$ are isomorphic as topological field theories. Furthermore, the effective Lagrangian descriptions for $\cA^{m,n}$, $\cA^{n,m}$, and $U(1)_{mn}$ are essentially identical. Technically, they can only be distinguished by coupling to appropriate background fields. Our choice of the identification is motivated by the field-theoretic results detailed in Appendix \ref{sec:app_complete_fusion_rules}.} Since the anomaly $p$ of the minimal TQFT is only defined modulo $M$, the quantity $2^{-1}$ denotes the unique multiplicative inverse of $2$ modulo $M$ assuming $\gcd(M,2)=1$. Importantly, it is an integer so that \eqref{eq:D7D7_fusion} is properly quantized. We now further utilize the equivalence of minimal TQFTs \cite{Hsin:2018vcg},
\beq
\cA^{M,p}\cong \cA^{M,pr^2}\,,
\eeq
by rescaling the generating line $a=e^{2\pi i \oint \alpha_1'}$ of a minimal TQFT as $a \to a^r$, for any $r$ such that $\gcd(M,r)=1$. Setting $r=2$ gives $\cA^{M,2^{-1}}\cong \cA^{M,2}$, and so we have fully reproduced the RHS of \eqref{eq:N1_N1_fusion}.

The question then arises of how to address the more general fusion $\cN_p\otimes \cN_{p'}$. We can repeat the above computation for stacks of $p$ and $p'$ coincident D5-branes, by viewing each stack as a single D7-brane with $p$ and $p'$ units of D5-brane charge respectively, i.e.~$\int_{S^2}f_2^1 = p$ and $\int_{S^2}f_2^2 = p'$. The reduction over $T^{1,1}$ now becomes
\begin{align}
\int_{M^4 \times T^{1,1}}\cI_{9,1}^{\rm D7} +\cI_{9,2}^{\rm D7} &= \int_{M^4} (p+p')dc_3 + \frac{M}{2}\left(pda_1^1da_1^1 + p'da_1^2da_1^2\right) + d(pa_1^1 + p'a_1^2)\cB_2 \nn\\
&= \int_{M^3} \frac{pp'(p+p')^{-1}M}{2}\,\alpha_1' d\alpha_1' + (p+p')\left(c_3 + \frac{M}{2}\,\alpha_1 d\alpha_1 + \alpha_1\cB_2\right)\, ,\label{eq:Np_fusion_rule_field_redefinition}
\end{align}
where the required field redefinition here is $(p+p')\alpha_1 = pa_1^1 + p'a_1^2$ and $\alpha_1' = -a_1^1 + a_1^2$. This exactly reproduces the general fusion in \eqref{eq:Np_Np'_fusion}, with the first term describing $\cA^{M,pp'(p+p')^{-1}}\cong \cA^{M,pp'(p+p')}$, and the terms in parentheses describing $\cN_{p+p'}$. Note that the latter can equivalently be described as a single D7-brane with $p+p'$ units of worldvolume flux.

\paragraph{Collective degrees of freedom.}

The conclusions above can be recasted from the perspective of the original stacks of $p$ and $p'$ D5-branes. We see that the gauge field $\alpha_1 \sim pa_1^1 + p'a_1^2$ defined on $\cN_{p+p'}$ is the overall {\it center-of-mass} mode for the combined stack of $p+p'$ D5-branes. Likewise, $a_1^1$ and $a_1^2$ are the center-of-mass modes of the individual stacks.

The other gauge field, $\alpha_1'$, should be interpreted as a mode describing the {\it relative} motion of the two stacks of $p$ and $p'$ D5-branes. This can be understood in the following way. The Chan-Paton bundle for the stack of $p$ D5-branes has a gauge group $U(p)$. It is known that the center-of-mass mode and the non-Abelian modes are decoupled from each other, so that the gauge group can be schematically considered as $SU(p)\times U(1)_{\rm CM}$. The same is true of course for the stack of $p'$ D5-branes as well as the final stack of $p+p'$ D5-branes. However, this final stack has another Abelian mode contained within it. Within $SU(p+p')$ there is a maximal subgroup given by
\beq
SU(p+p') \supset SU(p)\times SU(p')\times U(1)_{\rm rel}\,,
\eeq
where the $U(1)_{\rm rel}$ is generated by the Cartan element $\mathrm{diag}(p'\mathds{1}_p,-p\mathds{1}_{p'})$. It is this $U(1)_{\rm rel}$ that is describing the relative motion of the D5-brane stacks, which manifests itself as the decoupled minimal TQFT $\mathcal{A}^{M,pp'(p+p')}$ at the level of the fusion algebra \eqref{eq:Np_Np'_fusion}. Note that if we were to start with stacks of, say, $p+k$ and $p'-k$ D5-branes, then we would end up with a different $U(1)_{\rm rel}$, as is reflected by a different decoupled minimal TQFT $\mathcal{A}^{M,(p+k)(p'-k)(p+p')}$ in the fusion.

\begin{figure}[t!]
	\centering
	\includegraphics[width=\textwidth]{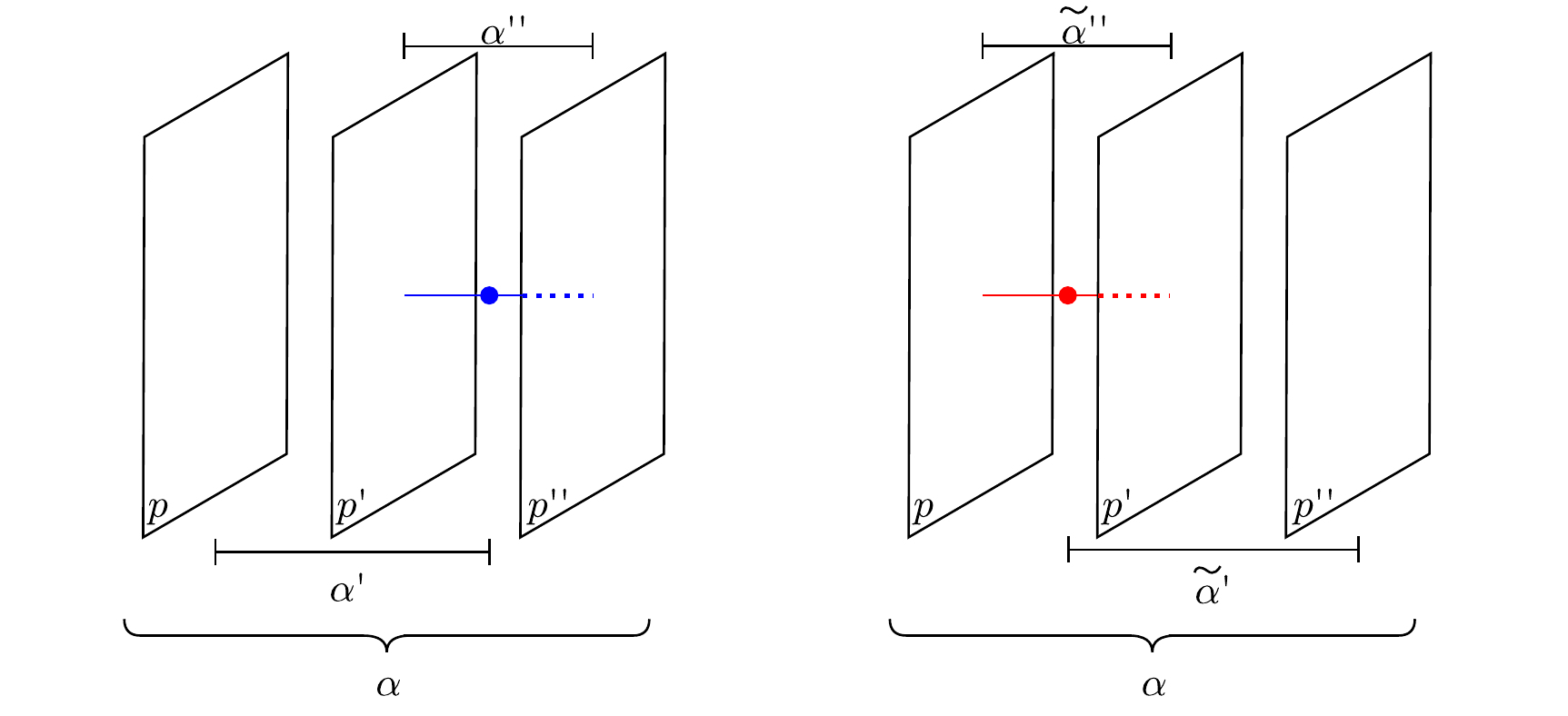}
	\caption{Fusion of three stacks of D5-branes. The left figure depicts the fusion $\cN_p\otimes(\cN_{p'}\otimes\cN_{p''})$, while the right depicts the fusion $(\cN_p\otimes\cN_{p'})\otimes\cN_{p''}$. In the former case, there is a center-of-mass mode $\alpha$ for the overall $p+p'+p''$ stack, and two other relative modes $\alpha'$ and $\alpha''$ respectively for the $p$ -- $(p'+p'')$ and $p'$ -- $p''$ stacks. The same applies to the figure on the right.} 
	\label{fig:triple_brane_fusion}
\end{figure}

We can carry out the same argument for more than two stacks of D5-branes. Consider, for instance, three stacks of $p$, $p'$, and $p''$ D5-branes as in Figure \ref{fig:triple_brane_fusion}. Each stack of D5-branes will itself have a center-of-mass mode describing the collective motion, i.e.~$a_1^1, a_1^2,$ and $a_1^3$, but we can repackage these degrees of freedom in different combinations. We may first consider an overall center-of-mass mode, $\alpha_1$, describing the collective motion of the three stacks. We can then consider a mode describing the relative motion of the leftmost D5-brane stack compared to the collective motion of the rightmost two D5-brane stacks given by $\alpha_1'$, as well as a mode describing the relative motion of the two rightmost D5-brane stacks given by $\alpha_1''$.
The explicit expressions for $\alpha_1,\alpha_1',\alpha_1''$ can be written as
\begin{align}
\alpha_1 &= \frac{1}{p+p'+p''}\left(pa_1^1 + p'a_1^2 + p''a_1^3\right)\,,\nn\\
\alpha_1' &= \frac{1}{p(p'+p'')}\left( (p'+p'')(pa_1^1) - p(p'a_1^2) - p(p''a_1^3)\right)\,,\\
\alpha_1'' &=  \frac{1}{p'p''}\left(p''(p'a_1^2) - p'(p''a_1^3)\right)\,.\nn
\end{align}
By regrouping our degrees of freedom in this way, we are effectively perfoming the fusion
\beq\label{eq:assoc_1}
\cN_p\otimes(\cN_{p'}\otimes\cN_{p''}) \cong \cA^{M,p'p''(p'+p'')}\otimes\cA^{M,p(p'+p'')(p+p'+p'')}\otimes\cN_{p+p'+p''}.
\eeq
The above describes the left side of Figure \ref{fig:triple_brane_fusion}. If we grouped the stacks of D5-branes in the other way as in the right side, we could define similar modes and would be performing the fusion
\beq\label{eq:assoc_2}
(\cN_p\otimes\cN_{p'})\otimes\cN_{p''} \cong \cA^{M,pp'(p+p')}\otimes\cA^{M,p''(p+p')(p+p'+p'')}\otimes\cN_{p+p'+p''}.
\eeq
Since the only difference between the two processes is the order in which we fuse the stacks of D5-branes, it is natural to argue that locality of the theory demands the fusion to be associative, i.e.~\eqref{eq:assoc_1} and \eqref{eq:assoc_2} should be physically equivalent. In Appendix \ref{sec:app_complete_fusion_rules}, we show that these two triple fusion rules are indeed isomorphic.

It is worth taking a moment to reassess the previous results. We are able to model the symmetry defects $\cN_p$ as stacks of $p$ D5-branes, which themselves can equivalently be described as a single D7-brane with $p$ units of worldvolume flux. The fusion \eqref{eq:Np_Np'_fusion} of the defects themselves does not seem to favor a particular approach, as it extracts only the Abelian information from the D$p$-branes. All the brane information captured by the fusion rules comes from a single center-of-mass mode as well as a single mode describing relative motion, or multiple relative modes in the case of more instances of fusion. In this sense, the dynamics of the non-Abelian modes in the worldvolume theory are ``washed out'' in the symmetry defects.

\subsection{$\cN_p \otimes \cN_{-p}$ Fusion and Tachyon Condensation}

\paragraph{Condensation defect.}

The case of $p'=-p \mod M$ requires special treatment. As discussed above, the $\widetilde{\mathcal{N}}_p(M^3)$ component of the defect is invertible and thus becomes the identity for the $\cN_p\otimes\cN_{-p}$ fusion. Regarding the fusion between the minimal TQFTs, we have, in terms of explicit worldvolume actions,
\begin{align}
	\mathcal{A}^{M,p}[\cB_2] \otimes \mathcal{A}^{M,-p}[\cB_2] & = \int \mathcal{D}a_1 \mathcal{D}a'_1 \exp\bigg(2\pi i \int_{M^3} \frac{pM}{2}a_1 da_1 - \frac{pM}{2}a'_1 da'_1 + p(a_1 - a'_1) \mathcal{B}_2\bigg)\nonumber\\
	& = \int \mathcal{D}\bar{a}_1 \mathcal{D}a'_1 \exp\bigg(2\pi i \int_{M^3} M \bar{a}_1 da'_1 +  \frac{p^{-1}M}{2} \bar{a}_1 d\bar{a}_1 + \bar{a}_1 \mathcal{B}_2\bigg)\nonumber\\
	& = (\mathcal{Z}_M)_0[\cB_2] \, ,\label{eq:Dijkgraaf-Witten_theory_coupled}
\end{align}
where $a_1$ and $a'_1$ respectively denote the auxiliary fields living on $\mathcal{A}^{M,p}$ and $\mathcal{A}^{M,-p}$, while $\bar{a}_1 = p (a_1 - a'_1)$. In the third line, we used the fact that if $M$ is odd, the Dijkgraaf-Witten (DW) term in $(\mathcal{Z}_M)_K$ is trivial when $K$ is an integer multiple of $M$, i.e.~$K \sim K + M$ (assuming $M^3$ is a spin manifold) \cite{cmp/1104180750}.\footnote{The continuum action for a 3d DW theory $(\mathcal{Z}_M)_K$ is given by $\int M a_1 da_1 + (k/2) a_1 dc_1$.} Integrating out $a'_1$ enforces $M\bar{a}_1 \in H^1(M^3;\mathbb{Z})$ to be an integral lift of a 1-cocycle in $H^1(M^3;\mathbb{Z}_M)$. We then rewrite \eqref{eq:Dijkgraaf-Witten_theory_coupled} as a discrete sum of the $\widehat{\mathbb{Z}}_M^{(1)}$ symmetry defects $\mathcal{U}_1(\Sigma^2)$ in \eqref{eq:U_defect},\footnote{For general $M$, there is an additional factor in the condensation defect coming from a choice of {\it discrete torsion}, or the choice to stack with an SPT phase when gauging. For $M$ odd, which we are assuming throughout this work, this discrete torsion factor will be trivial and thus not present in the condensation defect.}
\begin{equation}
	\mathcal{N}_p \otimes \mathcal{N}_{-p} = \frac{1}{|H^0(M^3;\mathbb{Z}_M)|} \sum_{\Sigma^2 \in H_2(M^3;\mathbb{Z}_M)} \mathcal{U}_1(\Sigma^2) \coloneqq \mathcal{C} \, ,\label{eq:condensate_defect}
\end{equation}
where $\Sigma^2$ is the Poincar\'{e} dual of the torsional reduction of $\bar{a}_1$ with respect to $M^3$. We will refer to $\cC$ as the ``condensation defect'' as is common in the literature. This nomenclature stems from the interpretation that the sum of operators signifies a ``higher gauging'' of the $\widehat{\bZ}_M^{(1)}$ symmetry \cite{Roumpedakis:2022aik}. The symmetry operators condense on the submanifold $M^3$ of spacetime, hence the name. Note that the factor of $|H^0(M^3;\mathbb{Z}_M)|$ is a standard gauge-theoretic normalization coming from the normalization of the path integral measure \cite{Gaiotto:2014kfa,Kaidi:2021xfk}. 

\paragraph{Tachyon condensation.}

The presence of the condensation defect can be understood completely in terms of the dynamical process of tachyon condensation, suggesting an interesting interpretation of non-invertible fusion in holographic setups. Here we review some relevant generic features of tachyon condensation \cite{Sen:1998sm,Sen:1998ii,Sen:1999mg}.

Our starting point is a D$p$\,-$\overline{{\rm D}p}$-brane pair wrapped on a submanifold $\cM^{p+1}$ of the total spacetime, where an anti-D$p$-brane is the orientation reversal of a D$p$-brane. Over the D$p$-brane we denote the Chan-Paton bundle by $\cE^+$, while over the anti-D$p$-brane we denote the bundle $\cE^-$. Without loss of generality, we may take $\cE^-  = \cK$ and $\cE^+ = \cL\otimes\cK$ for $\cL$ and $\cK$ complex line bundles.\footnote{The proper classification of D-branes is well-known to be formalized not by ordinary cohomology but by K-theory \cite{Minasian:1997mm,Witten:1998cd,Olsen:1999xx,Witten:2000cn}. In this formalism, the relevant quantity which classifies the D-branes is the K-theoretic ``difference'' between the bundles $\cE^\pm$. Hence, the choice of parametrization in the main text is for convenience.} It is known that the worldvolume theory for a brane-antibrane pair suffers from a tachyonic mode. The tachyon in this case should be viewed as a section of $\cE^+\otimes (\cE^-)^* \cong \cL$, where $(\cE^-)^*$ denotes the dual bundle, reflecting that it is charged under the $U(1)\times U(1)$ gauge symmetry of the worldvolume theory. We may consider vortex-like solutions correpsonding to a choice of tachyon profile that vanishes in a specified region of spacetime. To be precise, let us take the tachyon profile to be
\beq
T= c\cdot s\,,
\eeq
where $s$ is a section of $\cL$ which vanishes over a codimension-2 submanifold $\cM^{p-1}\subset \cM^{p+1}$, and $c$ is a constant chosen such that the tachyon takes its vev away from $\cM^{p-1}$. By construction, this localizes a codimension-2 solution to exist only along $\cM^{p-1}$, and far from $\cM^{p-1}$ the system is indistinguishable from the vacuum. In the above construction, we are only considering a single D$p$-brane and $\overline{{\rm D}p}$-brane producing a single D$(p-2)$-brane, but in general one may consider $2^{k-1}$ each of branes and antibranes to produce a D$(p-2k)$-brane \cite{Witten:1998cd}.

More explicitly, note that the worldvolume bundle $\cE$ has a natural $\bZ_2$ grading induced by the splitting between the brane and antibrane degrees of freedom, thereby giving the structure of a {\it superbundle}. Taking advantage of the superbundle structure, we may collect the worldvolume degrees of freedom into a single superconnection \cite{Quillen:1985vya},
\beq
\mathbb{A} = \begin{pmatrix}{}
\;a_1^+\; & \;iT \\
\;i\overline{T}\; & \;a_1^-
\end{pmatrix},
\eeq
where $a_1^\pm$ denote the Chan-Paton gauge fields over $\cE^\pm$, and $T$ is the tachyon field as a section. Using this superconnection, we may write an action that allows us to go directly from a D$p$\,-$\overline{{\rm D}p}$ configuration to a stable D$(p-2)$-brane. This action was proposed in \cite{Kennedy:1999nn}, later elaborated on in \cite{Alishahiha:2000du,Takayanagi:2000rz,Szabo:2001yd,Garousi:2007fk}, and can be written in the form
\beq\label{eq:brane-antibrane_action}
S_{{\rm WZ}}^{\text{D$p$\,-$\overline{\text{D}p}$}} = \mu_p\int_{\cM^{p+1}}C\wedge e^{B}\wedge {\rm Tr_s} e^{\mathbb{F}}\wedge\sqrt{\frac{\hat{\cA}(\cR_T)}{\hat{\cA}(\cR_N)}}\Bigg|_{(p+1)-{\rm form}}\,,
\eeq
where $\mathbb{F} = d\mathbb{A} + \mathbb{A}^2$ is the supercurvature given by
\beq
\mathbb{F} = \begin{pmatrix}{}
da_1^+ - {T}\overline{T} & iDT \\
i\overline{DT} & da_1^- - {T}\overline{T}
\end{pmatrix}\,.
\eeq
In the above, ${\rm Tr_s}$ is the $\bZ_2$-graded supertrace and the covariant derivatives $DT$ and $\overline{DT}$ are defined as
\beq
DT := d{T} + (a_1^+ - a_1^-) {T},\qquad \overline{DT} := d\overline{T} - (a_1^+ - a_1^-) \overline{T}\,.
\eeq
By writing $a_1^+ = a_{\cL} + a_{\cK}$ and $a_1^- = a_{\cK}$ to reflect the structure of the line bundles, we can see that the tachyon field $T$ couples only to $a_\cL$ and thus is tied only to the bundle $\cL$. The action \eqref{eq:brane-antibrane_action} may be rewritten as a WZ action for a D$(p-2)$-brane living inside the D$p$\,-$\overline{{\rm D}p}$ worldvolume, i.e.
\beq\label{eq:condensate_brane_WZ}
S_{{\rm WZ}}^{\text{D$p$\,-$\overline{\text{D}p}$}} = \mu_p\int_{\cM^{p+1}}C \, e^{B} \, {\rm ch}(\cK) \, ({\rm ch}(\cL) - 1) = \mu_{p-2}\int_{\cM^{p-1}}C \, e^{B+F_{\cK}} = S^{{\rm D}(p-2)}_{\rm WZ}\,,
\eeq
where in the final equality we integrated out the two directions transverse to the D$(p-2)$-brane's worldvolume $\cM^{p-1}$. Due to our choice of bundle structure, this is equivalent to integrating out the dependence on the line bundle $\cL$, and thus the tachyon.\footnote{It is worth pointing out that the factor involving ${\rm ch}(\cL)$ is encoding the K-theoretic difference alluded to in the previous footnote. Had we not made the previous simplification and instead chosen $\cE^+ = \cL_1\otimes\cK$ and $\cE^- = \cL_2\otimes\cK$ (we can always consider a common line bundle $\cK$), this term would have taken the form ${\rm ch}(\cL_1) - {\rm ch}(\cL_2)$. For the purpose of illustration, we assume in \eqref{eq:condensate_brane_WZ} that the integral of ${\rm ch}(\cL) - 1$ is equal to $2\pi$. In general, the ratio of D$p$-brane charges is given by $\mu_p/\mu_{p'} = (2\pi\sqrt{\alpha'})^{p'-p}$ \cite{Johnson:2003glb}, but we also adopt the convention that $2\pi\alpha'=1$, otherwise $B_2 + 2\pi\alpha'F_2$ should be the gauge-invariant combination instead.}

\paragraph{Condensation defect from tachyon condensation.}

Let us apply the general discussion above to our case where a D5-brane and an anti-D5-brane wrap the submanifold $M^3 \times S^3$ in the KT-KS geometry $\cW^5\times T^{1,1}$. When the branes are brought together there will be a solitonic D3-brane localized on a worldvolume $\cM^4 \subset M^3 \times S^3$, with a WZ action described by \eqref{eq:condensate_brane_WZ}.

To reconstruct the condensation defect $\mathcal{C} \sim \sum \mathcal{U}_1$ in \eqref{eq:condensate_defect}, we must first specify the submanifold $\cM^4$ that the solitonic D3-brane is wrapping. Equivalently, this amounts to specifying the bundle $\mathcal{L}$ defined along the two transverse directions of $\mathcal{M}^4$ with respect to $M^3 \times S^3$. The integration in \eqref{eq:condensate_brane_WZ} can be viewed as a two-step process: a fiber integration from the $S^3$ down to the Hopf base $\widetilde{S^2}$, and an application of Poincar\'e duality to land on $\Sigma^2 \in H_2(M^3;\mathbb{Z}_M)$. Both of these are encoded by the Chan-Paton flux $F_\cL$. Specifically, we can construct a $U(1)_R$\,-equivariant ansatz for the Chan-Paton flux as
\beq
F_{\cL} = f_2 + M\eta_1 \,\frac{D\psi}{2\pi} + a_0^i V_i\,,
\eeq
where $f_2,\eta_1,$ and $a_0^i$ are fields defined over the external spacetime, and $V_i$ are the volume forms of the spheres present in the $T^{1,1}$. Enforcing the Bianchi identity $dF_{\cL} = 0$ leads us to identify $a_0^1=a_0^2=a_0$ and $M\eta_1 = da_0$, so that
\beq\label{eq:D5_antiD5_Chan_Paton_flux_expansion}
F_\cL = M\eta_1 \,\frac{D\psi}{2\pi} - a_0\left(V_1+V_2 - \sfF_2\right) + \Lambda_2 = da_\cL + \Lambda_2\,,
\eeq
where $a_\cL = a_0 (D\psi/2\pi)$ is a locally defined gauge field associated with $\cL$. The 2-form $\Lambda_2$ is closed, and it plays the role of a basepoint flux for $F_\cL$, much like $f_2^i$ in the discussion surrounding \eqref{eq:I9_dielectric}.
Performing the integration in \eqref{eq:condensate_brane_WZ} using this field strength yields the desired WZ action for a solitonic D3-brane localized on $\mathcal{M}^4=\Sigma^2\times \widetilde{S^2}$. 

Note that the expression for $F_\cL$ give two possible sources for a D3-brane in the D5-$\overline{{\rm D}5}$ worldvolume. The first is as a soliton arising from tachyon condensation as discussed above. The second is from any worldvolume flux present on the original D5-brane and anti-D5-brane. The D5-branes used to construct $\cN_1(M^3)$ did not carry any worldvolume flux, so we may take the corresponding term to be zero if desired. 

At this point, one may try to reduce the anomaly polynomial $\mathcal{I}^\text{D3}_5$ for the D3-brane on the Hopf base $\widetilde{S^2}$, but end up finding an apparently vanishing result, i.e.
\beq\label{eq:vanishing_I3}
\int_{\widetilde{S^2}\subset S^3}\cI_5^{\rm D3} = \cI_3 \ \text{``}\!=\!\text{''} \ 0\,.
\eeq
However, this is too quick. The operator $\mathcal{U}_1(\Sigma^2)$ we are ultimately concerned with finding is a holonomy of a discrete gauge field $\mathcal{B}_2 \in H^2(M^3;\mathbb{Z}_M)$. The standard descent procedure falls short as the reduction above merely indicates that there is no nontrivial holonomy for {\it continuous} gauge fields; we may still have {\it discrete} holonomies. Roughly speaking, we can again think of $\cI_3$ as a ``field strength'' for torsional flux and thus would vanish when evaluated on integer-valued cycles. If not integer-valued, then what cycles {\it should} we evaluate this on? Since we aim to recover a $\bZ_M$-valued operator, we shall follow our nose and consider reducing on $\bZ_M$-valued {\it torsional} cycles.

The correct topological operator is then not the standard holonomy of $\cI_3$ on an integer-valued cycle, but a holonomy along a $\bZ_{M}$-valued cycle. Namely, the proper operator to consider is\footnote{We can alternatively phrase this in terms of the RR field strengths, by originally having a term in $G_5$ of the form $M\eta_3\wedge(V_1+V_2)$ with $\eta_3$ a $\bZ_M$-valued field strength. Due to the torsion, this term would formally be zero and na\"ively wouldn't be present in the expansion.}
\beq\label{eq:discrete_holonomy_argument}
\cU_1(\Sigma^2) = \exp\left(\frac{2\pi i}{M}\int_{\widetilde{\Sigma}^3} \cI_3\right) = \exp\left(\frac{2\pi i}{M} \,\big\langle\widetilde{\Sigma}^3,\cI_3 \big\rangle\right)\,,
\eeq
where $\widetilde{\Sigma}^3$ is a formal extension of $\Sigma^2$ such that $\d\widetilde{\Sigma}^3 = M\Sigma^2$. In the final expression we have rewritten the integral as a pairing between $\bZ_{M}$-valued cycles and cocycles, and should now consider everything to be valued in (co)homology with $\bZ_M$ coefficients. As we mentioned, $\cI_3$ should be viewed as a ``field strength'' for a discrete field, so we may express $\cI_3 = \beta'(\cB_2) \in H^3(M^3;\mathbb{Z}_M)$. This justifies our decision to consider the reduction on $\bZ_M$-valued cycles, as opposed to cycles valued in any other coefficients, as the only discrete 2-form gauge field present in the supergravity setup is $\cB_2 \in H^2(M^3;\mathbb{Z}_M)$.\footnote{By exactness of the short exact sequence $0 \to \mathbb{Z}_M \to \mathbb{Z}_{M^2} \to \mathbb{Z}_M \to 0$, the quantity $M\mathcal{B}_2$ is trivial, and so is $\beta'(M\mathcal{B}_2)$. This means that $\mathcal{U}_M(\Sigma^2) = \mathcal{U}_1(M\Sigma^2) = \exp\big(2\pi i \int_{\widetilde{\Sigma}^3} \mathcal{I}_3\big) = 0$ as we saw in \eqref{eq:vanishing_I3}.}

With this choice, we can evaluate the holonomy of $\cI_3$ as
\begin{equation}
	\exp\left(\frac{2\pi i}{M} \,\big\langle \widetilde{\Sigma}^3,\beta'\left(\cB_2\right)\big\rangle\right) = \exp\left(\frac{2\pi i}{M} \,\big\langle \beta'\big(\widetilde{\Sigma}^3\big) , \cB_2\big\rangle\right) = \exp\left(\frac{2\pi i}{M}\,\big\langle \Sigma^2, \cB_2\big\rangle\right)\,.
\end{equation}
In the first equality, we switched the Bockstein homomorphism in cohomology to its dual in homology, and in the second equality, we used the property that $\beta' = \d/M$ for the Bockstein homomorphism in homology (see Appendix \ref{sec:app_Bockstein_homomorphism} for details). This holonomy is nothing but the $\cU_1(\Sigma^2)$ operator we saw in the condensation defect \eqref{eq:condensate_defect}.

The arguments above regarding the D3-brane worldvolume action can be rephrased in the language of differential cohomology \cite{10.1007/BFb0075216,Hopkins:2002rd,Freed:2006yc,Cordova:2019jnf,Hsieh:2020jpj}. The standard characteristic class construction via the descent procedure constructs the curvature form for a (higher) anomaly bundle over the base spacetime. Na\"ively, this curvature form only captures the continuous data of a Cheeger-Simons differential cocycle; the torsional data is overlooked. What we see above is that the torsional data is not completely lost, but can be reconstructed through the presence of torsional flux.

Now that we have constructed an operator for the D3-brane wrapping $\Sigma^2\times\widetilde{S^2}$, there is one last thing to consider about the ${\rm D}3$-brane soliton arising from the D5-$\overline{{\rm D}5}$ worldvolume. The Chan-Paton field on the D5-$\overline{{\rm D}5}$ brane pair is dynamical and therefore should be summed over in a quantum theory. 
This includes a summing over the gauge field not only from the solitonic D3-brane itself, but also from $a_\cL$ along the other two transverse directions in the original D5-$\overline{{\rm D}5}$ worldvolume. A sum over $a_\cL$ configurations is equivalent to a sum over the configurations of $a_0$ defined in \eqref{eq:D5_antiD5_Chan_Paton_flux_expansion}. This in turn, as $a_0$ dictates the torsional 2-cycle $\Sigma^2$ that the D3-brane is wrapping (via Poincar\'{e} duality), yields a sum over all possible 2-cycles $\Sigma^2 \in H_2(M^3;\bZ_M)$. Hence, taking into account the proper gauge theoretic normalization, our final defect is given by
\beq
\cC = \frac{1}{|H^0(M^3;\bZ_{M})|} \sum_{\Sigma^2\in H_2(M^3;\bZ_{M})} \exp\left(\frac{2\pi i}{M}\int_{\Sigma^2}\cB_2\right)\,,
\eeq
which is in exact agreement with \eqref{eq:condensate_defect} for $p=1$. We have understood the condensation defect arising in the fusion $\cN_1\otimes \cN_{-1}$ as being a sum of all possible D3-brane configurations in the original ${\rm D}5$- and $\overline{{\rm D}5}$-brane system. It is worth noting that the $a_\cL$ that the condensation defect couples to is the center-of-mass mode for the brane-antibrane configuration. Even though it arises as a difference of gauge fields, this ``difference'' is really a sum with an orientation reversal coming from the anti-D5-brane. The condensation defect aligns with the earlier observation (in the case of $\mathcal{N}_p \otimes \mathcal{N}_{p'}$) that the symmetry defects are those coupling to the center-of-mass modes for the stacks of D$p$-branes.

\begin{figure}[t!]
	\centering
	\includegraphics[width=\textwidth]{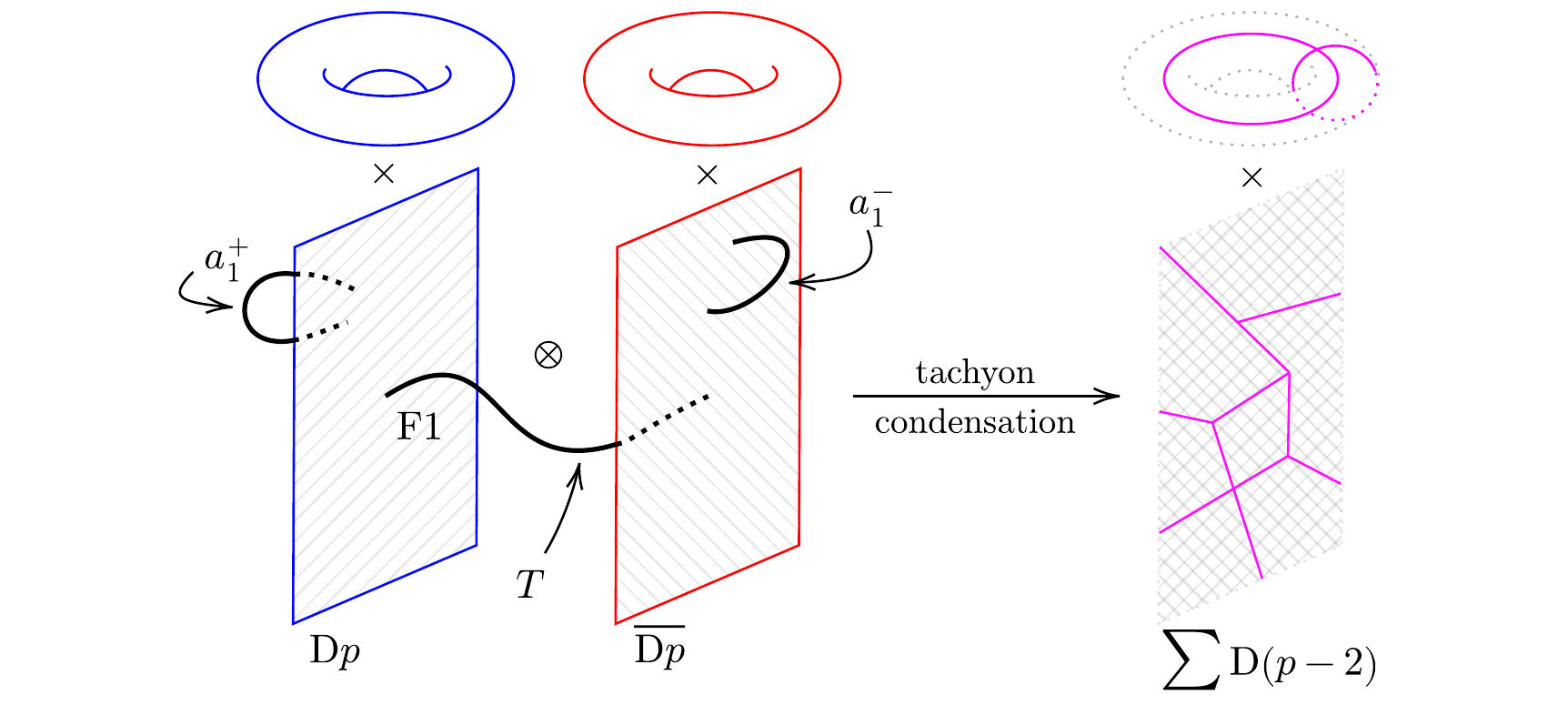}
	\caption{Condensation defect arising from tachyon condensation on a $\text{D$p$\,-$\overline{\text{D}p}$}$ system wrapping some internal manifold. We schematically decomposed the worldvolume of the branes as a direct product of internal and external manifolds. Chan-Paton modes exist both for open strings ending on the same brane or on different branes. The latter includes tachyonic modes, which, upon integrating out, lead to a condensate of lower branes within the (both internal and external) worldvolume of the original $\text{D$p$\,-$\overline{\text{D}p}$}$ system.}
	\label{fig:tachyon_condensation}
\end{figure}

We can extrapolate the above argument to the more general fusion of $\cN_p\otimes\cN_{-p}$. In this case, tachyon condensation will produce $p$ D3-branes defined on the $\Sigma^2\times \widetilde{S^2}$ submanifold. As we will elaborate upon later, this will cause each term in the sum over 2-cycles to have charge $p$ rather than charge one. As the sum is over {\it all} 2-cycles in $M^3$, this rescaling will simply reshuffle the terms in the sum and has no effect otherwise. We can then immediately see the brane realization of the more general fusion $\cN_p\otimes\cN_{-p} = \cC$ as derived in \eqref{eq:condensate_defect}.

There is one last comment to make regarding the computation of the condensation defect. When constructing the solitonic D3-brane on the D5-$\overline{{\rm D}5}$ worldvolume, the base $\widetilde{S^2}$ of the Hopf fibration was really one of the many possible choices on which to wrap the D3-brane. We could have chosen any possible linear combination of the two spheres in the base of the $T^{1,1}$ fibration and would have produced the same defect $\cU_q(\Sigma^2)$. From the point of view of the external spacetime, the defects are identical. In fact, one might argue that a more natural choice for the worldvolume of the D3-brane is one wrapping the equatorial $S^2$ of the $S^3$ cycle. For this choice the solitonic nature is more apparent as follows. When implementing tachyon condensation for a D5-$\overline{{\rm D}5}$ system, the tachyon profile is crucial in determining where the D3-brane is located. We could have picked the tachyon profile such that it approaches $T_0$, its minimum, at one pole of the $S^3$, and $-T_0$ at the other pole.\footnote{We would need to pick similar configurations in the external spacetime within $M^3$.} The tachyon will then have a profile such that it vanishes exactly along the equatorial $S^2$ specified above. The fact that this choice of internal directions for the D3-brane has no effect on the produced operator is an interesting phenomenon that we will encounter again later in this paper.

\subsection{Nature of the $\widehat{\bZ}_{M}^{(1)}$ Defect}

We have so far seen that the existence of the $\bZ_{2M}^{(0)}$ symmetry defect $\mathcal{N}_1(M^3)$ in the 5d bulk $\mathcal{W}^5$ can be attributed to a D5-brane wrapping $M^3 \times S^3$. Furthermore, the fusion of such a D5-brane with its orientation reversal produces a condensate of the $\widehat{\bZ}_M^{(1)}$ defects $\mathcal{U}_1(\Sigma^2)$ as solitonic D3-branes localized on $M^3$. A natural question to ask then is whether a $\mathcal{U}_1(\Sigma^2)$ defect defined on a generic $\Sigma^2 \not\subset M^3$ can be realized as a bona fide D3-brane existing on its own right.

The answer is actually baked in already in the presentation of \eqref{eq:condensate_brane_WZ}. Instead of considering solitonic D3-brane excitations on a D5-$\overline{{\rm D}5}$ worldvolume, we can consider a D3-brane on its own wrapping the Hopf base $\widetilde{S^2}\subset S^3$ inside the internal $T^{1,1}$, and an arbitrary external $\bZ_M$-valued cycle $\Sigma^2 \in H_2(\mathcal{W}^5;\mathbb{Z}_M)$. 
Following the identical arguments given previously, the D3-brane reproduces the $\widehat{\bZ}_M^{(1)}$ magnetic symmetry defect,
\beq\label{eq:ZMhat_generator}
\cU_1(\Sigma^2) = \exp\left(\frac{2\pi i}{M}\int_{\Sigma^2}\cB_2\right)\,.
\eeq

It is interesting to note that this symmetry defect can also be constructed by a closed fundamental (F1) string sitting in the external spacetime $\mathcal{W}^5$. As F1-strings do not couple to the Ramond-Ramond sector, the anomaly polynomial for this F1-string is simply given by
\beq
\cI_3^{\rm F1} = h_3 = db_2 - \widetilde{\beta}'(\mathcal{B}_2) \, ,
\eeq
where we used \eqref{eq:Bianchi_sols} and \eqref{eq:torsion_flux_definitions}. The fluctuating field $b_2$ is suppressed as we push the F1-string towards the conformal boundary, so what is left behind is a holonomy $\sim \exp\big(2\pi i \int_{\Sigma^2} \mathcal{B}_2/M\big)$ (up to a sign convention). We have thus showed that these F1-strings can also give rise to  the symmetry defect $\cU_1(\Sigma^2)$. 

An alternative understanding of this comes from a modification of the ``baryon vertex'' introduced in \cite{Witten:1998xy}. This utilizes the fact that F1-strings act as Wilson operators in the boundary field theory, which can be screened by operators carrying the appropriate charge. In the background we are concerned with, there are two possible causes for the screening of F1-strings: a D5-brane wrapping the entirety of the $T^{1,1}$, or a D3-brane wrapping only the $S^3\subset T^{1,1}$ \cite{Apruzzi:2021phx}. From the Bianchi identities,
\beq\label{eq:Baryon_Vertex}
\int_{T^{1,1}}dG_7 = -\int_{T^{1,1}} H_3 G_5 = -Ndb_2\,,\qquad \int_{S^3}dG_5 = -\int_{S^3}H_3G_3 = -Mdb_2\,,
\eeq
we can see that the former will screen $N$ F1-strings while the latter will screen $M$ F1-strings. The combined effect will be to screen $\gcd(N,M)=M$ F1-strings. This implies that the defect produced by an F1-string will generate a $\bZ_M$ (rather than $\bZ_N$) symmetry.

Due to the baryon vertex, a collection of $k$ F1-strings also generates the full $\widehat{\mathbb{Z}}_M^{(1)}$ symmetry as long as $\gcd(M,k) \neq 1$ (otherwise it will only generate a $\mathbb{Z}_{M/k}$ subgroup). With an eye towards the construction of the dual symmetry defect in Section \ref{sec:O_V_operators}, it is convenient for us to take $k=N/M$. One may then interpret the defect $\cU_1(\Sigma^2)$ to be either one D3-brane or $k$ coincident F1-strings. However, these descriptions are equivalent when considering the Myers Effect discussed earlier\footnote{We thank Oren Bergman for pointing this out to us.}. To see this, it is best to first consider the construction of \cite{Maldacena:2000yy} which describes an S-dual description to the KS/KT background. In that analysis, there exists nontrivial $H_3$ flux threading the $S^3$ of the internal geometry. If one imagines $k$ coincident D1-branes placed in the external spacetime of this background, the D1-branes will ``puff up'' into a single D3-brane with $k$ units of worldvolume flux through an $S^2\subset S^3$ via the standard Myers Effect. Transforming to the KS/KT background\footnote{This was described in \cite{Herzog:2001fq}, and a similar effect was discussed in \cite{Kachru:2002gs}.}, this tells us that a stack of $k$ F1-strings in the presence of a nontrivial $G_3$ background is equivalent to a D3-brane with $k$ units of worldvolume flux threading a transverse $S^2\subset S^3$, which we identify as the Hopf base $\widetilde{S^2}$ above. We thus see that the two descriptions presented above for $\cU_1(\Sigma^2)$ converge. We will use the D3-brane description, as opposed to the one involving F1-strings, in the discussion that follows.


Recall that earlier we constructed the $\mathbb{Z}_{2M}^{(0)}$ defects $\mathcal{N}_p(M^3)$ by stacking $p$ D5-branes on $M^3 \times S^3$, thanks to the fact that only the (Abelian) center-of-mass mode, but not the non-Abelian modes, of the stack contributes to this topological operator. Equivalently, we may take a D7-brane with $p$ units of worldvolume flux threading $S^2 \subset T^{1,1}$. It is straightforward to apply the same argument here to construct the general $\widehat{\mathbb{Z}}_M^{(1)}$ defects,
\begin{equation}
	\mathcal{U}_q(\Sigma^2) = \exp\bigg(\frac{2\pi i q}{M} \int_{\Sigma^2} \mathcal{B}_2\bigg) \, ,\label{eq:general_U_defects}
\end{equation}
from a stack of D3-branes wrapping $\Sigma^2 \times \widetilde{S^2}$. Again equivalently, this can also be realized by taking a D5-brane with $q$ units of worldvolume flux threading $S^2 \subset T^{1,1}$.

\paragraph{Defect fusion.}

Unlike the D5-brane operator $\mathcal{N}_p(M^3)$ on which the Chan-Paton fields form a nontrivial TQFT, the D3-brane operator $\mathcal{U}_q(\Sigma^2)$ has a relatively simple structure as seen in \eqref{eq:general_U_defects}. For two such defects wrapping the same external 2-cycle (mod $M$) $\Sigma^2 \subset \mathcal{W}^5$, their fusion obeys
\beq\label{eq:D3_general_fusion}
	\mathcal{U}_q(\Sigma^2) \otimes \mathcal{U}_{q'}(\Sigma^2) = \mathcal{U}_{q+q'}(\Sigma^2) =  \exp\bigg(\frac{2\pi i (q+q')}{M} \int_{\Sigma^2} \mathcal{B}_2\bigg) \, ,
\eeq
where the label $q$ of $\cU_q$ is only defined modulo $M$. In other words, the $\cU_q$ operators realize a group-like fusion algebra corresponding to a $\bZ_M$ group.




\section{$\mathbb{Z}_{M}^{(1)}$ and $\widehat{\bZ}_{2M}^{(2)}$ Defect Fusion}\label{sec:O_V_operators}

Let us proceed to discuss the remaining two symmetry defects resulting from the Gauss Law constraints presented in Section \ref{sec:symtft}, namely, $\cO_q$ and $\cV_p$, which respectively generate the $\bZ_M^{(1)}$ and $\widehat{\bZ}_{2M}^{(2)}$ symmetries. As we will show, these operators can also be realized by certain D-brane configurations in the KT-KS background. Remarkably, their fusion rules share some qualitative features with those we studied in the previous section, thus shedding light on the underlying brane kinematics and dynamics.


The $\bZ_M^{(1)}$ symmetry defect $\cO_q(W^2)$, as written in \eqref{eq:O_defect}, is composed of multiple 5d bulk gauge fields of different origins. Therefore, it is natural to expect that the string-theoretic construction is not given by a single type of D$p$-brane. In fact, we can construct the desired operator through a bound state of a single D3-brane and $k$ D1-branes.\footnote{We previously defined $k$ via $N=kM$. As long as $M^2\gg N \gg M$, there is no significant backreaction from the stack of $k$ probe D1-branes. We assume this limit for the above discussion.} Let us first consider a D3-brane wrapping the (homologically) nontrivial $S^2\subset T^{1,1}$. To reduce its WZ action, we have to carefully isolate the components of $G_5$ along the directions of the $S^2$. With \eqref{eq:RR_field_strength_expansion}, the full expression for $G_5 = (1+\star_{10})\cG_5$ is given by
\beq
G_5 = N\Omega_5 + g_2\omega_3 + Ng_5 - \frac{N}{2}(V_1+V_2)\star_5\sfF_2 + \star_5 g_2\omega_2\,,
\eeq
where $g_5$ is an external gauge field. By imposing the IIB Bianchi identities, we find that $\star_5 g_2$ can be solved as
\beq
\star g_2 = da_2 + \frac{1}{M}\,g_1g_2\,,
\eeq
for some globally well-defined $a_2$. Now that we have this expansion, the reduction of the D3-brane characteristic class over $S^2$ becomes
\beq
\int_{S^2}\cI_5^{\rm D3} = \int_{S^2}G_5 + G_3(B_2 + da_1) = da_2 + \frac{1}{M}\,g_1\cB_2 + g_1\left(da_1 + \frac{d\beta_1}{M}\right)\,,
\eeq
where we have made use of the expression for $g_2$ in \eqref{eq:Bianchi_sols}. We also hereafter absorb $\beta_1$ into the worldvolume Chan-Paton gauge field $a_1$ via a simple field redefinition. 

As argued previously in Section \ref{sec:N_U_operators}, the Chan-Paton field is dynamical and hence should be summed over in the worldvolume gauge theory. For the D3-brane operator at hand, performing the path integral enforces the condition $dg_1 = 0$, so we retain only the cohomological class (representative) $\cA_1$ of $g_1$ as in \eqref{eq:Bianchi_sols}.\footnote{The potential $c_0$ may be turned off via a gauge transformation.} The topological operator associated to the D3-brane wrapping $S^2\subset T^{1,1}$ is thus given by
\beq
\exp\bigg(2\pi i\int_{W^2}a_2 + \frac{2\pi i}{M} \int_{W^3} \cA_1\cB_2\bigg)\,,
\eeq
where $\d W^3 = W^2 \subset \mathcal{W}^5$.

In addition to the D3-brane configuration described above, we also consider a D1-brane in the external $\cW^5$, whose characteristic class is simply
\beq
\cI_3^{\rm D1} = G_3\big|_{\cW^5} = dc_2 \, .
\eeq
To reproduce the $\bZ_M^{(1)}$ symmetry generator, we need $k$ such D1-branes wrapping the $W^2$ of the D3-brane worldvolume, combining to give
\beq\label{eq:ZM_generator}
\cO_1(W^2,W^3) =  \exp\bigg(2\pi i\int_{W^2}(kc_2+a_2) + \frac{2\pi i}{M}\int_{W^3}\cA_1\cB_2\bigg)\,.
\eeq
Alternatively, we can derive the above expression for $\cO_1(W^2,W^3)$ by considering a D3-brane with $k$ units of worldvolume flux threading the $S^2\subset T^{1,1}$.

This combination of branes pairs nicely with the discussion of the $\cU_1(\Sigma^2)$ operator surrounding \eqref{eq:Baryon_Vertex}. There, a bound state of a single D3-brane and $k$ F1-strings gave rise to a $\widehat{\bZ}_M^{(1)}$ symmetry generator. It is natural to expect that the symmetry generator for the Pontryagin dual symmetry $\bZ_{M}^{(1)}$ should be generated by the S-dual brane configuration, i.e.~a bound state of a single D3-brane and $k$ D1-branes, which is precisely what was used to construct $\cO_1(W^2,W^3)$. 

As mentioned in Section \ref{sec:symtft}, this operator is only well-defined when $W^2 = \d W^3$. In order to find an operator that is well-defined even when $W^2$ is closed but is {\it not} a boundary, we may rewrite it in terms of auxiliary fields which we denote as $\lambda_1$ and $\varphi$,
\beq\label{eq:ZM_generator_proper}
\cO_1(W^2) = \int\cD\lambda_1\cD\varphi\exp\left(2\pi i\int_{W^2}\hat{c}_2  + \lambda_1\cA_1+ \varphi\cB_2 - M\varphi d\lambda_1\right)\,.
\eeq
We emphasize that, similarly to the worldvolume Chan-Paton field $a_1$ in $\mathcal{A}^{M,p}[\mathcal{B}_2]$ (see \eqref{eq:Z2M_generator}), the presence of the worldvolume gauge fields $\lambda_1$ and $\varphi$ can be understood in terms of anomaly inflow. The $\cA_1\cB_2/M$ term in \eqref{eq:ZM_generator} should be interpreted as an anomaly defined in one dimension higher than the D3-D1 worldvolume. Specifically, it describes a mixed anomaly between the $\mathbb{Z}_M^{(0)} \subset \mathbb{Z}_{2M}^{(0)}$ and $\mathbb{Z}_M^{(1)}$ global symmetries on the D3-D1 bound state. It is known in general that a theory with a {\it global} symmetry $G^{(k)}$ can be equivalently viewed as a {\it gauge} theory for the Pontryagin-dual symmetry $\widehat{G}^{(d-k-2)}$ \cite{Gaiotto:2014kfa}. In our case, this corresponds to a $\widehat{\bZ}_M^{(0)} \times \widehat{\bZ}_M^{(-1)}$ gauge theory on the D3-D1 worldvolume, which is exactly the TQFT $\mathcal{X}^{M,1}$ within $\cO_1(W^2)$.

\begin{figure}[t!]
	\centering
	\includegraphics[width=\textwidth]{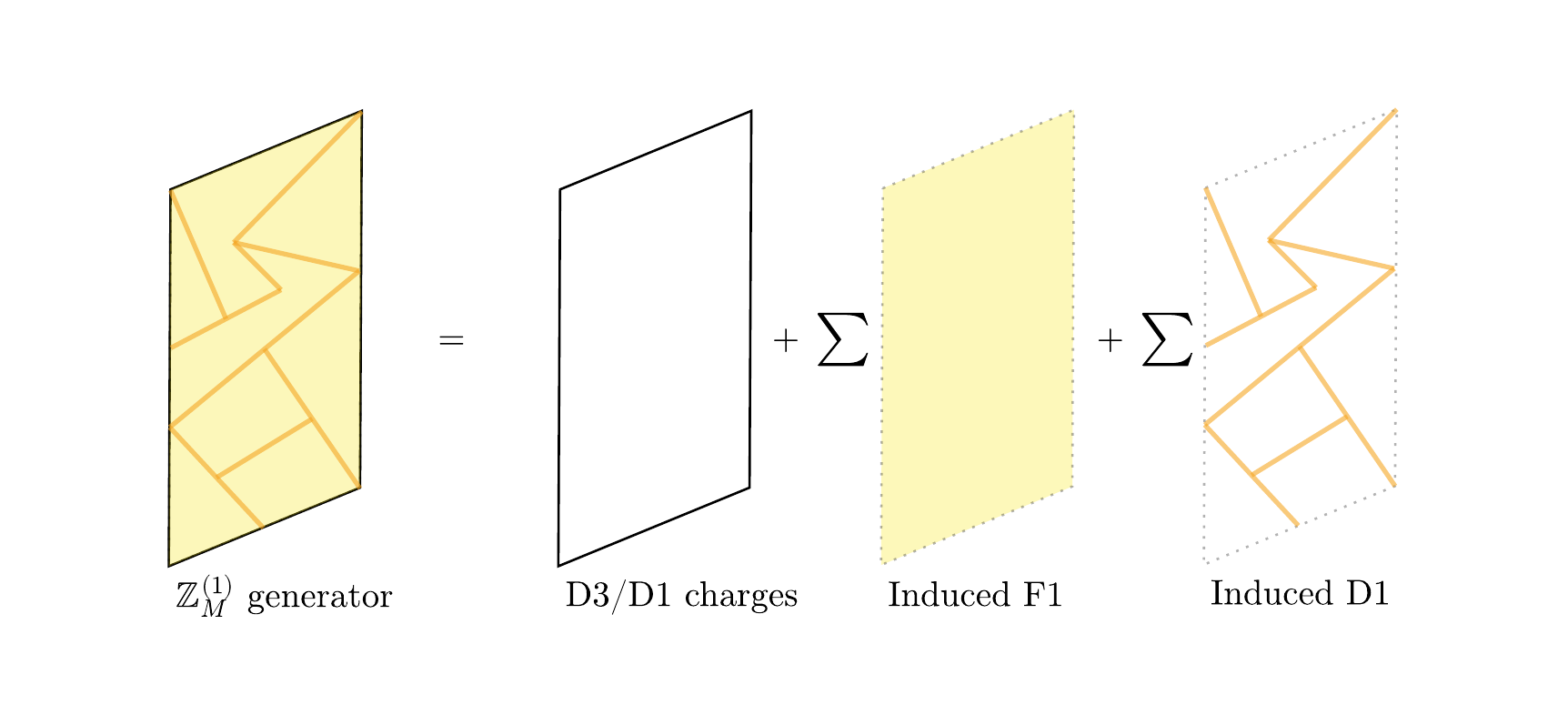}
	\caption{A schematic picture for the induced charges on the $\mathbb{Z}_M^{(1)}$ generator $\cO_q(W^2)$. In addition to the D3- and D1-brane charges from the constituent branes, there are additional induced charges describing F1-strings and D1-branes. The F1-charge is spread throughout the manifold $W^2$, while the D1-charge is localized to codimension-1 submanifolds.}
	\label{fig:induced_charge}
\end{figure}

Before proceeding, let us pause to examine the field content of the $\mathcal{O}_1(W^2)$ operator. The first term with $\hat{c}_2$ describes the D3- and D1-brane charges present in the bound state, while the final term is simply a BF term that Higgses the two gauge fields from $U(1)$ down to $\bZ_M$. On the other hand, the second and third terms can be interpreted as {\it induced} charges from F1-strings and D1-branes. To see this explicitly, note that $\varphi\in H^0(W^2;\bZ_M)$ and $\lambda_1\in H^1(W^2;\bZ_M)$ by virtue of the discussion in footnote \ref{footnote:XMq_field_content}. The coupling to $\cB_2$ in \eqref{eq:ZM_generator_proper} then implies that the D3-D1 worldvolume has a $\varphi$ number of induced F1-string charge, where $\varphi$ is summed from $0$ to $M-1$ in order to create a gauge-invariant operator. Similarly, the coupling to $\mathcal{A}_1$ in \eqref{eq:ZM_generator_proper} describes induced D1-brane charge localized on 1-cycles $\gamma^1$ which is Poincar\'e-dual to $\lambda_1$ (with respect to $W^2$). We will soon explain why $\mathcal{A}_1$ is associated with D1-branes when we discuss the $\mathcal{V}_1(\gamma^1)$ operator. See Figure \ref{fig:induced_charge} for a schematic picture of these induced charges on $\mathcal{O}_1(W^2)$.

\subsection{$\cO_q\otimes\cO_{q'}$ Fusion}

We are now interested in the fusion of these operators, which can be computed using methods similar to those utilized in Section \ref{sec:N_U_operators}. First we can consider the self-fusion of $\cO_1(W^2,W^3)$ where $W^2$ is a boundary such that $W^2 = \d W^3$. In this case, we may compute the fusion by taking the D3-D1 configuration to be a neutral D5-brane with the appropriate units of worldvolume flux. Specifically, we take two D5-branes each with worldvolume $W^2\times S^2\times \widetilde{S^2}$ and one unit of worldvolume flux threading the $\widetilde{S^2}$. Each D5-brane will have a characteristic class $\cI_{7,i}$ similar to what we used before in \eqref{eq:I9_dielectric}, that can be written as
\beq
\cI_{7,i}^{{\rm D}5} = G_5(B_2 + da_1^i + f_2^i) + \frac{1}{2}\,G_3(B_2 + da_1^i + f_2^i)^2\,,
\eeq
where $\int_{\widetilde{S^2}}f_2^i = 1$ {\it and} $\int_{S^2}f_2^i = k$ to account for the $k$ D1-branes in the bound state. Summing the two characteristic classes and performing the reduction over the internal geometry yields
\beq
\int_{S^2\times\widetilde{S^2}}\cI_{7,1}^{{\rm D}5}+\cI_{7,2}^{{\rm D}5} = 2\int_{S^2}G_5 + G_3(B_2 + da_1 + f_2^1 + f_2^2)\,,
\eeq
which is twice that of the reduction for a single D3-D1 bound state. We therefore find that the operators obey a group-like fusion,
\beq
\cO_1(W^2,W^3)\otimes\cO_1(W^2,W^3) = \cO_2(W^2,W^3)\,.
\eeq
The same can be shown to hold for general $\cO_q(W^2,W^3)$ and $\cO_{q'}(W^2,W^3)$. We emphasize that this fusion, and by extension this method, will only be valid in the case $W^2 = \d W^3$. If $W^2$ is a non-contractible closed manifold, we will need to take more care in handling the discrete worldvolume fields $\varphi$ and $\lambda_1$ discussed previously.

To account for these additional worldvolume fields, we can compute the fusion directly at the level of $\cO_1(W^2)$ or, in general, $\cO_q(W^2)$. We find, for $q'\neq -q$,
\beq\label{eq:OO_fusion}
\cO_q(W^2)\otimes\cO_{q'}(W^2) = \cX^{M,qq'(q+q')^{-1}}(W^2)\otimes\cO_{q+q'}(W^2)\,,
\eeq
where the coefficient is a decoupled 2d TQFT using the notation introduced in \eqref{eq:O1_TQFT}. The full details of this fusion can be found in Appendix \ref{sec:app_complete_fusion_rules}, but we highlight some important features here. It is first worth noting that the gauge fields $\varphi$ and $\lambda_1$ are localized worldvolume degrees of freedom coupled respectively to the 5d bulk fields $\mathcal{B}_2$ and $\mathcal{A}_1$. 
To arrive at the fusion rule \eqref{eq:OO_fusion}, we make a field redefinition analogously to what we did in \eqref{eq:Np_fusion_rule_field_redefinition}, i.e.
\begin{align}
(q+q')\widetilde{\varphi} &= q\varphi + q'\varphi'\,,& (q+q')\widetilde{\lambda}_1 &= q\lambda_1 + q'\lambda'_1\,,\nn\\
\widetilde{\varphi}' &= -\varphi + \varphi'\,,& \widetilde{\lambda}_1' &= -\lambda_1 + \lambda'_1 \, .
\end{align}
Ultimately, the $\widetilde{\varphi}$ and $\widetilde{\lambda}_1$ fields are associated to the $\cO_{q+q'}$ operator while the $\widetilde{\varphi}'$ and $\widetilde{\lambda}_1'$ fields comprise the decoupled TQFT coefficient. This lends itself to an interpretation reminiscent of that given for the fusion of the $\cN_p$ operators; the symmetry defects themselves couple to the center-of-mass modes of their corresponding brane stacks, whereas the topological sector that decouples from the system is encoded by the relative modes. In this case, the relative modes yield the 2d TQFT $\mathcal{X}^{M,qq'(q+q')^{-1}}$ arising as the fusion coefficient.

Despite the similarity, the mechanisms from which the 3d TQFT $\mathcal{A}^{M,pp'(p+p')}$ and the 2d TQFT $\mathcal{X}^{M,qq'(q+q')^{-1}}$ arise are slightly different. In the former case, since the Chan-Paton field $a_1$ is what couples to $\mathcal{B}_2$ in the 3d worldvolume action of $\mathcal{N}_p$, it captures
both the center-of-mass and relative modes of the brane stacks. However, in the latter case, these modes are instead captured by the pair of gauge fields $\varphi$ and $\lambda_1$, which respectively couple to $\mathcal{B}_2$ and $\mathcal{A}_1$ in the 2d worldvolume action of $\mathcal{O}_q$. 

\subsection{$\cO_q\otimes\cO_{-q}$ Fusion}

Due to the existence of the nontrivial 2d TQFT $\cX^{M,q}[\cA_1,\cB_2]$ within the $\cO_q(W^2)$ operator, one naturally expects it to be non-invertible. The fusion of $\cO_q$ with its orientation reversal produces another condensation defect $\widetilde{\cC}$ through
\begin{equation}
	\mathcal{O}_q(W^2) \otimes \mathcal{O}_{-q}(W^2) = \frac{1}{|H^0(W^2;\mathbb{Z}_M)|} \bigoplus_{\substack{\gamma^1 \in H_1(W^2;\mathbb{Z}_{2M}) \\ \Sigma^2 \in H_2(W^2;\mathbb{Z}_M)}} \mathcal{V}_2(\gamma^1) \otimes \mathcal{U}_1(\Sigma^2) \coloneq\widetilde{\cC}(W^2)\,.\label{eq:Oq_condensate}
\end{equation}
More technical details are provided in Appendix \ref{sec:app_complete_fusion_rules}. The presence of this sum over operators can be understood as a consequence of the discussion following \eqref{eq:ZM_generator_proper}, as it is simply enumerating the additional induced charges on the worldvolume $W^2$ of the D3-D1 operator.

We can interpret this condensation defect analogously to what we did to that in Section \ref{sec:N_U_operators}: through tachyon condensation between the D3-D1 bound state and its orientation reversal. In this case, the overall D3-brane charge and the overall D1-brane charge (from the $k$ D1-branes in $\mathcal{W}^5$) will annihilate completely, but the {\it induced} charges will not. This logic is the same as that of the non-invertible fusion of \cite{GarciaEtxebarria:2022vzq}. As the initial configuration is a product of {\it sums} of the induced states, the final result will be a sum over charges that have not been completely annihilated. The charges remaining in the (D3-D1)-($\overline{{\rm D}3\text{-}{\rm D}1}$) configuration are those of the F1-strings induced in the worldvolume, and some remaining D1-brane charge that can be seen coming from the coupling to $\lambda_1$ in the definition of $\cO_q(W^2)$. The reason why the sum includes terms of the form $\cV_2\otimes\cU_1$ can be understood from the BF term in \eqref{eq:ZM_generator_proper}. It acts as a constraint that the induced charges are equal for the F1-strings and D1-branes, so that we do not sum over {\it all} insertions of the $\widehat{\bZ}_{2M}^{(2)}$ symmetry generator but only those which satisfy the imposed charge constraint.

It is clear from the discussion following \eqref{eq:ZMhat_generator} that the F1-strings in the external spacetime represent the $\cU_1(\Sigma^2)$ factors in the condensation defect $\widetilde{\mathcal{C}}(W^2)$. The question remains of where the remaining D1-brane charge is localized. As stated previously, this D1-brane charge is induced on the original D3-D1 worldvolume and should not be thought of as a stand-alone D1-brane. We should interpret the induced charge as a D1-brane soliton on the worldvolume of the bound state, which is picked out by the tachyon condensation process. Specifically, we see that the net D1-brane charge is localized to a torsional 1-cycle $\gamma^1\in H_1(W^2;\bZ_M)$ in the external spacetime. In order for the soliton D1-brane to be stable, there must be a locus in the internal geometry where the tachyon profile vanishes. Analogously to the discussion in Section \ref{sec:N_U_operators} for the D5-$\overline{{\rm D}5}$ condensation, we can pick the tachyon profile such that it reaches its minimum $T_0$ at one pole of the internal $S^2$, and takes the value $-T_0$ at the other. This leaves a codimension-1 locus where the tachyon vanishes, and thus where a solitonic D1-brane can live. For simplicity, we take this to be the $\widetilde{S^1}$ equator of the $S^2$ cycle wrapped by the original D3-D1 bound states. The end result is then a collection of F1-string operators $\cU_1(\Sigma^2)$, and a solitonic D1-brane operator $\cV_2(\gamma^1)$.

We shall emphasize that the D1-brane operator $\cV_2(\gamma^1)$ resulting from tachyon condensation is stable only because of its solitonic nature. To construct a generic $\widehat{\mathbb{Z}}_{2M}^{(2)}$ defect from a stable D1-brane configuration, as we will do in the next subsection, we have to consider a different $S^1$ to wrap in the internal geometry of the KT-KS background. This is the same phenomenon we witnessed in constructing the condensation defect $\cC$ in Section \ref{sec:N_U_operators}, namely, when constructing symmetry operators from D$p$-branes, there is an ambiguity arising from which internal manifolds the branes wrap. In fact, multiple distinct configurations may lead to the same 5d bulk operator that cannot otherwise be distinguished. For the setup of interest, it is plausible that such a degeneracy can be lifted if we also consider the charges of the operators under the $SU(2) \times SU(2)$ isometries, corresponding to the $S^2 \times S^2$ base of $T^{1,1}$ in the KT-KS background.

\subsection{Nature of the $\widehat{\bZ}_{2M}^{(2)}$ Defect}

As alluded to earlier, one can construct a stand-alone $\widehat{\mathbb{Z}}_{2M}^{(2)}$ symmetry operator,
\beq\label{eq:Z2Mhat_generator}
\cV_1(\gamma^1) = \exp\left(\frac{2\pi i}{2M}\int_{\gamma^1}\cA_1\right)\,,
\eeq
with $\gamma^1 \in H_1(\mathcal{W}^5;\mathbb{Z}_{2M})$, by wrapping a D1-brane on the $S^1$ Hopf fiber of the $T^{1,1}$. The stability of this D1-brane can be attributed to the fact that the $S^1$ fiber is non-vanishing over the base $\widetilde{S^2}$, such that the brane is protected from shrinking to zero.

The explicit construction of $\mathcal{V}_1(\gamma^1)$ follows closely to that of $\mathcal{U}_q(\Sigma^2)$ in Section \ref{sec:N_U_operators}, i.e.~if we na\"ively reduce the characteristic class over the Hopf fiber, we find
\beq
\int_{S^1\subset S^3}\cI_3^{{\rm D}1} = \cI_2 \ \text{``}\!=\!\text{''} \ 0
\eeq
Much akin to the construction of the $\widehat{\bZ}_M^{(1)}$ operator, this is simply a side effect of the descent formalism overlooking information regarding discrete holonomies in our theory. The appropriate torsional flux that should appear this time can be anticipated, using the simple fact that the $S^1$ Hopf fiber is associated with the $U(1)$, or more precisely, $\mathbb{Z}_{2M}$ $R$-symmetry. Following the same arguments as presented around \eqref{eq:discrete_holonomy_argument}, we may express $\cI_2 = \widetilde{\beta}(\cA_1) \in H^2(\mathcal{W}^5;\mathbb{Z}_{2M})$, and by wrapping the D1-brane on a cycle $\gamma^1$ such that $\partial\widetilde{\gamma}^2 = 2M\gamma^1$, one can reproduce \eqref{eq:Z2Mhat_generator}. More generally, the operator
\beq
\cV_p(\gamma^1) = \exp\left(\frac{2\pi i p}{2M}\int_{\gamma^1}\cA_1\right) \, ,
\eeq
where $p \sim p+2M$, can be constructed from a stack of $p$ D1-branes.

It is straightforward to show that these operators obey a group-like fusion rule, i.e.
\begin{equation}
	\mathcal{V}_p(\gamma^1) \otimes \mathcal{V}_{p'}(\gamma^1) = \mathcal{V}_{p+p'}(\gamma^1) \, .
\end{equation}
Similarly to before, the fusion rule can be understood by considering the D1-brane used to construct $\cV_1(\gamma^1)$ as being instead a neutral D3-brane wrapping $\gamma^1\times S^1\times S^2$ with one unit of worldvolume flux through the sphere $S^2$. Bringing two such D3-branes together is equivalent to a single D3-brane with two units of worldvolume flux, which we can then extrapolate to arbitrary $p = 1,2,\dots,2M-1 \in \bZ_{2M}$.



\section{Cross-symmetry Defect Fusion and Braiding}\label{sec:cross_fusions}

\subsection{Dissolved Branes}

We discussed in Sections \ref{sec:N_U_operators} and \ref{sec:O_V_operators} the brane kinematics and dynamics underlying the parallel fusion rules for $\mathcal{N}_p \otimes \mathcal{N}_{p'}$, $\mathcal{O}_q \otimes \mathcal{O}_{q'}$, $\mathcal{U}_q \otimes \mathcal{U}_{q'}$, and $\mathcal{V}_p \otimes \mathcal{V}_{p'}$. These fusion rules alone, however, only form part of the full-fledged story involving the interplay between defects associated with different bulk gauge symmetries, as summarized in Table \ref{tab:fusion_rules} and derived in Appendix \ref{sec:app_complete_fusion_rules}. In the following, we will discuss the brane origin of the rest of these fusion rules.

Let us begin with the fusion rule,
\begin{equation}
    \mathcal{U}_q(\Sigma^2) \otimes \mathcal{N}_p(M^3) = \mathcal{N}_p(M^3) \, .\label{eq:U_N_fusion}
\end{equation}
The D-branes associated with $\mathcal{U}_q(\Sigma^2)$ and $\mathcal{N}_p(M^3)$ are respectively $q$ D3-branes wrapping $\Sigma^2 \times \widetilde{S^2}$ and $p$ D5-branes wrapping $M^3 \times S^3$, where $\Sigma^2 \subset M^3$ and $\widetilde{S^2} \subset S^3$. It now becomes clear why \eqref{eq:U_N_fusion} should be the way it is. Since the entire worldvolume of the D3-branes is a submanifold of the D5-branes, the former can be absorbed by the latter without a cost. This can be understood by the fact that D3-branes are dissolved in the D5-branes through the WZ coupling $da_1 C_4$, where we integrate over all configurations of the (dynamical) Chan-Paton field $a_1$. The effect of embedding extra D3-branes into the D5-branes then is merely to reshuffle the summands in the said integration.\footnote{If we instead interpret the defect $\mathcal{U}_q(\Sigma^2)$ to be coming from F1-strings (or more generally, a D3-F1 bound state), we can understand the fusion rule as the simple statement that a D5-brane freely absorbs F1-strings.}

Moreover, adding D3-branes to the condensation defect $\mathcal{C}(M^3)$, which is itself a sum of all D3-branes localized on $M^3 (\times S^3)$, has the identical effect of reshuffling the summands in the condensation defect. This naturally explains the fusion rule,
\begin{equation}
    \mathcal{U}_q(\Sigma^2) \otimes \mathcal{C}(M^3) = \mathcal{C}(M^3) \, .\label{eq:U_C_fusion}
\end{equation}
Alternatively, locality of the fusion allows us to interpret \eqref{eq:U_C_fusion} from a different but equivalent perspective, i.e.
\begin{equation}
    \mathcal{U}_q \otimes \mathcal{C} = \mathcal{U}_q \otimes (\mathcal{N}_p \otimes \mathcal{N}_{-p}) = (\mathcal{U}_q \otimes \mathcal{N}_p) \otimes \mathcal{N}_{-p} = \mathcal{N}_p \otimes \mathcal{N}_{-p} = \mathcal{C} \, .
\end{equation}
The expression above tells us that $\mathcal{U}_q \otimes \mathcal{C}$ can equivalently be thought of as a two-step process, namely, absorption of D3-branes by D5-branes, followed by tachyon condensation on the D5-$\overline{\text{D5}}$ system. Analogous arguments can be used to provide a brane interpretation of the following fusion rules,
\begin{equation}
    \begin{aligned}
        \mathcal{V}_p(\gamma^1) \otimes \mathcal{O}_q(W^2) & = \mathcal{O}_q(W^2) \, , & \qquad \mathcal{V}_p(\gamma^1) \otimes \widetilde{\mathcal{C}}(W^2) & = \widetilde{\mathcal{C}}(W^2) \, ,\\
        \mathcal{U}_q(\Sigma^2) \otimes \mathcal{O}_{q'}(W^2) & = \mathcal{O}_{q'}(W^2) \, , & \qquad \mathcal{U}_q(\Sigma^2) \otimes \widetilde{\mathcal{C}}(W^2) & = \widetilde{\mathcal{C}}(W^2) \, .
    \end{aligned}
\end{equation}

We now turn our focus to the fusion between the $\mathbb{Z}_{2M}^{(0)}$ defect $\mathcal{N}_p(M^3)$ with its associated condensation defect $\mathcal{C}(M^3)$. To motivate the subsequent explanation, we utilize the associativity of the fusion algebra again to rewrite the fusion in the following manner,
\begin{align}
    \mathcal{N}_p \otimes \mathcal{C} & = \mathcal{N}_p \otimes (\mathcal{N}_{p'} \otimes \mathcal{N}_{-p'})\nonumber\\
    & = (\mathcal{N}_p \otimes \mathcal{N}_{p'}) \otimes \mathcal{N}_{-p'}\nonumber\\
    & = \mathcal{A}^{M,pp'(p+p')} \otimes \mathcal{N}_{p+p'} \otimes \mathcal{N}_{-p'}\nonumber\\
    & = \mathcal{A}^{M,pp'(p+p')} \otimes \mathcal{A}^{M,-pp'(p+p')} \otimes \mathcal{N}_p \, ,\label{eq:N_C_fusion}
\end{align}
assuming $\gcd(M,p+p')=1$. As per the discussion in Section \ref{sec:N_U_operators}, there is a center-of-mass mode associated with each stack of D5-branes (or $\overline{\text{D5}}$-branes). Suppose we first fuse $p$ and $p'$ D5-branes. This results in a stack of $p+p'$ D5-branes, in addition to the relative mode $\mathcal{A}^{M,pp'(p+p')}$ between these two stacks. Next, we fuse this new stack of $p+p'$ D5-branes with $p'$ $\overline{\text{D5}}$-branes, which results in a stack of $p$ D5-branes, in addition to the relative mode $\mathcal{A}^{M,-pp'(p+p')}$ between the two original stacks. It turns out that these two minimal TQFTs, each of which is associated with an independent $\mathbb{Z}_M$ symmetry, combines to form a $\mathbb{Z}_M \times \mathbb{Z}_M$ 3d DW theory $(\mathcal{Z}_M)_0$ \cite{Hsin:2018vcg} (see Appendix \ref{sec:app_complete_fusion_rules} for details).

This result suggests that conceptually, rather than thinking of the condensation defect as a collection of D3-branes, it is more appropriate to think of the defect as a ``zero-charge'' D5-brane with solitonic D3-branes dissolved within, with an emphasis on ``D5-brane'' such that we can associate it with an appropriate notion of a center-of-mass mode. Accordingly, the decoupled DW theory in the fusion $\mathcal{N}_p \otimes \mathcal{C} = (\mathcal{Z}_M)_0 \otimes \mathcal{N}_p$ can be interpreted as the relative mode between the $p$ D5-branes and such a neutral D5-brane. Similar remarks apply to the remaining nontrivial fusion rules in Table \ref{tab:fusion_rules}, i.e.
\begin{equation}
    \mathcal{C} \otimes \mathcal{C} = (\mathcal{Z}_M)_0 \otimes \mathcal{C} \, , \quad \mathcal{O}_q \otimes \widetilde{\mathcal{C}} = (\mathcal{X}^{M,-1})^{\otimes 2} \otimes \mathcal{O}_q \, , \quad \widetilde{\mathcal{C}} \otimes \widetilde{\mathcal{C}} = (\mathcal{X}^{M,-1})^{\otimes 2} \otimes \widetilde{\mathcal{C}} \, .
\end{equation}

\subsection{Commutation Relations}

What we have worked out earlier in this paper is the full set of {\it parallel} fusion rules for the symmetry operators we constructed from D-branes. While the fusion algebra plays a crucial role in the interplay between the symmetry defects in the bulk, an equally important piece of information is the braiding rules of these defects. Here we study the case where there is a nontrivial linking between the manifolds on which the defects are defined. Depending on the field content of these defects, there may be nontrivial commutation relations, meaning that they are charged under each other.

To determine which pairs of operators are mutually charged, we first recall from the SymTFT \eqref{eq:SymTFT} that the conjugate momenta of the various 5d gauge fields are given by
\begin{equation}
    \begin{aligned}
        \Pi_{b_2} & = -2\pi M\hat{c}_2 \, , & \qquad \Pi_{\hat{c}_2} & = 2\pi M b_2 \to -2\pi\cB_2\,,\\
        \Pi_{\sfA_1} & = 4\pi Mc_3 \, , & \qquad \Pi_{c_3} & = -4\pi M \sfA_1 \to -2\pi\cA_1\,.
    \end{aligned}
\end{equation}
Importantly, we see that $(b_2,\hat{c}_2)$ and $(\sfA_1,c_3)$ are conjugate pairs. Note that at low energies, the fluctuating fields $Mb_2$ and $2M\sfA_1$ are respectively Higgsed to the discrete gauge fields $-\mathcal{B}_2 \in H^2(\mathcal{W}^5;\mathbb{Z}_M)$ and $\mathcal{A}_1 \in H^1(\mathcal{W}^5;\mathbb{Z}_{2M})$ via the St\"{u}ckelberg mechanism.

It follows that the commutation relation between the $\widehat{\mathbb{Z}}_M^{(1)}$ defect $\mathcal{U}_q(\Sigma^2)$ and the $\mathbb{Z}_M^{(1)}$ defect $\mathcal{O}_{q'}(W^2)$ is
\begin{equation}
    \mathcal{U}_q(\Sigma^2) \, \mathcal{O}_{q'}(W^2) = \exp\bigg(\frac{2\pi i qq'}{M} \, \ell(\Sigma^2,W^2)\bigg) \, \mathcal{O}_{q'}(W^2) \, \mathcal{U}_q(\Sigma^2) \, ,\label{eq:U_O_commutation_relation}
\end{equation}
where $\ell(\Sigma^2,W^2)$ denotes the linking number of $\Sigma^2$ and $W^2$ with respect to $\mathcal{W}^5$.\footnote{The linking number we are using in this context is defined in the sense of \cite{Bott1982}. By construction, $\ell(\Sigma^2,W^2)=0$ if $\Sigma^2 \subset W^2$, such that it is compatible with the parallel fusion rule $\mathcal{U}_q(\Sigma^2) \otimes \mathcal{O}_{q'}(W^2) = \mathcal{O}_{q'}(W^2)$.} As discussed in \cite{Moore:2004jv,Freed:2006yc,Freed:2006ya,Etheredge:2023ler}, the nontriviality of the commutation relation is very natural when considering the D$p$-branes associated with these operators. In Type II string theory, D$p$-branes are (electromagnetically) dual to D$(6-p)$-branes, and particularly, D3-branes are self-dual.\footnote{This duality is not to be confused with the $SL(2,\mathbb{Z})$ duality of Type IIB string theory, under which the pair $(B_2,C_2)$ transforms as a doublet while $C_4$ transforms as a singlet.} Therefore one does expect the commutation relation between $\mathcal{U}_q(\Sigma^2)$ and $\mathcal{O}_{q'}(W^2)$, both of which are constructed from D3-branes, to be nontrivial. By the same token, the $\widehat{\mathbb{Z}}_{2M}^{(2)}$ defect $\mathcal{V}_p(\gamma^1)$ (from D1-branes) and the $\mathbb{Z}_{2M}^{(0)}$ defect $\mathcal{N}_{p'}(M^3)$ (from D5-branes) do not commute. More explicitly, we find
\begin{equation}
    \mathcal{V}_p(\gamma^1) \, \mathcal{N}_{p'}(M^3)  = \exp\bigg(\!-\frac{2\pi i pp'}{2M} \, \ell(\gamma^1,M^3)\bigg) \, \mathcal{N}_{p'}(M^3) \, \mathcal{V}_p(\gamma^1) \,.\label{eq:V_N_commutation_relation}
\end{equation}
where $\ell(\gamma^1,M^3)$ again denotes the linking number of $\gamma^1$ and $M^3$ with respect to $\cW^5$.

For completeness, let us provide a broad-brush argument for the commutation relation between $\mathcal{N}_p(M^3)$ and $\mathcal{O}_q(W^2)$, as was also studied by \cite{Apruzzi:2022rei}. The corresponding manifolds typically do not link in $\mathcal{W}^5$, in which case these two operators commute. However, if we consider a refined case where $W^2 \cong S^1 \times L^1$, with $S^1 \subset M^3$, then heuristically the commutation relation takes the following form,
\begin{equation}
    \mathcal{N}_p(M^3) \, \mathcal{O}_q(S^1 \times L^1) \sim \exp\bigg(\!-2\pi i pq \, \ell(L^1,M^3) \int_{S^1} a_1\bigg) \, \mathcal{O}_q(S^1 \times L^1) \, \mathcal{N}_p(M^3) \, .\label{eq:N_O_commutation_relation}
\end{equation}
The ``charge'' here is not a c-number phase, but rather a holonomy of the Chan-Paton field of (multiple) open F1-strings, whose endpoints trace out the two copies of $S^1$ respectively embedded in $M^3$ and $W^2$. Via the open-closed string duality, one may equivalently interpret this process as the creation of a 2d worldsheet of closed F1-strings when the D3-brane operator $\mathcal{O}_q(S^1 \times L^1)$ is dragged through the D5-brane operator $\mathcal{N}_p(M^3)$. Note that the Chan-Paton field $a_1$ is a trivialization of the NSNS field $b_2$ on the boundary of this worldsheet (with $b_2 \to -\mathcal{B}_2/M$ at low energies), so we may schematically express the action of $\mathcal{N}_p(M^3)$ on $\mathcal{O}_q(S^1 \times L^1)$ as
\begin{equation}
    \mathcal{N}_p(M^3): \quad \mathcal{O}_q(S^1 \times L^1) \to \exp\bigg(\frac{2\pi i pq}{M} \, \ell(L^1,M^3) \int_{S^1 \times I} \mathcal{B}_2\bigg) \, \mathcal{O}_q(S^1 \times L^1)  \, ,
\end{equation}
where $I$ is the interval along the transverse direction between $\mathcal{N}_p(M^3)$ and $\mathcal{O}_q(S^1 \times L^1)$. Such a phenomenon is akin to the Hanany-Witten transition \cite{Hanany:1996ie}, where a D3-brane is created when a D5-brane crosses an NS5-brane (each wrapping manifolds along appropriate directions). By successively applying T and S dualities on this brane system, i.e.
\begin{equation*}
    (\text{NS5--D3--D5}) \xlongrightarrow{\text{T}^2} (\text{NS5--D1--D3}) \xlongrightarrow{\text{S}} (\text{D5--F1--D3}) \, ,
\end{equation*}
we arrive at the desired combination of branes in our construction.




\section{Boundary Conditions and Field Theory}\label{sec:boundary_conditions}

In the previous sections, we have been primarily concerned with the presence and action of the symmetry defects in the 5d bulk gauge theory, where all the discussed operators are realized and interact in nontrivial ways. While the bulk alone has an involved structure, there is another half of the story, namely, the symmetry structure of the 4d boundary field theory is determined by choices of (Neumann or Dirichlet) boundary conditions for the bulk gauge fields \cite{Maldacena:2001ss,Belov:2004ht,Gaiotto:2014kfa,Gaiotto:2020iye,Bah:2020uev}. Importantly, these choices also prescribe the spectrum of topological operators existing on the boundary.

\subsection{Allowed Choices from the SymTFT}

Owing to the canonical quantization imposed on conjugate gauge fields, one cannot simultaneously pick Neumann (or Dirichlet) boundary conditions on the pair of fields. In addition, there are potential obstructions from the anomaly terms in the SymTFT that forbids us from imposing Neumann boundary condition on a given gauge field. In the following, we are going to systematically study all the allowed choices of boundary conditions for the 5d bulk gauge fields in the KT-KS setup.

Recall that the SymTFT is given by
\beq\label{eq:SymTFT_BCs}
S_{\rm top} = 2\pi\int_{\cW^5}\bigg(Mb_2d\hat{c}_2 - 2Mc_3 d\sfA_1 + b_2\cA_1\cB_2 - \sfA_1\cB_2^2 - \frac{\mathcal{K}}{M}\,\cA_1\beta(\cA_1)\beta(\cA_1)\bigg)\,,
\eeq
where the first two terms are BF couplings respectively pairing $b_2$ and $\hat{c}_2$, and $c_3$ and $\sfA_1$. For concreteness, let us consider the former pair; the story for the latter pair is completely analogous. Varying the topological action, we find
\beq
\delta S_{\rm BF} = 2\pi\int_{\d\cW^5} Mb_2\delta\hat{c}_2 \,,
\eeq
where we have made a choice in having the variation on $\hat{c}_2$. By adding a boundary term we may equivalently view the variation as $\delta S_{\rm BF}' = 2\pi\int_{\d\cW^5}M\hat{c}_2\delta b_2$. Because the variation is first-order in derivatives, we must impose Dirichlet boundary condition on one of the gauge fields, and Neumann boundary condition on the other to have a well-defined variational problem.\footnote{These boundary conditions are {\it topological} boundary conditions as they do not require a metric on the boundary manifold, and are indeed invariant under diffeomorphisms of the boundary manifold. One can also discuss topological boundary conditions in theories other than BF theories. One such example is 3d Abelian Chern-Simons theory, where the existence of topological boundary conditions depends on the choice of the coefficient matrix for the theory \cite{Kapustin:2010hk}.} The difference between these two choices corresponds to different global structures for the gauge group of the 4d $\cN=1$ SYM field theory living on the boundary $\mathcal{W}^4 = \d\cW^5$. Specifically, an $SU(M)$ gauge theory has a $\bZ_M^{(1)}$ (electric) global symmetry, whereas a $PSU(M)$ gauge theory has a $\widehat{\bZ}_M^{(1)}$ (magnetic) global symmetry. In general, one may consider a larger family of boundary conditions when $M = kk'$ is not prime. This realizes a $\bZ_k\times\bZ_{k'}$ global symmetry on the boundary 4d $\mathcal{N}=1$ $SU(M)/\bZ_k$ SYM theory. We leave the study of such cases to future work.

The remaining cubic terms in \eqref{eq:SymTFT_BCs} place additional constraints on the allowed choices of boundary conditions for the bulk gauge fields. 
From the point of view of the boundary field theory, these interaction terms describe 't Hooft anomalies between different global symmetries. For example, the $\mathsf{A}_1 \mathcal{B}_2^2$ term is a mixed 't Hooft anomaly between the $\bZ_{2M}^{(0)}$ and $\bZ_M^{(1)}$ symmetries. Imposing Neumann boundary conditions on $b_2$ (and $\cB_2$ by extension) amounts to gauging the $\bZ_M^{(1)}$ symmetry on the boundary, which causes the $\bZ_{2M}^{(0)}$ symmetry to become non-invertible \cite{Kaidi:2021xfk,Apruzzi:2021phx,Apruzzi:2022rei}. Meanwhile, the self-anomaly term $\mathcal{A}_1 \beta(\mathcal{A}_1) \beta(\mathcal{A}_1)$ indicates $\sfA_1$ (and $\mathcal{A}_1$) {\it cannot} be given Neumann boundary conditions, reflecting the fact that one cannot gauge the anomalous $\bZ_{2M}^{(0)}$ R-symmetry in the 4d $\mathcal{N}=1$ SYM field theory.

\subsection{Boundary Conditions and Branes}

The choice of boundary conditions for the bulk supergravity fields determines the set of allowed configurations of the probe D$p$-branes in $\mathcal{W}^5$. {\it A priori}, there are $2^4 = 16$ choices of boundary conditions on the set of fields $(b_2,\hat{c}_2,c_3,\sfA_1)$. However, $\sfA_1$ only admits Dirichlet boundary condition due to its self-anomaly, so its conjugate, $c_3$, only admits Neumann boundary condition. This narrows the number of combinations down to 2. Table \ref{tab:boundary_conditions} summarizes the consequences of each of these two choices in the corresponding boundary field theories.

\renewcommand{\arraystretch}{1.5}
\begin{table}[t!]
\begin{center}
\begin{tabular}{|| c | c | c ||}
\hline
Boundary Conditions & (D,N,N,D) & (N,D,N,D)\\
\hline\hline
Gauge Group & $SU(M)$ & $PSU(M)$\\
Global Symmetries & $\bZ_{2M}^{(0)},\;{\bZ}_M^{(1)}$ & $\bZ_{2M}^{(0)},\;\widehat{\bZ}_M^{(1)}$\\
Gauged Symmetries & $\widehat{\bZ}_{2M}^{(2)},\;\widehat{\bZ}_M^{(1)}$ & $\widehat{\bZ}_{2M}^{(2)},\;{\bZ}_M^{(1)}$\\
Symmetry Generators & $\widetilde{\cN}_p(M^3),\;\widetilde{\cO}_q(W^2)$ & ${\cN}_p(M^3),\;\cU_q(\Sigma^2)$\\
Charged Operators & $\overline{\cV}(\overline{x}),\;\overline{\cU}(\overline{\gamma}^1)$ & $\overline{\cV}(\overline{x}),\;\overline{\cO}(\overline{W}^1)$\\
\hline
\end{tabular}
\caption{\label{tab:boundary_conditions} The allowed boundary conditions for the four bulk gauge fields $(b_2,\hat{c}_2,c_3,\sfA_1)$, and their effect on the boundary field theory. Here and in the main text ``N'' (``D'') denotes Neumann (Dirichlet) boundary condition imposed on a given gauge field.}
\end{center}
\end{table}

\paragraph{(D,N,N,D) boundary conditions.} To see the effect of boundary conditions concretely, let us again focus on the $(b_2,\hat{c}_2)$ BF pair discussed in the previous subsection, keeping in mind that $c_3$ and $\sfA_1$ must be given Neumann and Dirichlet boundary conditions respectively. We first consider the case where we impose Neumann boundary condition on $\hat{c}_2$. By the variational principle,
\beq\label{eq:variational_b2c2hat}
Mb_2\delta\hat{c}_2\big|_{\mathcal{W}^4} = 0\,,
\eeq
we must require that $Mb_2$, and by extension $\cB_2$, vanish on the conformal boundary $\mathcal{W}^4$. In terms of the symmetry structure of the boundary field theory, this is equivalent to choosing a gauge group of $SU(M)$ rather than $PSU(M)$.

With this choice of boundary conditions it is straightforward to see from \eqref{eq:ZMhat_generator} that $\cU_q(\Sigma^2)$ becomes a trivial operator when pushed to the boundary. This is expected for a theory with gauge group $SU(M)$, as the defects $\cU_q$ would generate the magnetic 1-form symmetry not present in this theory. 
Instead of having the D3-branes being pushed {\it parallel} to $\mathcal{W}^4$, we can consider {\it transverse} D3-branes such that one of their external dimensions is aligned with the radial coordinate of $\cW^5$. Using \eqref{eq:variational_b2c2hat}, we can write $b_2$ in terms of a $\bZ_M$-valued flat connection $\cB_1$ defined only on the conformal boundary, i.e.
\beq
Mb_2\big|_{\mathcal{W}^4} = d\cB_1 = 0\,,
\eeq
in which case the resultant operator becomes
\beq\label{eq:Wilson_line_op}
\overline{\cU}(\overline{\gamma}^1\subset\mathcal{W}^4) = \exp\left(\frac{2\pi i}{M}\int_{\overline{\gamma}^1}\cB_1\right)\,,
\eeq
with $\gamma^1$ closed. The commutation relation \eqref{eq:U_O_commutation_relation}, when localized to $\cW^4$, tells us that such a line operator living in the boundary field theory is charged under the $\bZ_M^{(1)}$ electric symmetry generated by the $\cO_q$ defects. Following the discussion below \eqref{eq:ZMhat_generator}, this could have been equivalently argued from an alternative construction of the $\cU_q$ defects using F1-strings. It is known in the literature that F1-strings ending on the conformal boundary trace out Wilson lines \cite{Maldacena:1998im}. The line operators $\overline{\cU}(\overline{\gamma}^1)$ should therefore be identified with these Wilson lines in the $SU(M)$ gauge theory.\footnote{More accurately, the $\overline{\cU}$ operators describe the {\it topological data} of a Wilson line in the picture of \cite{Gaiotto:2020iye,Freed:2022qnc}.}

Similarly to before, imposing Neumann boundary condition on $c_3$ leads to
\beq
2M\sfA_1\big|_{\cW^4} = d\cA_0 = 0\,,
\eeq
where $\cA_0 \in \{0,\dots,2M-1\}$ is a flat $\bZ_{2M}$-valued 0-form field. 
This corresponds to a local operator $\overline{\cV}(\overline{x}\in \cW^4)$, where $\overline{x} = \d\gamma^1$, that is charged under the $\mathbb{Z}_{2M}^{(0)}$ symmetry generated by the $\mathcal{N}_p$ defects according to \eqref{eq:V_N_commutation_relation}. 

The (D,N,N,D) boundary conditions also affect the non-invertibility of the $\bZ_M^{(1)}$ symmetry generator. From either \eqref{eq:ZM_generator} or \eqref{eq:ZM_generator_proper}, we see that $\cO_q$ simplifies dramatically when $\cA_1$ and $\cB_2$ have Dirichlet boundary conditions, i.e.
\beq
\widetilde{\cO}_q\left(W^2\subset\mathcal{W}^4\right) = \exp\left(2\pi i q\int_{W^2}\hat{c}_2\right)\,.
\eeq
Note that the $b_2 \mathcal{A}_1 \mathcal{B}_2$ anomaly trivializes as both fields $\cA_1$ and $\cB_2$ are turned off on $\mathcal{W}^4$, so the D3-D1 operator no longer couples to the TQFT $\mathcal{X}^{M,q}$ on which the auxiliary fields $\lambda_1$ and $\varphi$ live. Importantly, $\widetilde{\cO}_q$ is an invertible operator, which agrees with our expectation that this defect generates the (invertible) $\bZ_M^{(1)}$ center symmetry in the $SU(M)$ gauge theory. 
By the same token, the $\mathsf{A}_1 \mathcal{B}_2^2$ anomaly trivializes on $\mathcal{W}^4$, so the $\mathbb{Z}_{2M}^{(0)}$ defect also becomes an invertible operator \cite{Apruzzi:2022rei}, i.e.
\beq
\widetilde{\cN}_p\left(M^3\subset \mathcal{W}^4\right) = \exp\left(2\pi i p\int_{M^3}c_3\right)\,.
\eeq
This conforms with the fact that the $\bZ_{2M}^{(0)}$ $R$-symmetry is invertible in the $SU(M)$ gauge theory. 

\paragraph{(N,D,N,D) boundary conditions.} Let us now consider the case where $\hat{c}_2$ is chosen to have Dirichlet boundary conditions instead of $b_2$. This is equivalent to having a $PSU(M)$ gauge theory on the boundary. As has been discussed previously, this causes the $\cU_q$ operators to generate an invertible $\widehat{\bZ}_M^{(1)}$ symmetry and the $\cN_p$ operators to generate a non-invertible $\bZ_{2M}^{(0)}$ symmetry on the boundary. Furthermore, the bulk line operator $\cV_p$ ends on the boundary to become the local operator $\overline{\cV}(\overline{x}\in \cW^4)$. What remains to be discussed is the fate of the $\cO_q$ operators. When $\hat{c}_2$ is chosen to vanish on $\mathcal{W}^4$, the defect $\cO_q$ takes the form,
\beq
\cO_q\left(W^2\subset\mathcal{W}^4\right) = \int\cD\varphi\cD\lambda_1\exp\left(2\pi i\int_{W^2}\varphi\cB_2 - M\varphi d\lambda_1\right)\,,
\eeq
where we have kept in mind that $\cA_1\big|_{\mathcal{W}^4} = 0$ due to the self-anomaly term. In the above, $\lambda_1$ acts as a Lagrange multiplier causing $\varphi$ to take on integer values, such that the remaining operator turns into a (normalized) sum over trivial operators.\footnote{Alternatively, we can see that the $\cA_1\cB_2$ anomaly again vanishes on the conformal boundary due to the Dirichlet boundary condition on $\cA_1$. The associated TQFT involving $\varphi$ and $\lambda_1$ will then vanish and the operator $\cO_q$ will still be trivial.} This implies that the D3-D1 operator trivializes when placed {\it parallel} to the conformal boundary.

Nonetheless, we may consider D3-D1 bound states that instead end transversely to $\mathcal{W}^4$, i.e.~we take one of the directions in $W^2$ to align with the radial coordinate of $\cW^5$. From the variation \eqref{eq:variational_b2c2hat}, we can write on the boundary,
\beq
M\hat{c}_2\big|_{\mathcal{W}^4} = d\hat{\cC}_1 = 0\,,
\eeq
where $\hat{\cC}_1$ is a $\bZ_M$-valued flat connection. The associated operator on $\mathcal{W}^4$ is then given by
\beq
\overline{\cO}\big(\overline{W}^1\subset \mathcal{W}^4\big) = \exp\left(\frac{2\pi i}{M}\int_{\overline{W}^1}\hat{\cC}_1\right)\,,
\eeq
which is charged under the $\widehat{\bZ}_M^{(1)}$ magnetic symmetry generated by the $\mathcal{U}_q$ defects. Hence, the line operators $\overline{\cO}(\overline{W}^1)$ should be identified with 't Hooft lines in the $PSU(M)$ gauge theory.



\section{Discussion and Outlook}\label{sec:discussion}

In this paper, we have studied in detail the complete fusion algebra of {\it gauge} symmetry generators in the 5d bulk $\mathcal{W}^5$ of the KT-KS background (see Table \ref{tab:fusion_rules} for a summary). More precisely, the gauge symmetries of the bulk are encoded by the SymTFT \eqref{eq:SymTFT}, whose kinetic terms are suppressed near the conformal boundary $\mathcal{W}^4 \subset \mathcal{W}^5$. Via the Gauss Law constraints, the SymTFT indicates the existence of topological extended operators (near $\mathcal{W}^4$) which generate the gauge symmetries.

Intuitively, the candidates for these operators are nothing but the solitonic objects in Type II string theory, i.e.~D-branes, wrapping suitable internal submanifolds. We find that this is indeed the case, and particularly, the WZ coupling of D-branes to lower-degree RR fields naturally reproduces the mixed anomaly terms in the SymTFT. The agreement between the symmetry generators and the D-branes therefore serves as a nontrivial check for the consistency between the Type IIB supergravity action and the WZ action of D-branes. An even more stringent test can be carried out by further including the gravitational terms respectively in the 11-form Type IIB anomaly polynomial $\mathcal{I}_{11}$ and the WZ action of the D-branes.\footnote{We expect in this case that the mixed chiral-gravitational anomaly of 4d $\mathcal{N}=1$ $SU(M)$ SYM theory disguises as some nontrivial TQFT attached to the appropriate D-brane operator. The resultant defect should be similar to that constructed by \cite{Putrov:2023jqi}.} We leave a careful analysis of this to future work.

The consistency between the supergravity bulk and the D-branes teaches us three important lessons. Firstly, the construction of, say, non-invertible {\it global} symmetry generators in the dual field theory typically requires a manual stacking of individually non-topological operators (e.g.~\cite{Choi:2022jqy}), whereas we can obtain the full operator at once from a straightforward dimensional reduction of the D-brane WZ action. For example, the 3d TQFT $\mathcal{A}^{M,1}$ within the $R$-symmetry defect $\mathcal{N}_1$ comes for free when we reduce a D5-brane over $S^3$. The fact that the WZ action is compatible with the bulk supergravity action enables the former to capture the necessary anomaly data from the latter.

Secondly, the explicit fusion algebra of symmetry generators in the boundary field theory can be understood holographically as brane interactions in the bulk. For concreteness, in the case of the 4d $\mathcal{N}=1$ $PSU(M)$ SYM theory, charge-$p$ and charge-$p'$ $R$-symmetry defects fuse not only into a charge-$(p+p')$ defect, but also an extra decoupled TQFT. We argued in Section \ref{sec:N_U_operators} that this fusion rule is fully characterized by brane {\it kinematics}, i.e.~the former comes from the overall center-of-mass mode of the combined stack of branes, and the latter comes from the relative contributions from the center-of-mass modes of the individual stacks.

On the contrary, a condensation defect arises from the fusion between a charge-$p$ defect with its orientation reversal. We suggest that this can instead be interpreted as a {\it dynamical} process known as tachyon condensation. In fact, the dressing of an otherwise invertible operator with a nontrivial TQFT (as in \eqref{eq:N1_expression}) is indicative of lower-dimensional D3-branes being dissolved within the worldvolume of the D5-brane. As a proof of concept, we showed that by engineering a certain tachyon profile, the D5-$\overline{\text{D5}}$ fusion indeed results in a sum of (codimension-2) solitonic D3-brane operators that matches the field-theoretic fusion rule.

That being said, it is not entirely clear to us how unique the specific tachyon profile is. To illustrate this, recall that the worldvolume of the D5-branes is $M^3 \times S^3$, so there are at least three na\"{i}ve ways to construct a codimension-2 soliton, namely,
\begin{equation*}
    \gamma^1 \times S^3 \, , \qquad \Sigma^2 \times \widetilde{S^2} \, , \qquad M^3 \times S^1 \, ,
\end{equation*}
where $\gamma^1,\Sigma^2 \subset M^3$, and $\widetilde{S^2}$, $S^1$ are respectively the Hopf base and fiber of $S^3$. The first option amounts to integrating out a 2-cycle in $M^3$, and the second option amounts to integrating out a one-cycle in $M^3$ plus a fiber integration in $S^3$, whereas the third option does not seem to be viable because we cannot perform a ``base integration'' in $S^3$. Ruling out the first option requires some technical details, but essentially one can show that the resultant operator is incompatible with the fluxes present in the KT-KS background. This then singles out the second option, as far as the condensation defect is concerned. Nevertheless, there is still a degeneracy in the choice of the internal manifold on which the tachyon profile is defined. We may for example replace the Hopf base $\widetilde{S^2}$ with the equatorial $S^2$ of $S^3$, and one cannot tell the distinction between these two D3-brane configurations from the perspective of the 5d bulk operator. 

This degeneracy {\it may} be lifted by considering the isometries of the $S^2\times S^2$ base in the $T^{1,1}$ fibration. Including these isometries would introduce an additional $SU(2)_L\times SU(2)_R$ gauge symmetry in the bulk; in the dual field theory there are two sets of chiral superfields $A_r, B_u$ in bifundamental representations of the UV gauge group $SU(N+M)\times SU(N)$. Each set forms a doublet under an $SU(2)$ global symmetry, which gives a combined $SU(2)_L\times SU(2)_R$ global symmetry in the boundary field theory. In the supergravity theory, the effect of this symmetry can be seen in the following way. A D3-brane wrapping a submanifold of the $S^3$ cycle will act as a point particle on $S^2$ in the presence of a nonzero ``magnetic field'' arising from the $G_5$ flux threading $T^{1,1}$. This will cause the correpsonding operator to have a Landau-level label as discussed in \cite{Gaiotto:2009gz,Bah:2021iaa}. It is possible that this additional label will remove the degeneracy in the choice of internal manifold for the tachyon condensation. 

Moreover, as we saw in Section \ref{sec:N_U_operators}, the D3-brane operators $\mathcal{U}_q$ obey a group-like fusion rule, and notably, it annihilates with its orientation reversal. In other words, they do not undergo further tachyon condensation to form D1-branes. One may heuristically argue that there is a qualitative difference between the $\mathcal{N}_p \otimes \mathcal{N}_{-p}$ and $\mathcal{U}_q \otimes \mathcal{U}_{-q}$ fusions. The operator $\mathcal{N}_p$ consists of a 3d TQFT $\mathcal{A}^{M,p}$ on which Chan-Paton fields reside, whereas all the Chan-Paton modes are already integrated out in $\mathcal{U}_q$. Since the presence of worldvolume Chan-Paton modes are necessary (but not sufficient) for tachyon condensation to happen, as effective 5d bulk operators the former is non-invertible while the latter is not. It is likely that we can shed some light on these open questions by examining more constructions of condensation defects in other holographic setups.

The third lesson that we can learn is the bulk SymTFT contains a rich symmetry structure that encodes the data of multiple possible boundary field theories. As explained in Section \ref{sec:boundary_conditions}, the global symmetries present in the boundary field theory is not only specified by the SymTFT itself, but also by a choice of boundary conditions for the various supergravity gauge fields. Consequently, this determines the field content of a given D-brane operator when placed at the conformal boundary (transverse to the radial direction of $\mathcal{W}^5$). For instance, the electric $\mathbb{Z}_M^{(1)}$ symmetry defect $\mathcal{O}_q(W^2)$ can only be invertible if $W^2 \subset \mathcal{W}^4$, but once we move slightly into the bulk (but still near $\mathcal{W}^4$), the 2d TQFT $\mathcal{X}^{M,q}$ attached to $\mathcal{O}_q(W^2 \subset \mathcal{W}^5)$ renders the defect non-invertible. This illustrates a crucial point: a non-invertible {\it gauge} symmetry can be trivialized to an invertible {\it global} symmetry depending on the choice of boundary conditions.

There is a further subtlety that one should consider when using the Gauss Laws to derive the symmetry structure from the SymTFT. The analysis performed in this paper was only concerned with the {\it classical} Gauss Laws of the associated symmetries. In general there may be nontrivial and important effects from the {\it quantum} Gauss Laws as considered in \cite{Belov:2004ht}. These effects may contribute extra torsional phases to the symmetry generators.
It would be interesting to study the effect of these phases as well as their origin in holographic constructions.

Recently in the literature, the study of symmetry-generating D-brane operators has provided us with fruitful insights into the nature of non-invertible and/or higher-form global symmetries in quantum field theories. Multiple examples of non-invertible symmetries have indeed been uncovered in different string theory solutions using D-branes. On top of that, we have shown in this paper brane interactions in the bulk play a significant role in characterizing the fusion algebra of the boundary field theory. It is widely believed a fusion algebra involving defects of various dimensions furnishes part of the data of some higher fusion category (e.g.~\cite{Bhardwaj:2022yxj,Bhardwaj:2022lsg,Bhardwaj:2022kot,Bartsch:2022mpm,Bartsch:2022ytj,Copetti:2023mcq}). The data contained in the bulk theory should then be interpreted as describing the Drinfeld center of this higher fusion category \cite{Kong_2017,Ji:2019jhk,Chatterjee:2022kxb,Kong_2020,Zhang:2023wlu}. We therefore hope our work will offer a complementary understanding of the intricacies of such higher categorical structures.


\section*{Acknowledgments}

We are grateful to Oren Bergman, Federico Bonetti, T.~Daniel Brennan, Diego Garc\'ia-Sep\'ulveda, Pierre Heidmann, Patrick Jefferson, Ho Tat Lam, Ruben Minasian, Salvatore Pace, Sebastian Schulz, Hannah Tillim, and Peter Weck for interesting conversations and correspondence. We also thank the authors of \cite{Apruzzi:2023uma} for communicating about a related work and coordinating submission. IB, EL, and TW are supported by NSF grant PHY-2112699, and also in part by the Simons Collaboration on Global Categorical Symmetries.


\appendix



\section{SymTFT Computation}\label{sec:app_SymTFT_dualization}

We start with a brief review of the Klebanov-Tseylin (KT) \cite{Klebanov:2000nc} and Klebanov-Strassler (KS) setups \cite{Klebanov:2000hb} dual to the cascading field theory.  There are two relevant features. The first is a stack of $N$ D3-branes extending along flat space $\bR^{1,3}$ and placed at the tip of the conifold background as was studied in \cite{Klebanov:1998hh}. The conifold is a non-compact Calabi-Yau cone over a Sasaki-Einstein 5-manifold denoted $T^{1,1}$. The space $T^{1,1}$ is constructed as a Hopf fibration over $S^2\times S^2$ with Einstein metric given by \cite{Candelas:1989js,Cassani:2010na,Apruzzi:2022rei}
\beq
ds^2(T^{1,1}) = \frac{4}{9}\left(d\psi + \frac{1}{2}\sum_{i=1}^2\cos\theta_i d\phi_i\right)^{\!2} + \frac{1}{6}\sum_{i=1}^2\left(d\theta_i^2 + \sin^2\theta_i d\phi_i^2 \right)\,,
\eeq
where in the above $\{\theta_i,\phi_i\}$ denote the coordinates of the two $S^2$ geometries each with periodicity $2\pi$, and $\psi$ is the coordinate of the $S^1$ fiber with $4\pi$ periodic. The $T^{1,1}$ space admits the topology of $S^2\times S^3$ and cohomologically nontrivial  one 2-cycle, one 3-cycle, and 5-cycle that we denote by $\omega_2$, $\omega_3$, and $\Omega_5$ respectively. The near-horizon geometry of the stack of D3-branes in Type IIB string theory is correspond to $AdS_5\times T^{1,1}$, with $N$ units of $G_5$ flux.

We will be interested in gauging the $U(1)$ isometry of the fiber coordinate $\psi$ through the introduction of a connection $d\psi \to d\psi + 2\pi\sfA_1$, where the normalization is chosen such that  the flux of $\sfA_1$ has integral periods.  In general one may consider gauging the $SU(2)\times SU(2)$ isometries of the $S^2\times S^2$ base of $T^{1,1}$, in which case one would then have to incorporate global angular forms to account for the equivariant completion \cite{Bah:2018gwc,Bah:2018jrv,Bah:2019jts,Bah:2019vmq}. The effect of these forms do not play a role in this paper, so we refrain from gauging these isometries. We can construct a $U(1)$-equivariant completion of the volume forms mentioned earlier as
\beq\label{eq:omega23}
\omega_2 = -\frac{1}{2}(V_1 - V_2)\,,\qquad \omega_3 = (V_1-V_2)\frac{D\psi}{2\pi}\,,
\eeq
where $V_i$ denote the normalized volume forms of the 2-spheres in the base of the $T^{1,1}$ fibration. The normalization of these forms comes from demanding that $\int_{T^{1,1}}\omega_2\omega_3 = 1$. The $U(1)$-equivariant completion of $\Omega_5$ is written
\beq\label{eq:Omega5}
\Omega_5 = \omega_2\omega_3 + \frac{1}{2}\,\sfF_2(V_1+V_2)\frac{D\psi}{2\pi}\,,
\eeq
where $\sfF_2$ is the field strength of $\sfA_1$. 

The second main ingredient of the KT and KS setup is a stack of $M$ D5-branes wrapping the topological $S^2$ of the conifold and extending along the flat space $\bR^{1,3}$. The backreaction of this stack of branes has two effects on the supergravity setup, the first of which being that the conifold metric is replaced with that of the deformed conifold\footnote{There is another construction which considers the resolved conifold rather than the deformed conifold \cite{PandoZayas:2000ctr}. The difference is in which cycle is blown up, but the physical theory is still that of 4d $\cN = 1$ Super Yang-Mills. It would be interesting to explore the connections between the two theories with regard to the operators constructed out of branes.} where the $S^3$ cycle is blown up at the tip of the cone. The change in metric is felt very close to the original stack of $N$ D3-branes and similarly in the deep IR of the dual field theory, but far away from the origin the space is indistinguishable from the original conifold. In the near horizon limit, the effect of the stack of $M$ D5-branes is that the $G_5$ flux runs along the $AdS_5$ radius. The asymptotic region of the KS background corresponds to the KT \cite{Klebanov:2000nc}.  This latter  geometry is enough to determined the symmetry structure of interest in this paper.  

The dual field theory of the setup is $\cN=1$ Super Yang-Mills with gauge group $SU(N+M)\times SU(N)$ coupled to bifundamental matter \cite{Klebanov:2000nc}.  This is a non-conformal background where the gauge couplings run. As this theory flows to the IR a sequence of Seiberg duality transformations occurs, often referred to as the ``duality cascade''. For $N$ a multiple of $M$, this cascade ends with pure $\cN=1$ Super Yang-Mills theory with gauge group $SU(M)$. In this paper we will study the symmetry structure of  the $\cN=1\;SU(M)$ theory from the SymTFT derived from the dual gravitational background.  In particular we will use various brane probes of the supergravity to construct the possible symmetry generators and their fusion rules.  

%
%

To construct the SymTFT in this background, we first write down the 5d topological action using the nontrivial fluxes for $h_3$ and $\sfF_2$ as discussed in Section \ref{sec:symtft},
\begin{align}
2\pi\int_{W^5}\bigg[Nb_2g_3 &- \sfA_1(\cB_2)^2 + \cA_1b_2\cB_2 - \frac{1}{2M}\,\cA_1\widetilde{g}_2^2 + \frac{\mathcal{K}}{2M}\,\cA_1\beta(\cA_1)\beta(\cA_1)\\
&- \frac{1}{2M}\widetilde{g}_1\left(\widetilde{g}_2^2 + 2\widetilde{g}_2\cB_2 - \mathcal{K}\left((d\sfA_1)^2 + 3d\sfA_1\beta(\cA_1) + 3\beta(\cA_1)^2\right)\right) \bigg]\,,\nn
\end{align}
where the $\cA_1$ cubic coefficient is given by
\beq\label{eq:anomaly_coefficient}
\mathcal{K} = \frac{N(N+M)}{4} + \frac{M^2}{6} - \frac{1}{3}\,.
\eeq
The subleading corrections in $N$ may be computed directly from the UV quiver of the KT-KS setup. The full SymTFT is given by the above action augmented with standard kinetic terms for $g_1, g_2, g_3, h_3$, and $\sfF_2$ with associated couplings $\kappa_1,\kappa_2,\kappa_3, \kappa_h$, and $\kappa_{\sfF}$.

To elucidate the symmetry structure of the boundary field theory, we may dualize some of the fields in the above SymTFT. We first dualize $g_1$ by introducing a 3-form $c_3$ as a Lagrange multiplier to enforce the Bianchi identity.
\beq
S_{\rm mult} = 2\pi\int_{W^5}c_3\wedge(2Md\sfA_1 - dg_1)\,.
\eeq
The equation of motion for $g_1$ yields
\beq
\star_5 g_1 = \kappa_1\left(dc_3 - \frac{1}{2M}\left(\widetilde{g}_2^2 + 2\widetilde{g}_2\cB_2 - \mathcal{K}\left((d\sfA_1)^2 + 3d\sfA_1\beta(\cA_1) + 3\beta(\cA_1)^2\right)\right)\right) \coloneqq g_4\,.
\eeq
After integrating out $g_1$ by replacing it with its equation of motion, we may also dualize $g_2$ through an additional field $a_2$ acting as a Lagrange multiplier
\beq
S_{\rm mult}' = -2\pi\int_{W^5} a_2\wedge(dg_2 + Mdb_2)\,.
\eeq
The equation of motion for $g_2$ can then be found as 
\beq
\star_5 g_2 = -\kappa_2\left(da_2 + \frac{1}{M}\,\cA_1\cB_2\right)\,.
\eeq
After performing both dualizations, we may write our final action $S_{\rm top}$ as
\beq
2\pi\int_{W^5}\bigg(Mb_2d\hat{c}_2 - 2Mc_3 d\sfA_1 - \sfA_1\cB_2^2 + \cA_1b_2\cB_2 - \frac{\mathcal{K}}{M}\,\cA_1\beta(\cA_1)\beta(\cA_1)\bigg)\,,
\eeq
thus producing \eqref{eq:SymTFT}. In the low energy limit, the kinetic terms present in the full SymTFT are suppressed, so we often refer to $S_{\rm top}$ as ``the SymTFT'' unless otherwise specified.



\section{Brief Review of Bockstein Homomorphisms}\label{sec:app_Bockstein_homomorphism}

For a detailed introduction to Bockstein homomorphisms, we refer the reader to standard references in the literature, e.g.~\cite{Hatcher:478079,mccleary_2000}. Here we briefly review and explain the notation adopted in the main text. Let us first consider the following short exact sequence with $k$ prime,\footnote{If $k$ is not prime, then there are possibly additional torsion subgroups that we have to worry about, but we refrain from considering that in this appendix.}
\begin{equation}
	0 \to \mathbb{Z}_k \xrightarrow{\times \, k} \mathbb{Z}_{k^2} \xrightarrow{\text{mod} \, k} \mathbb{Z}_k \to 0 \, ,
\end{equation}
where the ``mod $k$'' map means reduction modulo $k$. Associated with it is a long exact sequence in homology,
\begin{equation}
	\cdots \to H_n(M;\mathbb{Z}_{k^2}) \xrightarrow{\text{mod} \, k} H_n(M;\mathbb{Z}_k) \xrightarrow{\beta} H_{n-1}(M;\mathbb{Z}_k) \xrightarrow{\times \, k} H_{n-1}(M;\mathbb{Z}_{k^2}) \to \cdots \, ,
\end{equation}
where the Bockstein homomorphism is defined to be a map
\begin{equation}
	\beta: H_n(M;\mathbb{Z}_k) \to H_{n-1}(M;\mathbb{Z}_k) \, .
\end{equation}
Explicitly, it acts on cycles (mod $k$), $\mathcal{C}^n \in H_n(M;\mathbb{Z}_k)$, as
\begin{equation}
	\beta(\mathcal{C}^n) = \frac{1}{k} \, \partial\mathcal{C}^n \, .
\end{equation}

A closely related construction starts with the short exact sequence
\begin{equation}
	0 \to \mathbb{Z} \xrightarrow{\times \, k} \mathbb{Z} \xrightarrow{\text{mod} \, k} \mathbb{Z}_k \to 0 \, ,
\end{equation}
which leads to the long exact sequence
\begin{equation}
	\cdots \to H_n(M;\mathbb{Z}) \xrightarrow{\text{mod} \, k} H_n(M;\mathbb{Z}_k) \xrightarrow{\widetilde{\beta}} H_{n-1}(M;\mathbb{Z}) \xrightarrow{\times \, k} H_{n-1}(M;\mathbb{Z}) \to \cdots \, .
\end{equation}
The codomain of this Bockstein homomorphism is different from that in the previous case, namely,
\begin{equation}
	\widetilde{\beta}: H_n(M;\mathbb{Z}_k) \to H_{n-1}(M;\mathbb{Z}) \, .
\end{equation}
On a cycle $\mathcal{C}^n \in H_n(M;\mathbb{Z}_k)$, the map similarly acts as
\begin{equation}
	\widetilde{\beta}(\mathcal{C}^n) = \frac{1}{k} \, \partial\mathcal{C}^n \, ,
\end{equation}
where the two connecting homomorphisms are related via a reduction map,
\begin{equation}
	\beta(\mathcal{C}^n) = (\text{mod} \, k) \circ \widetilde{\beta}(\mathcal{C}^n) \, .
\end{equation}
In summary, the relations between the various maps are depicted by the following commutative diagram.
\begin{equation}
	\begin{tikzcd}
		H_n(M;\mathbb{Z}) \arrow[d,"\text{mod} \, k^2"] \arrow[r,"\text{mod} \, k"] & H_n(M;\mathbb{Z}_k) \arrow[d,"\text{id}"] \arrow[r,"\widetilde{\beta}"] & H_{n-1}(M;\mathbb{Z}) \arrow[d,"\text{mod} \, k"] \arrow[r,"\times \, k"] & H_{n-1}(M;\mathbb{Z}) \arrow[d,"\text{mod} \, k^2"] \\
		H_n(M;\mathbb{Z}_{k^2}) \arrow[r,"\text{mod} \, k"] & H_n(M;\mathbb{Z}_k) \arrow[r,"\beta"] & H_{n-1}(M;\mathbb{Z}_k) \arrow[r,"\times \, k"] & H_{n-1}(M;\mathbb{Z}_{k^2})
	\end{tikzcd}
\end{equation}

We see that if $\partial\mathcal{C}^n = k \mathcal{C}^{n-1}$ for some $\mathcal{C}^{n-1} \in H_{n-1}(M;\mathbb{Z})$, i.e.~$\mathcal{C}^{n-1}$ is a torsion $n$-cycle of order $k$, then $\widetilde{\beta}(\mathcal{C}^n)=\mathcal{C}^{n-1}$. In addition, since the kernel of $\widetilde{\beta}$ is the image of the reduction map by exactness, we have $\widetilde{\beta}(k\mathcal{C}^n) = 0 \in H_{n-1}(M;\mathbb{Z})$ for any $\mathcal{C}^n \in H_n(M;\mathbb{Z}_k)$. Similar remarks apply to the $\beta$ map. Note that by construction, the Bockstein homomorphisms are nilpotent, e.g.
\begin{equation}
	\beta \circ \beta = 0 \, ,
\end{equation}
and they satisfy the Leibniz rule,
\begin{equation}
	\beta(a_m \cup b_n) = \beta(a_m) \cup n_n + (-1)^m \, a_m \cup \beta(b_n) \, .
\end{equation}

One can obtain dual Bockstein homomorphisms by constructing long exact sequences in cohomology instead, also denoted as $\beta$ and $\widetilde{\beta}$ with an abuse of notation. Again, the relevant maps are described by the following commutative diagram.
\begin{equation}
	\begin{tikzcd}
		H^n(M;\mathbb{Z}) \arrow[d,"\text{mod} \, k^2"] \arrow[r,"\text{mod} \, k"] & H^n(M;\mathbb{Z}_k) \arrow[d,"\text{id}"] \arrow[r,"\widetilde{\beta}"] & H^{n+1}(M;\mathbb{Z}) \arrow[d,"\text{mod} \, k"] \arrow[r,"\times \, k"] & H^{n+1}(M;\mathbb{Z}) \arrow[d,"\text{mod} \, k^2"] \\
		H^n(M;\mathbb{Z}_{k^2}) \arrow[r,"\text{mod} \, k"] & H^n(M;\mathbb{Z}_k) \arrow[r,"\beta"] & H^{n+1}(M;\mathbb{Z}_k) \arrow[r,"\times \, k"] & H^{n+1}(M;\mathbb{Z}_{k^2})
	\end{tikzcd}
\end{equation}
Just like how the Bockstein homomorphisms act on cycles as ``boundary operators,'' they act on cocycles as ``coboundary operators.'' More explicitly,
\begin{equation}
	\beta(c_n) = \frac{1}{k} \, \delta c_n
\end{equation}
for any $c_n \in H^n(M;\mathbb{Z}_k)$.

The Bockstein homomorphisms have important applications in physics, especially when torsional manifolds or fluxes are involved. One of them being the definition of the integral Stiefel-Whitney classes. For a degree-$n$ (generalized) Stiefel-Whitney class $w_n \in H^n(M;\mathbb{Z}_k)$, the corresponding integral Stiefel-Whitney class is defined as
\begin{equation}
	W_{n+1} \coloneqq \widetilde{\beta}(w_n) \in H^{n+1}(M;\mathbb{Z}) \, .
\end{equation}
If $n=k=2$, the degree-3 integral class $W_3 = \beta(w_2)$ measures the obstruction to having a $\text{spin}^c$ structure on the manifold $M$. Another example, used in our contruction, is the flat but topologically nontrivial part of the NSNS field $\mathcal{B}_2 \in H^2(\mathcal{W}^5;\mathbb{Z}_M)$, which corresponds to the cohomological class of the 't Hooft magnetic flux via $\mathcal{H}_3 = \widetilde{\beta}'(\mathcal{B}_2)$ as in \eqref{eq:torsion_flux_definitions}. In fact, these two examples are closely related through the Freed-Witten anomaly \cite{Freed:1999vc,Kapustin:1999di}.

Moreover, the Bockstein homomorphism for $k=2$ coincides with the first Steenrod square (see, e.g.~\cite{Diaconescu:2000wy}),
\begin{equation}
	\beta(c_n) = \text{Sq}^1(c_n) \coloneqq w_1(M) \cup c_n \in H^2(M;\mathbb{Z}_2) \, ,
\end{equation}
where $w_1(M) \in H^1(M;\mathbb{Z}_2)$ is the first Stiefel-Whitney class of $M$ (or more precisely, its normal bundle).

As an aside, we clarify that the Bockstein homomorphisms invoked in this paper shall be distinguished from that commonly used in ordinary differential cohomology, i.e.
\begin{equation}
	\widehat{\beta}: H^n(M;\mathbb{R}/\mathbb{Z}) \to H^{n+1}(M;\mathbb{Z}) \, ,
\end{equation}
which is associated with the short exact sequence, $0 \to \mathbb{Z} \to \mathbb{R} \to \mathbb{R}/\mathbb{Z} \to 0$. These two distinct Bockstein homomorphisms are in fact related to each other, as can be inferred from the commutative diagram \cite{Witten:1999vg},
\begin{equation}
	\begin{tikzcd}
		0 \arrow[d] \arrow[r] & \mathbb{Z} \arrow[d,"\text{id}"] \arrow[r,"\times \, k"] & \mathbb{Z} \arrow[d,"\times \, 1/k"] \arrow[r,"\text{mod} \, k"] & \mathbb{Z}_k \arrow[d,"\phi"] \arrow[r] & 0 \arrow[d] \\
		0 \arrow[r] & \mathbb{Z} \arrow[r,"i"] & \mathbb{R} \arrow[r] & \mathbb{R}/\mathbb{Z} \arrow[r] & 0
	\end{tikzcd}
\end{equation}
The map $\phi$ induces a homomorphism $\phi_\ast: H^n(M;\mathbb{Z}_k) \to H^n(M;\mathbb{R}/\mathbb{Z})$, so it follows that $\widetilde{\beta} = \widehat{\beta} \circ \phi_\ast$ by exactness.





\section{Derivation of Fusion Rules}\label{sec:app_complete_fusion_rules}

In this appendix, we are going to derive the nontrivial fusion rules between the symmetry defects. The reader may find similar derivations in \cite{Choi:2022zal}. We assume all manifolds are spin throughout the appendix.

\paragraph{\boldmath $\mathcal{N}_p \otimes \mathcal{N}_{p'}$ fusion.}

We already derived in Section \ref{sec:N_U_operators} the fusion rule for the case when $p' = -p \mod M$. Let us turn to the case when $p' \neq -p \mod M$ and assume $\gcd(M,p+p')=1$ (in addition to $\gcd(M,p)=\gcd(M,p')=1$). Recall that the defect $\mathcal{N}_p(M^3)$ can be decomposed as
\begin{equation}
	\mathcal{N}_p = \widetilde{\mathcal{N}}_p \otimes \mathcal{A}^{M,p}[\mathcal{B}_2] \, ,
\end{equation}
where $\widetilde{\mathcal{N}}_p = \exp(2\pi i \int_{M^3} c_3)$ obeys a group-like fusion rule,
\begin{equation}
	\widetilde{\mathcal{N}}_p \otimes \widetilde{\mathcal{N}}_{p'} = \widetilde{\mathcal{N}}_{p+p'} \, ,
\end{equation}
thus the only nontrivial part of the fusion $\mathcal{N}_p \otimes \mathcal{N}_{p'}$ is that between the minimal TQFTs, i.e.~$\mathcal{A}^{M,p}[\mathcal{B}_2] \otimes \mathcal{A}^{M,p'}[\mathcal{B}_2]$. An efficient way to compute the fusion rule is as follows.\footnote{We thank Ho Tat Lam for discussions about this point.}

The minimal TQFT $\mathcal{A}^{M,p}$ is a 3d theory with a $\mathbb{Z}_M$ 1-form symmetry generated by Wilson lines denoted by $a^s$ where $s=0,\dots,M-1$ \cite{Hsin:2018vcg}, and their topological spins are labeled by
\begin{equation}
	h[a^s] = \frac{ps^2}{2M} \, .
\end{equation}
The theory is modular given that its $S$-matrix,
\begin{equation}
	S(a^s,a^{s'}) = \frac{1}{\sqrt{M}} \, \exp\bigg(2\pi i \big\{h[a^s]+h[a^{s'}]-h[a^{s+s'}]\big\}\bigg) = \frac{1}{\sqrt{M}} \, \exp\bigg(\!-\frac{2\pi i pss'}{M}\bigg) \, ,
\end{equation}
is non-degenerate, i.e.~only the transparent line $a^0=\mathds{1}$ braids trivially with all other lines.\footnote{In a braided fusion category with fusion rules $a \otimes b = \sum_c N_{ab}^c \, c$, the $S$-matrix is given by $S_{ab} = \sum_c N_{\bar{a}b}^c (\theta_c / \theta_a \theta_b) (d_c/\mathcal{D})$, where $\theta_a = \exp(-2\pi i h)$ is the topological twist of the simple object (or anyon) $a$. If we represent a finite group $G$ with a fusion category, then the quantum dimension of any anyon is $d_a=1$, and the total quantum dimension of the category is $\mathcal{D}=\sqrt{\sum_a d_a^2}=\sqrt{|G|}$.} Similarly, $\mathcal{A}^{M,p'}$ has lines $a'^{\,r}$ with spins $h[a'^{\,r}]=p'r^2/2M$. We claim that for any $(p,p')$, there always exists an equivalent factorization of the minimal TQFTs of the form,
\begin{equation}
	\mathcal{A}^{M,p} \otimes \mathcal{A}^{M,p'} = \mathcal{A}^{M,n} \otimes \mathcal{A}^{M,p+p'} \, ,
\end{equation}
for some integer $n$ yet to be determined. Suppose we denote the sets of lines in $\mathcal{A}^{M,n}$ and $\mathcal{A}^{M,p+p'}$ respectively as
\begin{equation}
	b^s = (a^x a'^{\,y})^s = a^{xs} a'^{\,ys} \, , \qquad b'^{\,r} = (a a')^r = a^r a'^{\,r} \, ,
\end{equation}
then their spins are
\begin{equation}
	\begin{gathered}
		h[b^s] = \frac{p(xs)^2}{2M} + \frac{p'(ys)^2}{2M} = \frac{(px^2+p'y^2)s^2}{2M} \, ,\\
		h[b'^{\,r}] = \frac{pr^2}{2M} + \frac{p'r^2}{2M} = \frac{(p+p')r^2}{2M} \, .
	\end{gathered}
\end{equation}
One may see that $p+p'$ corresponds to the anomaly of $\mathcal{A}^{M,p+p'}$ as desired.

To impose the condition that $\mathcal{A}^{M,n}$ and $\mathcal{A}^{M,p+p'}$ are decoupled from each other, we demand the line $b^s$ to braid trivially with $b'^{\,r}$ for any $(s,r)$, i.e.
\begin{equation}
	h[b^s] + h[b'^{\,r}] - h[b^s b'^{\,r}] = -\frac{sr(px+p'y)}{M} = 0 \, .
\end{equation}
The simplest nontrivial integer solution for general $(p,p')$ is $(x,y)=(p',-p)$, so the spin of $b^s$ becomes
\begin{equation}
	h[b^s] = \frac{pp'(p+p')s^2}{2M} \, ,
\end{equation}
from which we can read off the anomaly label to be $(p+p')pp'$. We have therefore found that\footnote{This is consistent with the result of \cite{Choi:2022zal} where they chose the solution $(x,y)=(p^{-1},-p'^{-1})$, thus producing $\mathcal{A}^{M,p^{-1}+p'^{-1}} \cong \mathcal{A}^{M,(p+p')pp'}$. Here we have used the fact that $\mathcal{A}^{M,pr^2} \cong \mathcal{A}^{M,p}$ if $\gcd(M,r)=1$.}
\begin{equation}
	\mathcal{A}^{M,p} \otimes \mathcal{A}^{M,p'} = \mathcal{A}^{M,pp'(p+p')} \otimes \mathcal{A}^{M,p+p'} \, .
\end{equation}
Note that $\gcd(M,p+p')=\gcd(M,p)=\gcd(M,p')=1$ guarantees $\gcd(M,pp'(p+p'))=1$. Furthermore, since the lines $b'^{\,r} = a^r a'^{\,r}$ are those that are coupled to $\mathcal{B}_2$ in the product $\mathcal{A}^{M,p} \otimes \mathcal{A}^{M,p'}$ (whereas $b^s$ are decoupled), we conclude that the fusion rule between $\mathcal{N}_p$ and $\mathcal{N}_{p'}$ is given by
\begin{equation}
	\mathcal{N}_p \otimes \mathcal{N}_{p'} = \widetilde{\mathcal{N}}_{p+p'} \otimes \mathcal{A}^{M,pp'(p+p')} \otimes \mathcal{A}^{M,p+p'}[\mathcal{B}_2] = \mathcal{A}^{M,pp'(p+p')} \otimes \mathcal{N}_{p+p'} \, .\label{appeq:Np_fusion_rule}
\end{equation}

Alternatively, one may show this result using an effective Lagrangian description of $\mathcal{A}^{M,p}[\mathcal{B}_2] \otimes \mathcal{A}^{M,p'}[\mathcal{B}_2]$, in which case we have
\begin{equation}
	\int_{M^3} \bigg(\frac{pM}{2} \, a_1 da_1 + \frac{p'M}{2} \, a_1' da_1' + (p a_1 + p' a_1') \mathcal{B}_2\bigg) \, .
\end{equation}
Motivated by the ``correct'' answer from the previous approach, we define $pp'(p+p')\tilde{a}_1=p(p'a_1)+p'(-pa_1')$ and $(p+p')\tilde{a}_1'=pa_1+p'a_1'$. The Lagrangian then turns into
\begin{equation}
	pp'(p+p') \int_{M^3} \frac{M}{2} \, \tilde{a}_1 d\tilde{a}_1 + (p+p') \int_{M^3} \bigg(\frac{M}{2} \, \tilde{a}_1' d\tilde{a}_1' + \tilde{a}_1' \mathcal{B}_2\bigg) \, ,\label{appeq:Np_Np_field_redefinition}
\end{equation}
which can be identified as an effective Lagrangian for $\mathcal{A}^{M,(p+p')pp'} \otimes \mathcal{A}^{M,p+p'}[\mathcal{B}_2]$. Note that the former minimal TQFT is decoupled from $\mathcal{B}_2$, while the latter is associated with a stack of $p+p'$ D5-branes (reduced on $S^3$).

The fusion rule \eqref{appeq:Np_fusion_rule} is manifestly commutative, i.e.~$\mathcal{N}_p \otimes \mathcal{N}_{p'} = \mathcal{N}_{p'} \otimes \mathcal{N}_p$. We also claim that it is associative, and in fact any triple fusion admits the following factorization (assuming $\gcd(M,p+p'+p'')=1$),
\begin{equation}
	\mathcal{N}_p \otimes \mathcal{N}_{p'} \otimes \mathcal{N}_{p''} = \mathcal{A}^{M,pp'p''P(p+p'+p'')} \otimes \mathcal{A}^{M,P} \otimes \mathcal{N}_{p+p'+p''} \, ,\label{appeq:triple_Np_fusion_rule}
\end{equation}
where $P = (p+p')(p+p'')(p'+p'')$. To examine such a claim, we first apply \eqref{appeq:Np_fusion_rule} twice to write down
\begin{equation}
	(\mathcal{N}_p \otimes \mathcal{N}_{p'}) \otimes \mathcal{N}_{p''} = \mathcal{A}^{M,(p+p')pp'} \otimes \mathcal{A}^{M,(p+p'+p'')(p+p')p''} \otimes \mathcal{N}_{p+p'+p''} \, .
\end{equation}
Denoting the topological Wilson lines in $\mathcal{A}^{M,pp'(p+p')}$ and $\mathcal{A}^{M,(p+p')p''(p+p'+p'')}$ respectively as $a^s$ and $a'^{\,r}$, we define two sets of new lines,
\begin{equation}
	b^s = (a^{p''(p+p'+p'')} a'^{\,-pp'})^s \, , \qquad b'^{\,r} = (a a')^r \, ,
\end{equation}
whose spins are given by
\begin{align}
	h[b^s] & = \frac{pp'(p+p')[p''(p+p'+p'')s]^2}{2M} + \frac{(p+p')p''(p+p'+p'')[-pp's]^2}{2M}\nonumber\\
	& = \frac{pp'p''(p+p')(p+p'')(p'+p'')(p+p'+p'')s^2}{2M} \, ,\\
	h[b'^{\,r}] & = \frac{pp'(p+p')r^2}{2M} + \frac{(p+p')p''(p+p'+p'')r^2}{2M}\nonumber\\
	& = \frac{(p+p')(p+p'')(p'+p'')r^2}{2M} \, .
\end{align}
These two sets of lines braid trivially with each other, i.e.
\begin{equation}
	h[b^s] + h[b'^{\,r}] - h[b^s b'^{\,r}] = 0 \, ,
\end{equation}
so we have successfully reproduced the triple fusion rule \eqref{appeq:triple_Np_fusion_rule}. By the same token, we may write down
\begin{equation}
	\mathcal{N}_p \otimes (\mathcal{N}_{p'} \otimes \mathcal{N}_{p''}) = \mathcal{A}^{M,p'p''(p'+p'')} \otimes \mathcal{A}^{M,p(p'+p'')(p+p'+p'')} \otimes \mathcal{N}_{p+p'+p''} \, .
\end{equation}
In this case, one can check that the following redefinition of lines,
\begin{equation}
	b^s = (a^{p'p''} a'^{\,-p(p+p'+p'')})^s \, , \qquad b'^{\,r} = (a a')^r \, ,
\end{equation}
gives rise to \eqref{appeq:triple_Np_fusion_rule} as well. Since the fusion is also commutative, this triple fusion rule indeed holds for arbitrary permutations of $(p,p',p'')$ with an appropriate redefinition of lines. Holographically, this implies that we can freely reshuffle the three stacks of D5-branes without altering the end result, although the decoupled modes are sensitive to the number of branes in each individual stack.

\paragraph{\boldmath $\mathcal{N}_p \otimes \mathcal{C}$ fusion.}

The invertible operator $\widetilde{\mathcal{N}}_p$ essentially acts as a spectator, so it suffices to work out the fusion rule for $\mathcal{A}^{M,p}[\mathcal{B}_2] \otimes (\mathcal{Z}_M)_0[\mathcal{B}_2]$. In terms of the effective Lagrangian description, we have
\begin{equation}
	\int_{M^3} \bigg(\frac{pM}{2} \, a_1 da_1 + p a_1 \mathcal{B}_2\bigg) + \int_{M^3} \Big(M \Lambda_1 da_1' + \Lambda_1 \mathcal{B}_2\Big) \, .
\end{equation}
Defining $\hat{a}_1 = a_1 + p^{-1} \Lambda_1$ and $\hat{a}_1' = a_1' - \hat{a}_1$, we can rewrite the Lagrangian as
\begin{equation}
	\int_{M^3} \bigg(M \Lambda_1 d\hat{a}_1' + \frac{p^{-1}M}{2} \, \Lambda_1 d\Lambda_1\bigg) + \int_{M^3} \bigg(\frac{pM}{2} \, \hat{a}_1 d\hat{a}_1 + p \hat{a}_1 \mathcal{B}_2\bigg) \, ,
\end{equation}
which may be identified as $(\mathcal{Z}_M)_{p^{-1}M} \otimes \mathcal{A}^{M,p}[\mathcal{B}_2]$. Assuming $M$ is odd, the DW twist in $(\mathcal{Z}_M)_K$ has a periodicity $K \sim K+M$, so $(\mathcal{Z}_M)_{p^{-1}M} \cong (\mathcal{Z}_M)_0$ given $p^{-1} \in \mathbb{Z}_M$ is an integer. We thus conclude that
\begin{equation}
	\mathcal{N}_p \otimes \mathcal{C} = \mathcal{C} \otimes \mathcal{N}_p = (\mathcal{Z}_M)_0 \otimes \mathcal{N}_p \, ,
\end{equation}
where $(\mathcal{Z}_M)_0$ is decoupled from $\mathcal{B}_2$ and acts as a coefficient.

We can immediately verify that associativity (and commutativity) holds for the following triple fusion,
\begin{equation}
	\mathcal{N}_p \otimes \mathcal{N}_{p'} \otimes \mathcal{C} \, .
\end{equation}
More explicitly, performing the fusion in different orders yield the same answer,
\begin{equation}
	(\mathcal{N}_p \otimes \mathcal{N}_{p'}) \otimes \mathcal{C} = \mathcal{A}^{M,pp'(p+p')} \otimes (\mathcal{Z}_M)_0 \otimes \mathcal{N}_{p+p'} = \mathcal{N}_p \otimes (\mathcal{N}_{p'} \otimes \mathcal{C}) \, .
\end{equation}

\paragraph{\boldmath $\mathcal{C} \otimes \mathcal{C}$ fusion.}

The most efficient way to compute this fusion rule is to observe that
\begin{equation}
	\mathcal{C} \otimes \mathcal{C} = (\mathcal{N}_p \otimes \mathcal{N}_{-p}) \otimes \mathcal{C} = \mathcal{N}_p \otimes (\mathcal{N}_{-p} \otimes \mathcal{C}) = \mathcal{N}_p \otimes (\mathcal{Z}_M)_0 \otimes \mathcal{N}_{-p} = (\mathcal{Z}_M)_0 \otimes \mathcal{C} \, .
\end{equation}
Alternatively, one may derive this using a Lagrangian description, i.e.
\begin{equation}
	\int_{M^3} \Big(M \Lambda_1 da_1 + \Lambda_1 \mathcal{B}_2\Big) + \int_{M^3} \Big(M \Lambda_1' da_1' + \Lambda_1' \mathcal{B}_2\Big) \, .
\end{equation}
If we define $\hat{\Lambda}_1 = \Lambda_1 + \Lambda_1'$ and $\hat{\Lambda}_1' = a_1' - a_1$, then we will obtain
\begin{equation}
	\int_{M^3} M \hat{\Lambda}_1' d\Lambda_1' + \int_{M^3} \Big(M \hat{\Lambda}_1 da_1 + \hat{\Lambda}_1 \mathcal{B}_2\Big) \, ,
\end{equation}
which can be identified as $(\mathcal{Z}_M)_0 \otimes \mathcal{C}$. It can be readily checked that $\mathcal{N}_p \otimes \mathcal{C} \otimes \mathcal{C}$ and $\mathcal{C} \otimes \mathcal{C} \otimes \mathcal{C}$ are both commutative and associative.

\paragraph{\boldmath $\mathcal{U}_q(\Sigma^2) \otimes \mathcal{N}_p$ fusion.}

Suppose we define $\mathcal{U}_q$ on a 2-cycle $\Sigma^2 \in H_2(M^3;\mathbb{Z}_M)$, then we may use Poincar\'{e} duality to express
\begin{equation}
	\mathcal{U}_q(\Sigma^2) = \exp\bigg(\frac{2\pi i q}{M} \int_{\Sigma^2} \mathcal{B}_2\bigg) = \exp\bigg(\frac{2\pi i q}{M} \int_{M^3} \bar{\Lambda}_1 \mathcal{B}_2\bigg) = \exp\bigg(2\pi i q \int_{M^3} \Lambda_1 \mathcal{B}_2\bigg) \, ,
\end{equation}
where $\bar{\Lambda}_1 = \text{PD}(\Sigma^2) \in H^1(M^3;\mathbb{Z}_M)$ and $M \Lambda_1 = \bar{\Lambda}_1$. Hence, the worldvolume Lagrangian of $\mathcal{U}_q(\Sigma^2) \otimes \mathcal{N}_p$ can be written as
\begin{align}
	\int_{M^3} \bigg(q \Lambda_1 \mathcal{B}_2 + \frac{pM}{2} \, a_1 da_1 + p a_1 \mathcal{B}_2\bigg) & = \int_{M^3} \bigg(\frac{q^2 p^{-1} M}{2} \, \Lambda_1 d\Lambda_1 + \frac{pM}{2} \, a_1' da_1' + p a_1' \mathcal{B}_2'\bigg)\nonumber\\
	& = \int_{M^3} \bigg(\frac{q^2 p^{-1}}{2} \, \bar{\Lambda}_1 \beta(\bar{\Lambda}_1) + \frac{pM}{2} \, a_1' da_1' + p a_1' \mathcal{B}_2'\bigg) \, ,
\end{align}
where $a_1' = a_1 + q p^{-1} \Lambda$ and $\mathcal{B}_2' = \mathcal{B}_2 - q p^{-1} M d\Lambda_1 = \mathcal{B}_2$. Note that the last two terms can be identified as $\mathcal{A}^{M,p}[\mathcal{B}_2]$. The Bockstein homomorphism is associated with the short exact sequence $0 \to \mathbb{Z}_M \to \mathbb{Z}_{M^2} \to \mathbb{Z}_M \to 0$, and acts explicitly as $\beta(\bar{\Lambda}_1) = d\bar{\Lambda}_1/M$. Suppose we define the shorthand notation,
\begin{equation}
	Q(\Sigma^2) = \int_{M^3} \bar{\Lambda}_1 \beta(\bar{\Lambda}_1) \, ,
\end{equation}
then we may express
\begin{equation}
	\mathcal{U}_q(\Sigma^2) \otimes \mathcal{N}_p = \mathcal{N}_p \otimes \mathcal{U}_q(\Sigma^2) = (-1)^{q^2 p^{-1} Q(\Sigma^2)} \otimes \mathcal{N}_p \, .
\end{equation}
Similarly to before, the fusions $\mathcal{U}_q(\Sigma^2) \otimes \mathcal{U}_{q'}(\Sigma^2) \otimes \mathcal{N}_p$ and $\mathcal{U}_q(\Sigma^2) \otimes \mathcal{N}_p \otimes \mathcal{N}_{p'}$ are both commutative and associative. If $M$ is odd, the phase $(-1)^{Q(\Sigma^2)}$ is always trivial \cite{Choi:2022zal}, so we recover the fusion rule presented in Section \ref{sec:N_U_operators}.

Since the condensation defect $\mathcal{C}$ is essentially a sum of magnetic symmetry defects, i.e.
\begin{equation}
	\mathcal{C} = \frac{1}{|H^0(M^3;\mathbb{Z}_M)|} \bigoplus_{\Sigma^2 \in H_2(M^3;\mathbb{Z}_M)} \mathcal{U}_1(\Sigma^2) \, ,\label{appeq:condensate_defect_composition}
\end{equation}
we find that
\begin{align}
	\frac{1}{|H^0(M^3;\mathbb{Z}_M)|} \bigoplus_{\Sigma^2 \in H_2(M^3;\mathbb{Z}_M)} \mathcal{U}_1(\Sigma^2) \otimes \mathcal{N}_p & = \frac{1}{|H^0(M^3;\mathbb{Z}_M)|} \bigoplus_{\Sigma^2 \in H_2(M^3;\mathbb{Z}_M)} (-1)^{p^{-1} Q(\Sigma^2)} \otimes \mathcal{N}_p\nonumber\\
	& = (\mathcal{Z}_M)_{p^{-1}M} \otimes \mathcal{N}_p\nonumber\\
	& = (\mathcal{Z}_M)_0 \otimes \mathcal{N}_p \, .
\end{align}
This result is consistent with the previously derived fusion rule for $\mathcal{C} \otimes \mathcal{N}_p$.

\paragraph{\boldmath $\mathcal{U}_q(\Sigma^2) \otimes \mathcal{C}$ fusion.}

We can again use \eqref{appeq:condensate_defect_composition} to deduce that
\begin{align}
	\mathcal{U}_q(\Sigma^2) \otimes \mathcal{C} & = \frac{1}{|H^0(M^3;\mathbb{Z}_M)|} \bigoplus_{\widehat{\Sigma}^2 \in H_2(M^3;\mathbb{Z}_M)} \mathcal{U}_q(\Sigma^2) \otimes \mathcal{U}_1(\widehat{\Sigma}^2)\nonumber\\
	& = \frac{1}{|H^0(M^3;\mathbb{Z}_M)|} \bigoplus_{\widetilde{\Sigma}^2 \in H_2(M^3;\mathbb{Z}_M)} \mathcal{U}_1(\widetilde{\Sigma}^2) = \mathcal{C} \, ,
\end{align}
where $\widetilde{\Sigma}^2 = \widehat{\Sigma}^2 + q \Sigma^2 \in H_2(M^3;\mathbb{Z}_M)$. In other words, the effect of such a fusion is merely to relabel the 2-cycles (mod $M$) that are being summed over in the condensation defect. We can again easily verify that this fusion rule is compatible with the previous ones that we have derived.

\paragraph{\boldmath $\mathcal{O}_q \otimes \mathcal{O}_{q'}$ fusion.}

Let us consider first the case where $q' \neq -q \mod M$. The worldvolume Lagrangian of the $\mathcal{O}_q \otimes \mathcal{O}_{q'}$ fusion reads
\begin{equation}
	\int_{W^2} \Big((q+q') \hat{c}_2 + (q\lambda_1 + q'\lambda_1') \mathcal{A}_1 + (q\varphi + q'\varphi') \mathcal{B}_2 - qM \lambda_1 d\varphi - q'M \lambda_1' d\varphi'\Big) \, .
\end{equation}
Rewriting the Lagrangian above in terms of the redefined fields, $ \widetilde{\lambda}_1 = -\lambda_1 + \lambda_1'$, $ \widetilde{\varphi} = -\varphi + \varphi'$, $(q+q') \widetilde{\lambda}_1' = q \lambda_1 + q' \lambda_1'$, $(q+q') \widetilde{\varphi}' = q \varphi + q' \varphi'$,\footnote{Note the resemblance of these field redefinitions to those in the discussion preceding \eqref{appeq:Np_Np_field_redefinition}.} we obtain
\begin{equation}
	-qq'(q+q')^{-1} \int_{W^2} M \widetilde{\lambda}_1 d\widetilde{\varphi} + (q+q') \int_{W^2} \Big(\hat{c}_2 + \widetilde{\lambda}_1' \mathcal{A}_1 + \widetilde{\varphi}' \mathcal{B}_2 - M \widetilde{\lambda}_1' d\widetilde{\varphi}'\Big) \, ,
\end{equation}
Consequently, we may express
\begin{equation}
	\mathcal{O}_q \otimes \mathcal{O}_{q'} = \mathcal{X}^{M,qq'(q+q')^{-1}} \otimes \mathcal{O}_{q+q'} \, .
\end{equation}

On the other hand, if $q'=-q \mod M$, then the Lagrangian simply reduces to
\begin{equation}
	q \int_{W^2} \Big(\widetilde{\lambda}_1 \mathcal{A}_1 + \widetilde{\varphi} \mathcal{B}_2 - M \lambda_1' d\widetilde{\varphi} - M \widetilde{\lambda}_1 d\varphi\Big) \, .
\end{equation}
Integrating out $\lambda_1'$ and $\varphi$ respectively enforces $M \lambda_1' \in H^1(W^2;\mathbb{Z})$ and $M\varphi \in H^0(W^2;\mathbb{Z})$. Applying similar arguments in the discussion following \eqref{eq:Dijkgraaf-Witten_theory_coupled}, we conclude that
\begin{align}
	\mathcal{O}_q \otimes \mathcal{O}_{-q} & = \frac{1}{|H^0(W^2;\mathbb{Z}_M)|} \sum_{\substack{\gamma^1 \in H_1(W^2;\mathbb{Z}_M) \\ \Sigma^2 \in H_2(W^2;\mathbb{Z}_M)}} \exp\bigg(\frac{2\pi i}{M} \int_{\gamma^1} \mathcal{A}_1\bigg) \exp\bigg(\frac{2\pi i}{M} \int_{\Sigma^2} \mathcal{B}_2\bigg)\nonumber\\
	& = \frac{1}{|H^0(W^2;\mathbb{Z}_M)|} \bigoplus_{\substack{\gamma^1 \in H_1(W^2;\mathbb{Z}_{2M}) \\ \Sigma^2 \in H_2(W^2;\mathbb{Z}_M)}} \mathcal{V}_2(\gamma^1) \otimes \mathcal{U}_1(\Sigma^2) \coloneqq \widetilde{\mathcal{C}} \, .\label{appeq:Oq_condensate}
\end{align}
Note that if $W^2$ is connected, then $\Sigma^2 = s W^2$ for $s=1,\dots,M$. The condensation defect $\widetilde{\mathcal{C}}$ is essentially a sum of all the $\widehat{\mathbb{Z}}_{2M}^{(2)}$ and $\widehat{\mathbb{Z}}_M^{(1)}$ defects localized on $W^2$.\footnote{Technically, $\mathcal{V}_p$ is defined on 1-cycles $\gamma^1 \in H_1(W^2;\mathbb{Z}_{2M})$. We assume  that there are no extra torsion elements in $H_1(W^2;\mathbb{Z}_{2M})$ compared to $H_1(W^2;\mathbb{Z}_M)$, so that the sum over $\gamma^1 \in H_1(W^2;\mathbb{Z}_M)$ in \eqref{appeq:Oq_condensate} makes sense.}

\paragraph{\boldmath $\mathcal{O}_q \otimes \widetilde{\mathcal{C}}$ and $\widetilde{\mathcal{C}} \otimes \widetilde{\mathcal{C}}$ fusions.}

The worldvolume Lagrangian corresponding to this fusion is
\begin{equation}
	\int_{W^2} \Big(q \hat{c}_2 + q \lambda_1 \mathcal{A}_1 + q \varphi \mathcal{B}_2 - q M \lambda_1 d\varphi + \widetilde{\lambda}_1 \mathcal{A}_1 + \widetilde{\varphi} \mathcal{B}_2 - M \lambda_1' d\widetilde{\varphi} - M \tilde{a} d\varphi'\Big) \, .
\end{equation}
Defining $\widehat{\lambda}_1 = \lambda_1 + q^{-1} \widetilde{\lambda}_1$, $\widehat{\varphi} = \varphi + q^{-1} \widetilde{\varphi}$, $\widehat{\lambda}_1' = \lambda_1 - \lambda_1' + q^{-1} \widetilde{\lambda}_1$, $\widehat{\varphi}\phantom{\,}' = \varphi - \varphi'$, we can rewrite the Lagrangian above as
\begin{equation}
	\int_{W^2} \Big(M \widetilde{\lambda}_1 d\widehat{\varphi}\phantom{\,}' + M \widehat{\lambda}_1' d\widetilde{\varphi}\Big) + q \int_{W^2} \Big(\hat{c}_2 + \widehat{\lambda}_1 \mathcal{A}_1 + \widehat{\varphi} \mathcal{B}_2 - M \widehat{\lambda}_1 d\widehat{\varphi}\Big) \, .
\end{equation}
Therefore, we find that
\begin{equation}
	\mathcal{O}_q \otimes \widetilde{\mathcal{C}} = \mathcal{X}^{M,-1} \otimes \mathcal{X}^{M,-1} \otimes \mathcal{O}_q \, .\label{appeq:Oq_C_fusion}
\end{equation}
It also follows immediately from \eqref{appeq:Oq_condensate} and \eqref{appeq:Oq_C_fusion} that
\begin{equation}
	\widetilde{\mathcal{C}} \otimes \widetilde{\mathcal{C}} = \mathcal{X}^{M,-1} \otimes \mathcal{X}^{M,-1} \otimes \widetilde{\mathcal{C}}_q \, .
\end{equation}

\paragraph{\boldmath $\mathcal{V}_p(\gamma^1) \otimes \mathcal{O}_q$, $\mathcal{U}_q(\Sigma^2) \otimes \mathcal{O}_{q'}$, $\mathcal{V}_p(\gamma^1) \otimes \widetilde{\mathcal{C}}$, and $\mathcal{U}_q(\Sigma^2) \otimes \widetilde{\mathcal{C}}$ fusions.}

Analogously to the earlier derivation for $\mathcal{U}_q(\Sigma^2) \otimes \mathcal{N}_p$, one can show that
\begin{equation}
	\mathcal{V}_p(\gamma^1) \otimes \mathcal{O}_q = \mathcal{O}_q \, , \qquad \mathcal{U}_q(\Sigma^2) \otimes \mathcal{O}_{q'} = \mathcal{O}_{q'} \, ,
\end{equation}
assuming $\gamma^1 \in H_1(W^2;\mathbb{Z}_{2M})$ and $\Sigma^2 \in H_2(W^2;\mathbb{Z}_M)$. We stress again that here we are only considering parallel fusion, in which case $\Sigma^2$ and $W^2$ do not link, thus there are no nontrivial commutation relations involved. Using \eqref{appeq:Oq_condensate}, we also have
\begin{equation}
	\mathcal{V}_p(\gamma^1) \otimes \widetilde{\mathcal{C}} = \widetilde{\mathcal{C}} \, , \qquad \mathcal{U}_q(\Sigma^2) \otimes \widetilde{\mathcal{C}} = \widetilde{\mathcal{C}} \, .
\end{equation}





\bibliographystyle{./ytphys}
\bibliography{./refs}

\end{document}